\documentclass[a4paper,11pt]{article}
\pdfoutput=1
\usepackage[a4paper,top=3cm,bottom=2.6cm,left=2.6cm,right=2.6cm]{geometry}
\usepackage{amsmath,amssymb,amsfonts,mathtools,bm}
\usepackage{graphicx,booktabs,array,xcolor,enumitem}
\usepackage{tikz}
\usetikzlibrary{arrows.meta,decorations.pathmorphing,calc,patterns}
\usepackage[font=small,labelfont=bf]{caption}
\usepackage{microtype}
\usepackage[section]{placeins}
\usepackage[colorlinks=true,linkcolor=blue!50!black,citecolor=green!40!black,urlcolor=red!50!black]{hyperref}
\hypersetup{pdftitle={The 't Hooft model on (A)dS2},pdfauthor={Panos Betzios}}
\usepackage{orcidlink}
\usepackage{needspace}
\usepackage[capitalise,nameinlink]{cleveref}
\numberwithin{equation}{section}
\setlist{itemsep=2pt,topsep=4pt}
\makeatletter
\renewcommand\section{\@startsection{section}{1}{\z@}{-3.5ex \@plus -1ex \@minus -.2ex}{2.3ex \@plus.2ex}{\normalfont\Large\bfseries}}
\renewcommand\subsection{\@startsection{subsection}{2}{\z@}{-3.25ex\@plus -1ex \@minus -.2ex}{1.5ex \@plus .2ex}{\normalfont\large\bfseries}}
\makeatother

\newcommand{\gs}{\mathsf{g}}            
\newcommand{\ms}{\mathsf{m}}            
\newcommand{\gb}{\bar g}                

\newcommand{\tr}{\operatorname{tr}}
\newcommand{\PV}{\mathrm{P}\!\!\int}
\newcommand{\cP}{\mathcal{P}}
\newcommand{\cD}{\mathcal{D}}
\newcommand{\cK}{\mathcal{K}}
\newcommand{\cC}{\mathcal{C}}
\newcommand{\cL}{\mathcal{L}}
\newcommand{\cT}{\mathcal{T}}
\newcommand{\Htf}{H_{\rm tH}}
\newcommand{\dd}{\mathrm{d}}
\newcommand{\ii}{\mathrm{i}}

\newcommand{\SLtwo}{SL(2,\mathbb{R})}
\newcommand{\half}{\tfrac12}
\newcommand{\vev}[1]{\langle #1\rangle}
\newcommand{\BS}{\mathsf{K}}             
\newcommand{\Par}{\mathsf{P}}            
\newcommand{\pmax}{p_{\max}}

\begin{document}

\thispagestyle{empty}
\begin{flushleft}
\vspace*{0.4cm}
{\LARGE\sffamily\bfseries The 't~Hooft model on (A)dS$_2$\par}
\vspace{0.9cm}
{\large\sffamily\bfseries
Panos Betzios\,\orcidlink{0000-0002-5350-9404}\par}
\vspace{0.5cm}
{\small\itshape
\href{https://www.ugent.be/we/physics-astronomy/en}{Department of Physics and Astronomy}, Ghent University,\\
Krijgslaan, 281-S9, 9000 Gent, Belgium\par}
\vspace{0.4cm}
{\small\textit{E-mail:} \href{mailto:panos.betzios@ugent.be}{\texttt{panos.betzios@ugent.be}}\par}
\end{flushleft}

\vspace{0.7cm}
\noindent\textsc{Abstract:} We solve large-$N$ two-dimensional QCD with fundamental quarks---the 't~Hooft model---on rigid AdS$_2$ and dS$_2$. Their isometries preserve the null gauge $A_-=0$ and in affine coordinates the curvature enters the light-front Hamiltonian only through the special-conformal generator of the null line. The $\SLtwo$ Casimir then results in an exact meson equation: 't~Hooft's equation plus a Legendre operator whose sign is determined by the curvature. On AdS$_2$ it is related to equations proposed in the context of folded/segmented strings and light-front holography. The pion is massless in the chiral limit and obeys a Gell-Mann--Oakes--Renner relation that fixes the chiral condensate, the dimensions cross over from Regge to logarithmic growth once the flux tube exceeds the curvature radius, and the flat-space integrable structure survives. We also solve the problem of a quark interacting with an antiquark placed at the boundary: beyond a critical coupling the string pulls it back to the boundary, resulting in a scattering phase. On dS$_2$ and due to the expansion of space, the mesons are unstable resonances even at $N=\infty$ and eventually dissociate. Their widths are given by Boltzmann factors at the de~Sitter temperature.

\vspace{0.45cm}
\noindent\textsc{Keywords:} Field Theories in Lower Dimensions, $1/N$ Expansion, Confinement, Integrable Field Theories, de Sitter space

\newpage
\hrule
\vspace{-0.3cm}
\tableofcontents
\vspace{0.4cm}
\hrule
\vspace{0.6cm}

\section{Introduction}
\label{sec:intro}

Quantum field theories on rigid curved backgrounds of maximal symmetry are a useful laboratory for strongly coupled dynamics. In anti-de~Sitter space (AdS) the curvature acts as an infrared regulator that preserves as many symmetries as flat space: massive excitations cannot reach the timelike boundary, the spectrum becomes discrete, and the physical data can also be thought of as scaling dimensions and couplings of a conformal theory living on the boundary \cite{Callan:1989em,Paulos:2016fap,Carmi:2018qzm}\footnote{We should emphasize that in this work we are directly studying strongly coupled dynamics of a QFT in \emph{the bulk} and will not concern ourselves with any holographic ramifications.}. In this language one can ask sharp questions about confinement \cite{Aharony:2012jf,Ciccone:2024guw,Ciccone:2025qcd}, about the fate of a mass gap as the radius is varied \cite{Copetti:2023sya}, and about how the flat-space $S$-matrix emerges as $R\to\infty$ \cite{Paulos:2016fap}. De~Sitter space (dS), the other maximally symmetric background, describes an expanding universe with no boundary, no $S$-matrix and no positive conserved energy, and interacting quantum field theory on a rigid de~Sitter background has attracted renewed interest \cite{Sleight:2019hfp,Sleight:2020obc,Hogervorst:2021uvp,DiPietro:2021sjt,Penedones:2023uqc,Loparco:2023rug,Anninos:2024fty,Katsevich:2026kgc}. Strongly coupled models that are at the same time exactly solvable are scarce in both settings, and almost all of them involve scalars or fermions without gauge fields \cite{Carmi:2018qzm,Hogervorst:2021spa,Copetti:2023sya,Giombi:2021cnr}.

The canonical solvable model of confinement is two-dimensional QCD with quarks in the fundamental representation, in the limit of a large number of colours $N$ with $g^2N$ fixed: the 't~Hooft model \cite{tHooft:1974pnl}. Its solution uses a null coordinate $x^+$ as time and the light-cone gauge $A_-=0$. The gluons then have no propagating degrees of freedom and produce only an instantaneous linear potential between colour charges, and at large $N$ the mesons are pure quark--antiquark states. Their masses $\mu_n$ follow from an integral equation for the distribution $\varphi(x)$ of the light-front momentum fraction $x\in[0,1]$ carried by the quark \cite{tHooft:1974pnl,Callan:1975ps,Einhorn:1976uz,Bars:1977ud,Brower:1978wm},
\begin{equation}\label{eq:intro-tHooft}
\mu^2\varphi(x)=\big(m^2-\gb^2\big)\Big(\frac1x+\frac1{1-x}\Big)\varphi(x)-\gb^2\,\PV_0^1\!\dd x'\,\frac{\varphi(x')}{(x-x')^2},\qquad \gb^2=\frac{g^2N}{2\pi} \, .
\end{equation}
The spectrum corresponds to a Regge trajectory, $\mu_n^2\simeq\pi^2\gb^2n$, set by the string tension $T_s=\pi\gb^2/2$ of the flux tube, and about which much is known analytically \cite{Fateev:2009jf,Ambrosino:2023dik,Litvinov:2024riz}. The method of deriving this equation relies on two flat-space symmetries: translations along the light front, which make the light-front momentum conserved, and boosts, which leave the gauge invariant. Neither AdS$_2$ nor dS$_2$ has a null Killing vector, so it is not obvious that a light-cone solution can survive on these backgrounds.

In this paper we settle this question in the affirmative, in both cases.
Our construction involves three steps. (I) Both spaces are conformally flat, and their isometries act on each of the two null coordinates separately, by M\"obius maps, which preserves the null gauge $A_-=0$. (II) We quantize along the null geodesics using their affine parameter $\lambda$. In these coordinates $(y,\lambda)$ the metric is $\dd s^2=\dd y\,\dd\lambda\pm\lambda^2\dd y^2/4R^2$ (here and below, the upper sign is for AdS$_2$ and the lower for dS$_2$ and $R$ is their radius). Light-front evolution is generated by an isometry, and the remaining isometries act kinematically on each null line. (III) Two-dimensional Yang--Mills theory depends on the metric only through the area form \cite{Witten:1991we}, and in the coordinates $(y,\lambda)$ the area form, $\sqrt{-g}\,\dd y\,\dd\lambda=\half\,\dd y\,\dd\lambda$, is the same as in flat light-cone coordinates. The Yang--Mills action, and with it the instantaneous linear Coulomb interaction between the quarks that it produces in the null gauge, is therefore identical to the one of flat space, with $\lambda$ in the role of the flat null coordinate $x^-$ (\cref{sec:gauge}). The curvature enters only through the quark kinetic term, and the exact light-front Hamiltonian is
\begin{equation}\label{eq:intro-H}
H=\Htf\pm\frac{\cK}{4R^2},\qquad \cK=\int\!\dd\lambda\;\psi^\dagger\,\ii\big(\lambda^2\partial_\lambda+\lambda\big)\psi
\end{equation}
Here $\Htf$ is \emph{literally} 't~Hooft's flat-space Hamiltonian, and $\cK$ is the special-conformal generator of the null line, a one-body operator independent of the coupling. The quadratic Casimir of the isometry group $\SLtwo$, whose eigenvalue determines the scaling dimension $\Delta$ of a state, then splits into $\pm R^2$ times the flat invariant-mass operator plus the collinear conformal Casimir of the constituents \cite{Braun:2003rp}, see \eqref{eq:casimir-split} in \cref{sec:casimir}. This is the curved-space replacement of $P^2=M^2$. Two further facts make the reduction exact: the light-front Fock vacuum is invariant under all three isometries, i.e.\ it is the $\SLtwo$-invariant vacuum of AdS$_2$, as simple as in flat space (\cref{sec:vacuum}); and the Casimir commutes with the kinematic generators, so that the centre-of-mass motion decouples. On a quark--antiquark pair it gives the AdS$_2$ 't~Hooft equation, derived in \cref{sec:equation} (eq.~\eqref{eq:AdS-tHooft}),
\begin{equation}\label{eq:intro-AdS}
\Delta(\Delta-1)\,\varphi(x)=\big(\ms^2-\gs\big)\Big(\frac1x+\frac1{1-x}\Big)\varphi(x)-\gs\,\PV_0^1\!\dd x'\,\frac{\varphi(x')}{(x-x')^2}-\frac{\dd}{\dd x}\Big[x(1-x)\frac{\dd\varphi}{\dd x}\Big],
\end{equation}
with $\gs=\gb^2R^2$ and $\ms=mR$ the two dimensionless parameters of the model. This is 't~Hooft's equation with $\mu^2R^2\to\Delta(\Delta-1)$ plus a Legendre operator. On dS$_2$ the Legendre term simply changes sign, and $\Delta(1-\Delta)=\mu^2R^2$ now corresponds to the de~Sitter mass. There the light front is a generator of the bifurcate Killing horizon of a static patch, and the equation is an eigenvalue equation for the Casimir rather than for an energy, which is why it makes sense although de~Sitter space has no conserved positive energy (\cref{sec:ds-halfspace}).

\Needspace{5\baselineskip}
\paragraph{Main results.}
Let us now list our main findings together with the corresponding sections (conventions, details on the collinear Casimir and the numerical methods can be found in the appendices):

\begin{itemize}
\item \emph{AdS$_2$ spectrum} (\cref{sec:equation} and Appendix~\ref{sec:numerics}). Equation~\eqref{eq:intro-AdS} interpolates between the description of two free AdS$_2$ fermions and 't~Hooft's equation in flat space. All dimensions satisfy $\Delta_n\ge1$, with equality only for the massless pion. For a large radius the dimension $\Delta_n$ of the $n$-th meson and its flat-space mass $\mu_n$ are related by $\Delta_n(\Delta_n-1)=R^2\mu_n^2+\vev{\cL}_n+O(R^{-2})$, where the first curvature correction $\vev{\cL}_n$ is the expectation value of the Legendre operator $\cL=-\frac{\dd}{\dd x}x(1-x)\frac{\dd}{\dd x}$ in 't~Hooft's flat-space wavefunction $\varphi_n$ (\cref{sec:limits}). The pion is exactly massless at $m=0$, with $\Delta_\pi=1$: it is the ``tilt'' operator of the axial symmetry, which the boundary condition breaks. It obeys an AdS$_2$ Gell-Mann--Oakes--Renner relation $\Delta_\pi(\Delta_\pi-1)\simeq2\ms\sqrt{1+\pi^2\gs/3}$. The latter identifies the chiral condensate on AdS$_2$ as $\vev{\bar\psi\psi}=-N\sqrt{\gb^2/12+1/4\pi^2R^2}$: the AdS boundary condition breaks the axial symmetry explicitly and contributes a condensate of its own, which adds in quadrature to the flat-space one (\cref{sec:special}). Highly excited mesons cross over at $n\sim\gs$ from the Regge trajectory to $\Delta_n\simeq n+2\gs\log n$, once the flux tube is longer than the curvature radius. For heavy quarks the equation reduces to a Schr\"odinger problem for two particles in the AdS$_2$ harmonic trap, bound by the linear potential (\cref{sec:rapidity}). We solve the equation numerically with two independent methods, that have been checked to reproduce the flat-space spectrum to twelve digits.
\item \emph{Exact structure} (\cref{sec:rapidity} and Appendix~\ref{sec:integrable}). Using relative rapidity variables for the constituents the equation becomes a P\"oschl--Teller problem with a relativistic kinetic term. It defines a Birman--Schwinger operator \cite{Birman:1961,Schwinger:1961zz}, that is trace class \cite{Simon:2005}, resulting in exact sum rules, a Baxter equation and a closed-form large-$n$ spectrum. The integrable structure found by Fateev, Lukyanov and Zamolodchikov (FLZ) in flat space \cite{Fateev:2009jf} survives: we prove that the ratio of the even and odd spectral determinants is a special value of a $Q$-function, and give the spectral sums in closed form as sums over the roots of the endpoint (indicial) equation, which reduce to their known values in the flat space limit. The same argument proves the flat-space ratio relations of \cite{Fateev:2009jf,Litvinov:2024riz} for all quark masses. The curvature also regularizes the spectral problem: the spectral sums converge without regularization, and $\gs=0$ is an exactly solvable point around which perturbation theory is regular.
\item \emph{Boundary conditions} (\cref{sec:bc}). AdS boundary conditions for the quarks become conditions on the endpoint behaviour of $\varphi(x)$ and we discuss their properties. With the 't~Hooft interaction switched off ($\gs=0$) and a mixed boundary condition, the equation reproduces the spectrum of fermion bilinears of the SYK model with $q$-body interactions, in which the elementary fermion has dimension $\Delta=1/q$; this spectrum was obtained from studies of a folded string at imaginary $R^2$ \cite{Vegh:2024fsa,Vegh:2025syk}. In our language it is that of two \emph{non-interacting} AdS$_2$ fermions of mass $\ms=\half-1/q$, with a particular mixed condition between their two possible boundary falloffs (a self-adjoint extension of the free equation).
\item \emph{A quark tied to the boundary} (\cref{sec:anchored}). A quark joined by a flux tube to a static antiquark on the boundary is solved exactly: the string acts as an $\SLtwo$-invariant electric field and only shifts the quark mass, giving $(\Delta-\half)^2=R^2(m^2-\gb^2-T_s^2R^2)$. When the tension exceeds $T_s^2R^2>m^2-\gb^2$, i.e.\ $\ms^2<\gs+\pi^2\gs^2/4$ (always the case for $m<\gb$), the dimension becomes complex, $\Delta=\half\pm\ii\varkappa$ with $\varkappa=R\sqrt{T_s^2R^2-m^2+\gb^2}$, so that the Casimir takes a value in the range of the principal series of $\SLtwo$; the string then pulls the quark back to the boundary, and we compute in closed form the amplitude for a quark sent in from the boundary to return back. Its time delay equals the classical round-trip time, and completing the dynamics with a boundary condition for the returning quark produces a geometric, Efimov-like tower of bound states. This boundary condition breaks the dilatation symmetry of the AdS$_2$ boundary, and the situation is the confining-string counterpart of a charged particle in a supercritical AdS$_2$ electric field, whose boundary dimension also becomes complex \cite{Pioline:2005pf,Faulkner:2009wj,Kaplan:2009kr}.
\item \emph{De~Sitter space} (\cref{sec:desitter}). On dS$_2$ the curvature term, the cosmological horizon and the acceleration, lead to a repulsive potential. Even at $N=\infty$ the mesons are therefore resonances, decaying as the de~Sitter expansion overwhelms attraction and separates quarks and antiquarks (even if at $N=\infty$ the string connecting them cannot break). We define the resonances by analytic continuation of the AdS$_2$ spectrum in $R^2$, which is equivalent to requiring that the constituents only leave through future infinity, and compute them by complex scaling. Their widths are Boltzmann factors at the de~Sitter temperature $T_{\rm dS}=1/2\pi R$, $\Gamma_n\propto e^{-(V_{\rm top}-\mu_n)/T_{\rm dS}}$, where $V_{\rm top}$ is the maximal energy of a static quark--antiquark pair joined by a string in the static patch, i.e.\ the top of the barrier the pair has to cross in order to dissociate. Mesons sufficiently lighter than $V_{\rm top}$ are therefore long-lived. Since $V_{\rm top}\simeq\pi\gb^2R$ at strong coupling and $\mu_n^2\simeq\pi^2\gb^2n$, there are about $\gb^2R^2$ such narrow mesons, those whose size is smaller than the de~Sitter horizon; $V_{\rm top}$ is thus a maximal static-patch energy for confined mesons, although the static-patch Hamiltonian itself has unbounded spectrum. A heavy stable particle in dS$_2$ would carry a principal-series representation, $\Delta=\half+\ii\nu$; the meson resonances sit slightly off this line, with ${\rm Re}\,\Delta_n-\half$ measuring their width. The quark--antiquark sector also contains exact $\Delta=1$ states, in the discrete series, at every coupling and mass; they describe a constituent leaving through future infinity while the other arrives from past infinity, and they are the only discrete-series states that survive the interaction (\cref{sec:ds-remarks}). At $m=0$ the pion is exactly massless at all couplings, but for small non-zero quark mass a light bound pion exists only above a critical string tension $T_s=3/2\pi R^2$. Correspondingly the chiral condensate is reduced by the expansion, $\vev{\bar\psi\psi}=-N\sqrt{\gb^2/12-T_{\rm dS}^2}$, and vanishes when the de~Sitter temperature reaches the flat-space condensate per colour.
\end{itemize}

\paragraph{Relation to previous work.} To our knowledge the planar meson spectrum of QCD$_2$ has not been studied on AdS$_2$ or dS$_2$ before. The Schwinger model has been analysed on AdS$_2$ \cite{Bharadwaj:2026sch,Sadjadi:2000qf} and solved exactly on dS$_2$ \cite{Anninos:2024fty}. Other exactly solvable interacting models on dS$_2$ include the $p$-Schwinger model, whose $p$ degenerate vacua give rise to distinct de~Sitter-invariant states \cite{AguileraDamia:2026pU}, and the Ising field theory deformed by its thermal operator, solved through fermionization \cite{Galati:2026ising}. Massless multiflavour QCD$_2$ at a large number of flavours was treated on a static background of the global AdS$_2$ form in \cite{Sadjadi:2000qq}. Pure gauge theories on rigid AdS$_2$ appear in \cite{Mezei:2017kmw}, free Dirac fields and their boundary conditions in \cite{SerranoBlanco:2022}, and the AdS$_2$ worldsheet of the higher-dimensional Yang--Mills flux tube in AdS in \cite{Gabai:2025flux,Gabai:2026flux}. Closest to our work is the light-cone quantization of a folded string in AdS$_2$ by Vegh \cite{Vegh:2023qsc}, which leads to a generalized 't~Hooft equation that coincides exactly with \eqref{eq:intro-AdS} at the special point $m^2=\gb^2$. Our equation for a general quark mass corresponds in fact to one proposed by Forshaw and Sandapen \cite{Forshaw:2025pion,Forshaw:2025fs} as a phenomenological model of the pion in light-front holography, combining the 't~Hooft kernel with the longitudinal confinement operator of basis light-front quantization \cite{Li:2015zda}. Our derivation identifies this equation as the exact planar meson equation of QCD$_2$ on AdS$_2$; the comparison is discussed in detail in \cref{sec:discussion}. The relation between folded strings and 't~Hooft's equation goes back to \cite{Bardeen:1975gx}, and light-front holographic models are reviewed in \cite{Brodsky:2014yha}. A relation between double-scaled SYK, proposed to describe the de~Sitter static patch, and a strongly coupled 't~Hooft model in the flat-space limit was put forward in \cite{Sekino:2025bsc,Miyashita:2025rpt} and it would be interesting to revisit these works in light of our findings.

\section{Setup}
\label{sec:setup}

\subsection{Geometry and isometries}

\begin{figure}[t]
\centering
\begin{tikzpicture}[scale=1.45,>=Stealth]
  \fill[blue!4] (-1,-2.35) rectangle (1,2.35);
  \fill[blue!13] (1,-2) -- (-1,0) -- (1,2) -- cycle;
  \draw[thick] (-1,-2.35) -- (-1,2.35) node[above,font=\small] {$\sigma=-\frac\pi2$};
  \draw[thick] (1,-2.35) -- (1,2.35) node[above,font=\small] {$\sigma=+\frac\pi2$};
  \draw[dashed,blue!60!black] (1,-2) -- (-1,0) -- (1,2);
  \node[font=\footnotesize,blue!50!black] at (0.55,-0.75) {Poincar\'e};
  \draw[very thick,red!70!black,->] (-1,-0.8) -- (1,1.2);
  \fill[red!70!black] (-0.6,-0.4) circle (0.03);
  \node[font=\footnotesize,red!60!black,anchor=north west] at (-0.55,-0.42) {$\lambda=0$};
  \node[font=\footnotesize,red!60!black,anchor=west] at (1.05,1.2) {$\lambda\to+\infty$};
  \node[font=\footnotesize,red!60!black,anchor=east] at (-1.05,-0.8) {$\lambda\to-\infty$};
  \node[font=\footnotesize,red!60!black,rotate=45] at (-0.05,0.42) {$y={\rm const}$};
  \draw[->] (1.9,-1.9) -- (1.9,-1.3) node[above,font=\footnotesize] {$\tau$};
  \draw[->] (1.9,-1.9) -- (2.5,-1.9) node[right,font=\footnotesize] {$\sigma$};
\end{tikzpicture}
\caption{Global AdS$_2$ (strip) with a Poincar\'e patch (shaded wedge, bounded by the dashed Poincar\'e horizons). The light front is a complete null geodesic $x^+=y={\rm const}$, which runs from one boundary component to the other. Its affine parameter $\lambda=2R^2/z$ ranges over $(0,\infty)$ inside the Poincar\'e patch, with $\lambda\to+\infty$ at the boundary point $x^-=x^+$; it continues through the horizon ($\lambda=0$) to the other boundary at $\lambda\to-\infty$.}
\label{fig:geometry}
\end{figure}
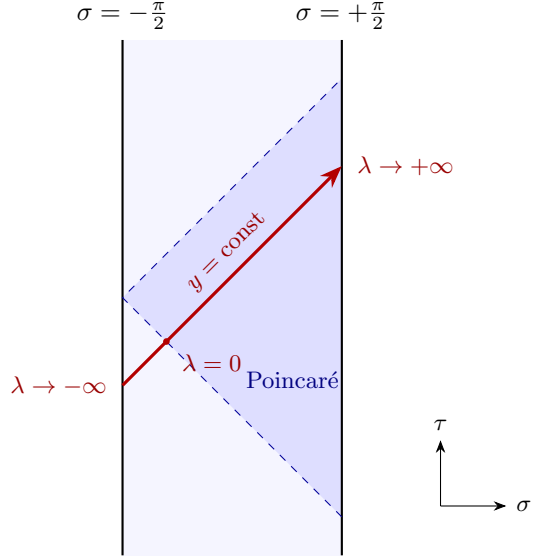

 Two-dimensional anti-de~Sitter space of radius $R$ can be covered by several coordinate systems, and in what follows we shall use two of them. The Poincar\'e patch is
\begin{equation}\label{eq:poincare}
\dd s^2=\frac{R^2}{z^2}\big(\dd t^2-\dd z^2\big)=\Omega^2\,\dd x^+\dd x^-,\qquad
x^\pm=t\pm z,\qquad \Omega=\frac Rz=\frac{2R}{x^+-x^-},
\end{equation}
i.e.\ the region $x^+>x^-$ of a flat plane with conformal factor $\Omega$. The coordinate $z>0$ measures the distance from the conformal boundary at $z=0$, which in the null coordinates is the diagonal $x^+=x^-$; the Poincar\'e horizons, beyond which these coordinates do not extend, lie at $x^-\to-\infty$ and $x^+\to+\infty$. Global AdS$_2$, which covers the whole spacetime, is the strip
\begin{equation}\label{eq:global}
\dd s^2=\frac{R^2}{\cos^2\sigma}\big(\dd\tau^2-\dd\sigma^2\big),\qquad -\tfrac\pi2<\sigma<\tfrac\pi2,
\end{equation}
with two disconnected boundary components at $\sigma=\pm\pi/2$; the Poincar\'e patch covers a wedge of it that touches one of them (\cref{fig:geometry}). The global time $\tau$ is dimensionless, and energies conjugate to it are measured in units of $1/R$.

The Killing vectors in light-cone coordinates are
\begin{equation}\label{eq:killing}
L_{-1}=\partial_++\partial_-,\qquad L_0=x^+\partial_++x^-\partial_-,\qquad L_1=(x^+)^2\partial_++(x^-)^2\partial_-,
\end{equation}
i.e.\ $L_{-1}=\partial_t$, $L_0=t\partial_t+z\partial_z$ and $L_1=(t^2+z^2)\partial_t+2tz\partial_z$, with $[L_m,L_n]=(n-m)L_{m+n}$. Each is of the form $a(x^+)\partial_++a(x^-)\partial_-$ with $a\in\{1,x,x^2\}$. In other words, \emph{the isometry group is the diagonal subgroup $\SLtwo_{\rm diag}\subset\SLtwo_+\times\SLtwo_-$ of the global conformal group of the flat plane}, acting by the same M\"obius map on $x^+$ and on $x^-$. Since $f(a)-f(b)=(a-b)\sqrt{f'(a)f'(b)}$ for a M\"obius map $f$, the conformal factor transforms as a primary with weights $(\half,\half)$, which is the statement that the metric $\Omega^2\dd x^+\dd x^-$ is invariant. None of the Killing vectors is null, so there is no symmetry that translates along a light front. This is the basic obstacle to a naive light-cone treatment, and it is what we overcome in \cref{sec:gauge,sec:symmetry}.

\subsection{Action and the Weyl frame}
\label{sec:action}

The 't~Hooft model on AdS$_2$ is
\begin{equation}\label{eq:action}
S=\int\dd^2x\sqrt{-g}\,\Big[-\frac1{2g^2}\tr F_{\mu\nu}F^{\mu\nu}+\bar\Psi\big(\ii\gamma^ae_a{}^\mu\nabla_\mu-m\big)\Psi\Big],\qquad \nabla_\mu=\partial_\mu+\tfrac14\omega_\mu{}^{ab}\gamma_{ab}-\ii A_\mu ,
\end{equation}
with gauge group $U(N)$ or $SU(N)$, $A_\mu=A_\mu^aT^a$, $\tr T^aT^b=\half\delta^{ab}$, one flavour of quarks $\Psi$ in the fundamental representation, and $e_a{}^\mu$, $\omega_\mu{}^{ab}$ the zweibein and spin connection of the AdS$_2$ metric. We take the 't~Hooft limit, $N\to\infty$ with
\begin{equation}\label{eq:gbar}
\gb^2\equiv\frac{g^2N}{2\pi}
\end{equation}
fixed. In the normalization of 't~Hooft's original paper \cite{tHooft:1974pnl}, $g_{\rm tH}^2=g^2/2$ and $\gb^2=g_{\rm tH}^2N/\pi$, which is the unit of mass squared in which his equation is usually written (Appendix~\ref{app:conventions}). The physical string tension, the energy per unit length of the electric flux between a quark and an antiquark, is $T_s=g^2C_F/2\simeq\pi\gb^2/2$, with $C_F=(N^2-1)/2N\to N/2$ the quadratic Casimir of the fundamental representation. Unequal quark masses can be studied using a trivial generalization, which we indicate where relevant. The model has two dimensionless parameters,
\begin{equation}\label{eq:couplings}
\gs\equiv\gb^2R^2=\frac{g^2NR^2}{2\pi},\qquad \ms\equiv mR .
\end{equation}

It is useful to strip off the conformal factor. If one writes $g_{\mu\nu}=\Omega^2\eta_{\mu\nu}$ and rescales the quark field as $\Psi=\Omega^{-1/2}\Psi_{\rm f}$, the spin connection drops out of the massless Dirac operator, which is Weyl covariant in two dimensions. The Yang--Mills and mass terms are not Weyl invariant, and pick up explicit factors of $\Omega$:
\begin{equation}\label{eq:weyl}
S=\int\dd^2x\,\Big[-\frac{\Omega^{-2}}{2g^2}\tr F_{\mu\nu}F^{\mu\nu}\big|_\eta+\bar\Psi_{\rm f}\big(\ii\gamma^\mu D_\mu-m\,\Omega\big)\Psi_{\rm f}\Big],\qquad D_\mu=\partial_\mu-\ii A_\mu .
\end{equation}
The 't~Hooft model on AdS$_2$ is therefore the 't~Hooft model on the half-plane $z>0$ with position-dependent couplings $1/g^2(z)=z^2/g^2R^2$ and $m(z)=mR/z$, supplemented by boundary conditions at $z=0$. These are also the scalings required for the dilatation $L_0$ to be a symmetry. We refer to \eqref{eq:weyl}, in which the metric is flat and the couplings vary in space, as the Weyl frame action.

\subsection{What it means to solve the model}
\label{sec:solve}

There is no $S$-matrix in AdS: particles cannot be separated to infinity, because the boundary is at finite distance in global time. The physical data are instead the boundary conformal data, the scaling dimensions and operator-product coefficients of the operators of the boundary conformal theory, each of which corresponds to a state of the bulk theory. The states of the bulk theory (each bulk particle, for instance a meson, together with its excitations) organize into representations of the global Hamiltonian
\begin{equation}\label{eq:Hglob}
H_{\rm glob}=\tfrac12\Big(R\,H+\frac KR\Big),
\end{equation}
which generates $\tau$ translations: they form unitary positive-energy (lowest-weight) representations of $\SLtwo$ (the discrete series $D^+_\Delta$)\footnote{In such a representation the spectrum of $H_{\rm glob}$ is bounded from below. We call these representations lowest-weight, as in the mathematics literature, because the weight (the $H_{\rm glob}$ eigenvalue) of the primary state is the lowest one. In the CFT literature the same representations are often called highest-weight representations, the primary being the state annihilated by the operator that lowers the energy. The two names refer to the same objects.}. Each such representation consists of a state of lowest energy $\Delta$, which corresponds to a boundary operator of scaling dimension $\Delta$ (a ``primary''), and a tower of excitations with energies $\Delta+k$, $k=1,2,\dots$, which correspond to its derivatives. Here $H$, $D$ and $K$ are the conserved charges associated with $L_{-1}$, $L_0$ and $L_1$, normalized in \cref{sec:symmetry}, eq.~\eqref{eq:charges}, so that
\begin{equation}\label{eq:sl2}
[H,D]=\ii H,\qquad [K,D]=-\ii K,\qquad [H,K]=2\ii D ,
\end{equation}
the algebra of conformal quantum mechanics \cite{deAlfaro:1976vlx}. With $L_\pm=\half(RH-K/R)\pm\ii D$ one has $[H_{\rm glob},L_\pm]=\mp L_\pm$, so that $L_-$ raises the global energy by one unit ($L_\pm$ should not be confused with the Killing vectors $L_{\pm1}$), and the quadratic Casimir
\begin{equation}\label{eq:casimir}
\cC_2=\tfrac12\big(HK+KH\big)-D^2
\end{equation}
takes the value $\Delta(\Delta-1)$ on the representation with lowest weight $\Delta$. The Casimir commutes with all three generators, so it can be computed in any frame, and this is how we shall determine $\Delta$ without ever diagonalizing $H_{\rm glob}$ directly. At large $N$ the single-meson states are created by colour-singlet quark bilinears, and solving the model at leading order means computing the meson dimensions $\{\Delta_n\}$ as functions of $(\gs,\ms)$. We do this in \cref{sec:equation}. For a free bulk scalar of mass $\mu$ one has $\Delta(\Delta-1)=\mu^2R^2$, and this relation will be recovered for mesons in the flat-space limit.

\section{Null gauge and the affine light-front frame}
\label{sec:gauge}

\subsection{\texorpdfstring{An $\SLtwo$-invariant null gauge}{An SL(2,R)-invariant null gauge}}

We impose
\begin{equation}\label{eq:gauge}
A_-=0 ,
\end{equation}
i.e.\ $n^\mu A_\mu=0$ with $n=\partial_-$ tangent to the null geodesics (light rays) $x^+={\rm const}$. (Null geodesics of a conformally flat metric are those of the flat metric.) The flat-space properties of light-cone gauge survive unchanged in the Weyl frame: $F_{+-}=-\partial_-A_+$ is linear in the gauge field, so there is no gluon self-interaction; $\partial_+A_+$ does not appear in the action, so $A_+$ is a Lagrange multiplier and the gluon exchange is instantaneous in $x^+$; and the residual gauge transformations are $x^-$-independent.

The most important property is that \eqref{eq:gauge} \emph{is invariant under all isometries}. For a Killing vector $\xi=a(x^+)\partial_++a(x^-)\partial_-$,
\begin{equation}
(\mathcal{L}_\xi A)_-=\xi^\mu\partial_\mu A_-+A_-\,a'(x^-)+A_+\,\partial_-a(x^+)=\xi^\mu\partial_\mu A_-+A_-\,a'(x^-),
\end{equation}
which vanishes when $A_-=0$. Gauge-fixed correlators are therefore $\SLtwo$-covariant, and the isometries can be used in the gauge-fixed theory exactly as boosts are used in flat light-cone gauge. This is special to null gauges. For the temporal gauge $A_t=A_++A_-=0$, for instance, one finds $(\mathcal L_{L_1}A)_++(\mathcal L_{L_1}A)_-=2(x^+-x^-)A_+\neq0$, so special conformal transformations do not preserve it and the model becomes much harder to study.

\subsection{Affine null coordinates}
\label{sec:affine}

A null geodesic has no proper length, but it has a preferred parametrization, its affine parameter: the parameter in which the geodesic equation takes its force-free form $\ddot x^\mu+\Gamma^\mu_{\nu\rho}\dot x^\nu\dot x^\rho=0$. It is unique up to the affine maps $\lambda\to a\lambda+b$. For the null lines $x^+=y$ of the metric \eqref{eq:poincare} it is defined by $\dd\lambda=\Omega^2\dd x^-$, and we fix it by
\begin{equation}\label{eq:lambda}
\lambda=\frac{4R^2}{x^+-x^-}=\frac{2R^2}{z},\qquad \Omega=\frac{\lambda}{2R},\qquad \partial_{x^-}\big|_{x^+}=\Omega^2\partial_\lambda .
\end{equation}
In the coordinates $(y,\lambda)$, with $y\equiv x^+$, the metric becomes
\begin{equation}\label{eq:affine-metric}
\dd s^2=\dd y\,\dd\lambda+\frac{\lambda^2}{4R^2}\,\dd y^2,\qquad \sqrt{-g}=\tfrac12 .
\end{equation}
This is AdS$_2$ in coordinates of Eddington--Finkelstein type, adapted to a family of null geodesics. The lines $y={\rm const}$ are affinely parametrized null geodesics ($g_{\lambda\lambda}=0$ and $g_{y\lambda}$ is constant), and the volume form is the flat one. Inside the Poincar\'e patch $\lambda\in(0,\infty)$: $\lambda\to\infty$ at the boundary point $x^-=x^+$ where the geodesic ends, and $\lambda\to0$ on the past Poincar\'e horizon. The geodesic continues into global AdS$_2$. Along a null line of the global metric \eqref{eq:global} the affine parameter is an affine function of $\tan\sigma$, so the complete null geodesic, from one boundary component to the other, is the whole real line $\lambda\in\mathbb R$ (\cref{fig:geometry}). The light front of the AdS$_2$ theory is this complete null line, exactly as the light front of the flat theory is the line $x^+={\rm const}$.

Since $\partial_{x^+}|_{x^-}=\partial_y-\frac{\lambda^2}{4R^2}\partial_\lambda$, the Killing vectors \eqref{eq:killing} become
\begin{equation}\label{eq:killing-affine}
L_{-1}=\partial_y,\qquad L_0=y\,\partial_y-\lambda\,\partial_\lambda,\qquad L_1=y^2\partial_y+\big(4R^2-2y\lambda\big)\partial_\lambda .
\end{equation}
Three facts about \eqref{eq:killing-affine} drive everything that follows. In the rest we call a generator \emph{kinematic} if it does not involve the Hamiltonian, so that its action on states is fixed by the free theory, and \emph{dynamical} otherwise.
\begin{enumerate}[label=(\roman*)]
\item The Poincar\'e time translation $\partial_t$ is $\partial_y$ at fixed $\lambda$. Light-front evolution in the affine frame \emph{is} Poincar\'e time evolution, and its generator is a conserved charge, whereas in the Weyl frame the light-front Hamiltonian depends explicitly on $x^+$ through $\Omega$.
\item The $\lambda$-components of the Killing vectors are at most linear in $\lambda$. On each null line the isometries act by translations and dilatations of the affine parameter, never by special conformal maps of it.
\item On the null line $y=0$ the generators $L_0|_{y=0}=-\lambda\partial_\lambda$ and $L_1|_{y=0}=4R^2\partial_\lambda$ do not involve $\partial_y$. They act \emph{kinematically}, while $L_{-1}$ is the only dynamical generator. This is the AdS$_2$ counterpart of the flat statement that $P_-$ and the boost are kinematic while $P_+$ is dynamical.
\end{enumerate}

\paragraph{The same construction in other frames.} The coordinates $(y,\lambda)$ are Eddington--Finkelstein coordinates adapted to the Poincar\'e patch: one coordinate, $y$, labels the light rays, and the other, $\lambda$, measures affine distance along each ray, normalized so that it depends only on the Poincar\'e radial coordinate $z$. It is this normalization that makes $\partial_y$ at fixed $\lambda$ the Poincar\'e time translation. Nothing singles out the Poincar\'e patch, however, and the same construction can be carried out for any static frame of AdS$_2$. As an example take the global coordinates \eqref{eq:global}, the global null coordinate $u=\tau-\sigma$, and along each ray $u={\rm const}$ the affine parameter normalized to depend on $\sigma$ only, $\lambda_g=2R^2\tan\sigma$. The metric becomes
\begin{equation}\label{eq:global-EF}
\dd s^2=\dd u\,\dd\lambda_g+\Big(R^2+\frac{\lambda_g^2}{4R^2}\Big)\dd u^2 ,
\end{equation}
and $\partial_u$ at fixed $\lambda_g$ is now the global time translation $\partial_\tau$. The two frames describe the same light rays, since $x^+$ is a function of $u$ alone, and on each ray the two affine parameters are related by an affine map, $\lambda_g=a(y)\lambda+b(y)$, i.e.\ by a translation and a rescaling of the ray, which by point (ii) are kinematic operations. Everything that is built on a single light ray is therefore the same in the two frames: the Hilbert space, the vacuum, the flat gauge sector (the area form is flat in any affine null coordinates), the Coulomb interaction, the Casimir and, with it, the meson equation of \cref{sec:equation}. What changes is the operator that moves from one light ray to the next, the light-front Hamiltonian. The constant $R^2$ in \eqref{eq:global-EF} couples to the energy-momentum tensor through the translation generator of the ray, and the resulting Hamiltonian is precisely the global Hamiltonian \eqref{eq:Hglob}, $H_{\rm glob}=\frac12(RH+K/R)$, as one also sees from $\partial_u|_{\lambda_g}=\partial_\tau=\frac12(RL_{-1}+L_1/R)$. In the global frame the light-front energies are thus the global energies $\Delta_n+k$, a discrete set, and the eigenstates are the lowest-weight states constructed in \cref{sec:tHooftAdS} together with their descendants. More generally, rescaling $y$ and $\lambda$ in opposite ways preserves $\dd y\,\dd\lambda$ and the $\lambda^2$ term, so the constant in $g_{yy}$ can always be brought to $0$, $+R^2$ or $-R^2$. These are the three types of Killing vectors of AdS$_2$ (parabolic, elliptic and hyperbolic): the Poincar\'e time, the global time, and the time of the AdS$_2$ black hole (Rindler-AdS$_2$), for which $g_{yy}=\lambda^2/4R^2-R^2$ vanishes at the two horizons $\lambda=\pm2R^2$ crossed by the light ray and the light-front Hamiltonian is $\frac12(RH-K/R)$. The model is solved in all three frames by the same equation; the frames differ only in which combination of the charges $H$, $D$ and $K$ generates the evolution. We use the Poincar\'e frame because there the Hamiltonian is 't~Hooft's plus a single term, and the dilatation $D$ is a manifest symmetry, $[H,D]=\ii H$, which organizes the scattering problems of \cref{sec:anchored}. The null line $y=0$ is, in addition, a Killing horizon of the dilatation $L_0$, which acts on it kinematically; restricted to the half-line $\lambda>0$ the Fock vacuum is thermal with respect to $D$ at inverse temperature $2\pi$, the AdS$_2$ counterpart of the static-patch statement of \cref{sec:ds-halfspace}.

\subsection{The gauge sector is flat in the affine frame}
\label{sec:gaugeflat}

In the gauge \eqref{eq:gauge} one has $A=A_+\dd x^+=A_y\dd y$ with $A_y=A_+$ and $A_\lambda=0$. With the measure $\dd^2x=\half\dd x^+\dd x^-$ and $F_{\mu\nu}F^{\mu\nu}=-8F_{+-}^2$, the gauge and interaction terms of \eqref{eq:weyl} are
\begin{equation}\label{eq:SYM}
S_{\rm YM}+S_{\rm int}=\int\dd x^+\dd x^-\Big[\frac{\Omega^{-2}}{g^2}\big(\partial_-A_+^a\big)^2+A_+^aJ^a\Big]
=\int\dd y\,\dd\lambda\Big[\frac1{g^2}\big(\partial_\lambda A^a_y\big)^2+A^a_y\,j^a\Big],
\end{equation}
where $J^a\,\dd x^-=j^a\,\dd\lambda$ is the colour charge on the null line (defined in \eqref{eq:current} below). The second form is identical to the flat-space light-cone action in the variables $(y,\lambda)$. This is a reflection of a general fact: two-dimensional Yang--Mills theory depends on the metric only through the area form $\sqrt{-g}\,\dd^2x$ \cite{Witten:1991we}, and by \eqref{eq:affine-metric} the coordinates $(y,\lambda)$ are area preserving. Consequently, $F_{\mu\nu}F^{\mu\nu}=-8F_{y\lambda}^2$ in the metric \eqref{eq:affine-metric}, independently of $g_{yy}$. Gauss's law reads $\frac2{g^2}\partial_\lambda^2A_y^a=j^a$. With no electric flux at the ends of the null line (the possible boundary conditions are discussed in \cref{sec:bc}) the solution is $A^a_y=\frac{g^2}{4}\int\dd\lambda'\,|\lambda-\lambda'|\,j^a(\lambda')$, and eliminating $A_y$ produces the instantaneous Coulomb term
\begin{equation}\label{eq:coulomb}
H_{\rm C}=-\frac{g^2}{8}\int\dd\lambda\,\dd\lambda'\;j^a(\lambda)\,|\lambda-\lambda'|\,j^a(\lambda').
\end{equation}
For a colour-singlet quark--antiquark pair this is a linear potential $\frac{g^2C_F}{4}|\lambda_1-\lambda_2|\simeq\frac{\pi\gb^2}{4}|\lambda_1-\lambda_2|$, \emph{linear in the affine separation}. In the Weyl frame the same kernel reads $\Omega(x^-)\Omega(x'^-)|x^--x'^-|$, since $|\lambda-\lambda'|=\Omega\Omega'|x^--x'^-|$.

\subsection{The quark sector}
\label{sec:quarks}

With $\gamma^0=\sigma^1$, $\gamma^1=\ii\sigma^2$ and $\Psi_{\rm f}=(\eta,\chi)^T$, the quark Lagrangian of \eqref{eq:weyl} in the gauge \eqref{eq:gauge} is 
\begin{equation}
\mathcal L_q=\ii\chi^\dagger D_+\chi+\ii\eta^\dagger\partial_-\eta-\frac{m\Omega}{2}\big(\chi^\dagger\eta+\eta^\dagger\chi\big).
\end{equation}
As always on the light front, half of the spinor components are not dynamical: $\eta$ has no $x^+$ derivative and obeys the constraint $\ii\partial_-\eta=\frac{m\Omega}2\chi$. Eliminating it gives the light-front mass term
\begin{equation}
H_m=\frac{m^2}{4}\int\dd x^-\,\chi^\dagger\,\Omega\,\frac1{\ii\partial_-}\,\Omega\,\chi ,
\end{equation}
where $(\ii\partial_-)^{-1}$ has the antisymmetric kernel $\frac1{2\ii}{\rm sgn}(x^--x'^-)$. We now pass to the affine frame and define the null-line field
\begin{equation}\label{eq:psi-affine}
\psi(y,\lambda)\equiv\Omega^{-1}\chi\,,\qquad \{\psi_i(\lambda),\psi_j^\dagger(\lambda')\}=\delta_{ij}\,\delta(\lambda-\lambda'),
\end{equation}
corresponding to a density of weight $\half$ on the null line, with $\psi^\dagger\psi\,\dd\lambda=\chi^\dagger\chi\,\dd x^-$. The colour current is
\begin{equation}\label{eq:current}
j^a(\lambda)=\psi^\dagger T^a\psi .
\end{equation}
Using $\Omega^{-2}\partial_-=\partial_\lambda$ one finds $\Omega(\ii\partial_-)^{-1}\Omega\,\chi=\Omega\,(\ii\partial_\lambda)^{-1}\psi$, so the mass term is also flat:
\begin{equation}
H_m=\frac{m^2}{4}\int\dd\lambda\;\psi^\dagger\,\frac1{\ii\partial_\lambda}\,\psi .
\end{equation}
The only place where the information about the background geometry survives is the symplectic term, the term with the time derivative that fixes the canonical structure. With $\partial_{x^+}|_{x^-}=\partial_y-\frac{\lambda^2}{4R^2}\partial_\lambda$ and $\chi=\Omega\psi$ (note $\partial_y\Omega|_\lambda=0$),
\begin{equation}\label{eq:symplectic}
\int\dd x^-\,\ii\chi^\dagger\partial_{x^+}\chi=\int\dd\lambda\,\ii\psi^\dagger\partial_y\psi-\frac1{4R^2}\int\dd\lambda\;\psi^\dagger\,\ii\big(\lambda^2\partial_\lambda+\lambda\big)\psi .
\end{equation}

\subsection{The exact light-front Hamiltonian}

Collecting the various terms, the Lagrangian on the null line is $\int\dd\lambda\,\ii\psi^\dagger\partial_y\psi-H$ with
\begin{equation}\label{eq:H-exact}
\boxed{\;\begin{aligned}
H&=\Htf+\frac{\cK}{4R^2},\qquad \cK=\int\dd\lambda\,\psi^\dagger\,\ii\big(\lambda^2\partial_\lambda+\lambda\big)\psi ,\\
\Htf&=\frac{m^2}{4}\int\dd\lambda\,\psi^\dagger\frac1{\ii\partial_\lambda}\psi-\frac{g^2}{8}\int\dd\lambda\,\dd\lambda'\,j^a(\lambda)\,|\lambda-\lambda'|\,j^a(\lambda') \, .
\end{aligned}\;}
\end{equation}
(the issue of normal ordering is discussed in \cref{sec:twobody}). This set is exact: no approximation in $1/N$ or in $R$ has been made. $\Htf$ is precisely the light-front Hamiltonian of the flat 't~Hooft model, with $\lambda$ in the role of $x^-$. The operator $\cK$ is hermitian (the term $\lambda$ symmetrizes $\lambda^2\partial_\lambda$), quadratic in the fields and independent of the coupling. As we show in \cref{sec:symmetry}, it is the special-conformal generator of the null line. Since $H$ generates $y$-translations at fixed $\lambda$, by point (i) above it is the Poincar\'e Hamiltonian of AdS$_2$. Equation~\eqref{eq:H-exact} can also be read covariantly: the metric \eqref{eq:affine-metric} is the flat light-cone metric plus $g_{yy}=\lambda^2/4R^2$, and $\cK/4R^2$ is the coupling of this component to $T_{\lambda\lambda}$.

In momentum space, $\psi(\lambda)=\int_0^\infty\frac{\dd k}{2\pi}\big[b(k)e^{-\ii k\lambda}+d^\dagger(k)e^{\ii k\lambda}\big]$, and on single-particle wavefunctions $f(k)$ one has
\begin{equation}\label{eq:momentum-rep}
\ii\partial_\lambda\to k,\qquad \lambda\to-\ii\partial_k,\qquad \ii\big(\lambda^2\partial_\lambda+\lambda\big)\to-\partial_k\,k\,\partial_k .
\end{equation}
Single-particle states thus carry positive light-front momentum $k>0$ conjugate to $\lambda$, exactly as in flat space, and $\cK$ is a positive second-order operator in $k$.

\section{Isometries on the null-line Hilbert space}
\label{sec:symmetry}

The Hamiltonian \eqref{eq:H-exact} generates evolution in $y$. To make use of the AdS$_2$ symmetry we also need the other two isometries as operators on the light-front Hilbert space, the Fock space built on the null line by the modes $b(k)$, $d(k)$ of \eqref{eq:momentum-rep}. In this section we construct these operators, verify that they close into the algebra \eqref{eq:sl2}, identify the vacuum, and derive the decomposition of the Casimir that replaces the flat-space mass-shell condition.

\subsection{Kinematic and dynamical generators}

Under the flow of a Killing vector $\xi=a(y)\partial_y+b(y,\lambda)\partial_\lambda$ of the form \eqref{eq:killing-affine}, the field \eqref{eq:psi-affine} transforms as a density of weight $\half$ in $\lambda$ and as a scalar in $y$,
\begin{equation}
\delta\psi=-\big(a\,\partial_y+b\,\partial_\lambda+\tfrac12\partial_\lambda b\big)\psi .
\end{equation}
This follows from the transformation of the flat-frame component $\chi$, a primary of weight $(0,\half)$, together with $\psi=\Omega^{-1}\chi$. On the null line $y=0$ we trade $\partial_y$ for $-\ii H$ and read off the charges ($\delta\psi=-\ii[Q,\psi]$). It is convenient to introduce the three one-body operators
\begin{equation}\label{eq:collinear}
\cP=\int\dd\lambda\,\psi^\dagger\,\ii\partial_\lambda\,\psi,\qquad
\cD=\int\dd\lambda\,\psi^\dagger\,\ii\big(\lambda\partial_\lambda+\tfrac12\big)\psi,\qquad
\cK=\int\dd\lambda\,\psi^\dagger\,\ii\big(\lambda^2\partial_\lambda+\lambda\big)\psi ,
\end{equation}
which generate translations, dilatations and special conformal transformations of the null line, $\lambda\to(a\lambda+b)/(c\lambda+d)$, on constituents of weight $\half$. They satisfy $[\cD,\cP]=-\ii\cP$, $[\cD,\cK]=\ii\cK$ and $[\cP,\cK]=2\ii\cD$, in this context the algebra of the \emph{collinear conformal group} of the line \cite{Braun:2003rp}. They are quadratic in the fields and independent of the interactions and of $R$: they are purely kinematic. The AdS$_2$ charges are then
\begin{equation}\label{eq:charges}
\boxed{\;H=\Htf+\frac{\cK}{4R^2},\qquad D=-\cD,\qquad K=4R^2\,\cP\;}
\end{equation}
Only $H$ contains the interaction, i.e.\ only $H$ is dynamical. The special conformal generator $K$ of AdS$_2$ is ($4R^2$ times) the total light-front momentum of the constituents, and the AdS$_2$ dilatation is the dilatation of the null line.

The algebra \eqref{eq:sl2} follows from the collinear algebra and two properties of $\Htf$: translation invariance in $\lambda$, $[\cP,\Htf]=0$, and scaling weight one, $[\cD,\Htf]=\ii\Htf$. Both are manifest in \eqref{eq:H-exact}, since $(\ii\partial_\lambda)^{-1}$ and $|\lambda-\lambda'|$ are translation invariant and scale like a length. Explicitly,
\begin{align}
[H,K]&=4R^2[\Htf,\cP]+[\cK,\cP]=-2\ii\cD=2\ii D,\nonumber\\
[H,D]&=-[\Htf,\cD]-\tfrac1{4R^2}[\cK,\cD]=\ii\Htf+\tfrac{\ii}{4R^2}\cK=\ii H,\\
[K,D]&=-4R^2[\cP,\cD]=-\ii K .\nonumber
\end{align}
The radius thus enters only through the single kinematic term $\cK/4R^2$, which disappears as $R\to\infty$, where $\SLtwo$ contracts to the Poincar\'e algebra with $K/4R^2\to P_-$.

\subsection{The vacuum}
\label{sec:vacuum}

In flat space the light-front vacuum is simple for a well-known reason. Every constituent carries positive light-front momentum $k>0$, every term in the Hamiltonian conserves the total momentum $\cP$, and the only state with $\cP=0$ is the Fock vacuum $|0\rangle$, which is annihilated by all $b(k)$ and $d(k)$. Hence no interaction can mix $|0\rangle$ with other states, and the Fock vacuum is the ground state, up to subtleties from modes with exactly $k=0$ (zero modes). On AdS$_2$ there is a new ingredient, the special conformal generator $\cK$, which does not commute with $\cP$. One must therefore ask whether $\cK$ creates particles out of $|0\rangle$. The question matters because the vacuum of a theory on AdS$_2$ should be invariant under all three isometries, $H|0\rangle=D|0\rangle=K|0\rangle=0$\footnote{In the boundary conformal theory it corresponds to the identity operator, with $\Delta=0$.}.

Pair creation from the vacuum can only come from the terms $b^\dagger d^\dagger$ in a bilinear operator. For $\cK$ the coefficient of $b_i^\dagger(k)d_i^\dagger(k')$ follows from the terms $\psi^\dagger \sim b^\dagger(k)e^{\ii k\lambda}$ and $\psi \sim  d^\dagger(k')e^{\ii k'\lambda}$,
\begin{equation}\label{eq:pair-kernel}
\int\dd\lambda\;e^{\ii k\lambda}\,\ii\big(\lambda^2\partial_\lambda+\lambda\big)e^{\ii k'\lambda}=2\pi\Big[k'\,\delta''(k+k')+\delta'(k+k')\Big] \, .
\end{equation}
This is a distribution supported at $k+k'=0$; using $q\,\delta''(q)=-2\delta'(q)$ it can be written as $\pi(k'-k)\delta''(k+k')$, which is odd under the exchange of quark and antiquark. Since $k,k'\ge0$, the support is the single corner $k=k'=0$ of the two-particle momentum space. For $\cP$ and $\cD$ the corresponding kernels, $\propto(k'-k)\delta(k+k')$ and $(k'-k)\delta'(k+k')$, are supported at the same corner but are less singular, and their overlap with any normalizable two-particle state vanishes; only $\cK$, whose kernel contains $\delta''$, is sensitive to the boundary behaviour. The physical meaning is clear: $\cK\propto\lambda^2$ is dominated by large $|\lambda|$, i.e.\ by the neighbourhood of the AdS boundary (\cref{sec:gauge}), and vanishing light-front momentum is the momentum-space counterpart of the presence of the boundary. In other words, whether the corner contributes depends on how the wavefunctions behave at small momenta, and this is precisely what the AdS boundary conditions control (we analyse them in \cref{sec:bc}).

Since $\cK$ is a colour singlet, the pair it creates is a colour-singlet pair, so it suffices to compute its overlap with the colour-singlet two-particle states. These are spanned by the meson states $|\phi\rangle$ of \cref{sec:twobody}, whose wavefunctions have the form \eqref{eq:lowest-weight}, $\phi(k_1,k_2)=P^{\Delta-1}e^{-4RP}\varphi(x)$ with $P=k_1+k_2$ and $x=k_1/P$, together with their descendants, which differ by polynomials in $P$. (This form is derived in \cref{sec:equation}; all that matters here is that the wavefunction vanishes as $P^{\Delta-1}$ at small total momentum.) The overlap is most transparent in position space. The two-body amplitude $\Phi(\lambda_1,\lambda_2)=\int\frac{\dd k_1\dd k_2}{(2\pi)^2}\phi(k_1,k_2)e^{-\ii k_1\lambda_1-\ii k_2\lambda_2}$ is, for the wavefunction \eqref{eq:lowest-weight},
\begin{equation}
\Phi(\lambda_1,\lambda_2)=\frac{\Gamma(\Delta+1)}{(2\pi)^2}\int_0^1\dd x\;\frac{\varphi(x)}{\big[4R+\ii\lambda_2+\ii x(\lambda_1-\lambda_2)\big]^{\Delta+1}} ,
\end{equation}
and the pair-annihilation part of $\cK$ acts at coincident points,
\begin{equation}\label{eq:pair-overlap}
\begin{aligned}
\langle0|\cK|\phi\rangle&\propto\int_{-\infty}^{\infty}\dd\lambda\;\Big[\ii\big(\lambda^2\partial_{\lambda_1}+\lambda\big)\Phi(\lambda_1,\lambda_2)\Big]_{\lambda_1=\lambda_2=\lambda}\\
&=\frac{\Gamma(\Delta+1)}{(2\pi)^2}\int_{-\infty}^{\infty}\dd\lambda\,\Big[\frac{(\Delta+1)\lambda^2\int_0^1x\varphi}{(4R+\ii\lambda)^{\Delta+2}}+\frac{\ii\lambda\int_0^1\varphi}{(4R+\ii\lambda)^{\Delta+1}}\Big].
\end{aligned}
\end{equation}
The integrand is analytic in the lower half $\lambda$-plane, because the wavefunction is supported on positive momenta, and it decays as $\lambda^{-\Delta}$ at large $|\lambda|$, because the wavefunction vanishes as $P^{\Delta-1}$ at small momentum. For $\Delta>1$ the contour can be closed in the lower half-plane and the overlap vanishes. For $\Delta=1$ the integrand decays only as $1/\lambda$, with a coefficient proportional to $\int\varphi-2\int x\varphi$; with a symmetric cutoff the arc at infinity contributes, and
\begin{equation}
\langle0|\cK|\phi\rangle\propto\int_0^1\dd x\,(1-2x)\,\varphi(x)\qquad(\Delta=1),
\end{equation}
which vanishes because the only state with $\Delta=1$ (it exists only at $\ms=0$ and is the massless pion of \cref{sec:limits}) is even under charge conjugation $x\to1-x$. The antisymmetry of the kernel \eqref{eq:pair-kernel} under quark--antiquark exchange is the structural reason: the pair that $\cK$ creates is odd under charge conjugation. We have confirmed these statements by direct numerical integration of \eqref{eq:pair-overlap}. The descendants carry additional powers of $P$, which only improve the decay at large $\lambda$, so the same conclusion holds for them. With the standard boundary conditions (those for which the meson wavefunctions vanish at the endpoints, so that $\Delta\ge1$; see \cref{sec:bc}), therefore, $\cK$ does not create pairs out of the Fock vacuum: $\cP|0\rangle=\cD|0\rangle=\cK|0\rangle=0$.

The interacting part is the flat-space operator $\Htf$, and here the flat-space argument applies: after normal ordering, every term in $\Htf$ conserves $\cP$, so $\Htf|0\rangle=0$ at large $N$, up to the zero-mode subtleties familiar from flat space \cite{Bars:1977ud,Brodsky:1997de}. Hence $H|0\rangle=D|0\rangle=K|0\rangle=0$, and $|0\rangle$ plays the role of the $\SLtwo$-invariant vacuum of global AdS$_2$, with $H_{\rm glob}|0\rangle=0$. The light-front vacuum is as simple on AdS$_2$ as in flat space, up to the well-known freedom in choosing different AdS boundary conditions.

Two comments are in order. First, a trivial light-front vacuum is compatible with a non-vanishing chiral condensate. In flat space the 't~Hooft model has $\langle\bar\psi\psi\rangle\neq0$ at $N=\infty$ \cite{Zhitnitsky:1985um}, and in the light-front formulation this condensate is encoded in the zero modes and, equivalently, in the endpoint behaviour of the wavefunctions. On AdS$_2$ a constant one-point function $\langle\bar\psi\psi\rangle$ is allowed by the isometries, and we expect it to be encoded in the same way; its value is discussed in \cref{sec:special}. Second, on AdS$_2$ the question is further complicated by the fact that the quark boundary conditions break the axial symmetry explicitly, as discussed in \cref{sec:limits}. A complete treatment of the zero modes on AdS$_2$ is beyond the scope of this paper; we list it among the open problems in \cref{sec:discussion}.

\subsection{\texorpdfstring{Single-particle sector: the AdS$_2$ fermion}{Single-particle sector: the AdS2 fermion}}
\label{sec:oneparticle}

Before including interactions, consider a single free quark ($g=0$). On a single-particle wavefunction $f(k)$, \eqref{eq:momentum-rep} gives
\begin{equation}
H=\frac{m^2}{4k}-\frac1{4R^2}\,\partial_kk\,\partial_k,\qquad K=4R^2k,\qquad D=\ii\big(k\partial_k+\tfrac12\big),
\end{equation}
which is a realization of the de~Alfaro--Fubini--Furlan conformal quantum mechanics \cite{deAlfaro:1976vlx}. In terms of $s=4Rk$ the global Hamiltonian \eqref{eq:Hglob} is $H_{\rm glob}=\half\big[\ms^2/s-\partial_ss\,\partial_s+s\big]$, and its lowest-weight states (the states of lowest energy, from which the others follow by acting with $L_-$) are
\begin{equation}
f(k)=k^{\pm\ms}\,e^{-4Rk},\qquad H_{\rm glob}=\tfrac12\pm\ms .
\end{equation}
These are the two admissible quantizations of a Dirac fermion of mass $m$ in AdS$_2$, $\Delta_\psi=\half\pm mR$ \cite{Henningson:1998cd,Mueck:1998iz}: a field in AdS has two possible fall-offs at the boundary, and the choice of which one is allowed to fluctuate is part of the definition of the theory. Normalizability in $L^2(\dd k)$ requires $2(\pm\ms)>-1$, so the ``alternate'' quantization $\Delta_\psi=\half-\ms$ exists exactly for $\ms<\half$, which is the usual unitarity window. The Casimir \eqref{eq:casimir} evaluates to $\cC_2=m^2R^2-\frac14$, and indeed $\Delta_\psi(\Delta_\psi-1)=\ms^2-\frac14$ for both signs. Two lessons follow. First, the light-front construction knows about AdS boundary conditions. Second, it encodes them in the \emph{small-$k$} behaviour of the light-front wavefunction: small light-front momentum means large $|\lambda|$, i.e.\ the vicinity of the boundaries.

\subsection{\texorpdfstring{The AdS$_2$ Casimir replaces $P^2=M^2$}{The AdS2 Casimir replaces the mass-shell condition}}
\label{sec:casimir}

In flat space the mass of a state is the Casimir of the Poincar\'e group, $P^\mu P_\mu=M^2$, which in light-front variables reads $M^2=4P_+P_-$ in our conventions (Appendix~\ref{app:conventions}). The light-front Hamiltonian $P_+$ is then $M^2/4P_-$, and since $P_-$ is conserved, the dynamics reduces to an eigenvalue problem for the invariant-mass operator $\hat M^2=4P_-P_+$, which acts only on the internal wavefunction. This is the structure behind 't~Hooft's equation. On AdS$_2$ there is no $P^2$, but there is the quadratic Casimir $\cC_2$ of $\SLtwo$, eq.~\eqref{eq:casimir}, which plays the same role: it commutes with all three isometries, and its eigenvalue $\Delta(\Delta-1)$ determines the boundary dimension $\Delta$, the quantity we want.

We now evaluate $\cC_2$ on the light-front Hilbert space. Substituting \eqref{eq:charges} into \eqref{eq:casimir},
\begin{equation}
\tfrac12(HK+KH)=2R^2\big(\Htf\cP+\cP\Htf\big)+\tfrac12\big(\cK\cP+\cP\cK\big),\qquad D^2=\cD^2 ,
\end{equation}
and using $[\cP,\Htf]=0$ in the first term we obtain
\begin{equation}\label{eq:casimir-split}
\boxed{\;\cC_2=4R^2\,\cP\,\Htf+\Big[\tfrac12\big(\cK\cP+\cP\cK\big)-\cD^2\Big]=R^2\,\hat M^2_{\rm flat}+\cC_{\rm coll}\;}
\end{equation}
The two pieces have simple meanings.
\begin{itemize}
\item $\hat M^2_{\rm flat}=4\cP\Htf$ is the invariant-mass operator of the \emph{flat} 't~Hooft model: $\cP$ is the total light-front momentum and $\Htf$ the flat light-front Hamiltonian, so this is $4P_-P_+$ written in the variable $\lambda$. Its eigenvalues are 't~Hooft's meson masses $\mu_n^2$.
\item $\cC_{\rm coll}=\half(\cK\cP+\cP\cK)-\cD^2$ is the quadratic Casimir of the collinear conformal algebra \eqref{eq:collinear}. It is built from the kinematic generators only, so it is independent of the coupling and of $R$: it is a group-theoretical operator that acts on the constituents. On a single constituent it is the number $-\frac14$, the Casimir $h(h-1)$ of a representation of weight $h=\half$. On two constituents it is the Casimir of the tensor product of two weight-$\half$ representations, which decomposes into irreducible representations of weight $\ell+1$, $\ell=0,1,2,\dots$ (the familiar decomposition $\half\otimes\half=\bigoplus_{\ell\ge0}(\ell+1)$ of lowest-weight representations); on each of these $\cC_{\rm coll}=\ell(\ell+1)$. This is the operator that governs light-ray operators and evolution kernels in QCD \cite{Braun:2003rp}.
\end{itemize}
Equation \eqref{eq:casimir-split} is the AdS$_2$ replacement of the flat-space mass-shell condition: \emph{the Casimir of a bound state is $R^2$ times its flat invariant mass squared plus the collinear Casimir of its constituents.} This leads to three immediate consequences.

First, the two pieces do not commute with each other, because $\hat M^2_{\rm flat}$ commutes with $\cP$ and $\cD$ but not with $\cK$. They must be diagonalized together, and the meson spectrum is not simply the flat spectrum shifted by group theory. Only in the limit $R\to\infty$, where the first term dominates, do we recover $\Delta(\Delta-1)\simeq R^2\mu_n^2$, the familiar relation between the dimension and the mass of a bulk particle, and only for free massless quarks when $\hat M^2_{\rm flat}=0$, is the spectrum dictated by pure group theory i.e. $\Delta=\ell+1=2\Delta_\psi+\ell$ with $\Delta_\psi=\half$, the two-particle primaries of two free massless fermions.

Second, both pieces commute with $\cP$ and with $\cD$. The total light-front momentum $P$ is therefore a spectator, as in flat space, and the eigenvalue problem for $\cC_2$ involves only the relative motion of the constituents, i.e.\ only the momentum fraction $x$ (\cref{sec:twobody}). This is what reduces the AdS$_2$ problem to a one-dimensional equation.

Third, the whole effect of the curvature is the presence of a single term $\cC_{\rm coll}$, which is exactly computable. On a quark--antiquark pair it becomes the Legendre differential operator in $x$, as we show next.

\section{\texorpdfstring{The AdS$_2$ 't~Hooft equation}{The AdS2 't Hooft equation}}
\label{sec:equation}

\subsection{\texorpdfstring{Two-body reduction at large $N$}{Two-body reduction at large N}}
\label{sec:twobody}

At $N=\infty$ the single-meson states are colour-singlet quark--antiquark states,
\begin{equation}
|\phi\rangle=\frac1{\sqrt N}\int_0^\infty\frac{\dd k_1\,\dd k_2}{(2\pi)^2}\;\phi(k_1,k_2)\,b_i^\dagger(k_1)\,d_i^\dagger(k_2)|0\rangle ,
\end{equation}
and $H$ closes on this sector: there are no dynamical gluons, and at $N=\infty$ the amplitude to create an additional quark pair is suppressed by $1/N$, so the number of constituents is conserved. The flat part $\Htf$ acts exactly as in the flat light-front treatment \cite{tHooft:1974pnl,Bars:1977ud,Brodsky:1997de}, and we should now recall two well-known subtleties. Normal ordering the Coulomb term produces a one-body self-energy of the quark, which shifts $m^2\to m^2-\gb^2$ and contains an infrared-divergent piece, reflecting the infinite energy of an isolated colour charge in a confining theory. That piece cancels against the divergence of the exchange term in a colour singlet, and the finite remainder is 't~Hooft's principal-value prescription for the kernel. The Fourier transform of the linear potential is $|\lambda|\to-2/q^2$, and with $P=k_1+k_2$ and $x=k_1/P$,
\begin{equation}\label{eq:tHooft-operator}
4P\,\Htf\,\phi=\cT\phi,\qquad
\cT\varphi(x)=\big(m^2-\gb^2\big)\Big(\frac1x+\frac1{1-x}\Big)\varphi(x)-\gb^2\,\PV_0^1\dd x'\,\frac{\varphi(x')}{(x-x')^2},
\end{equation}
where $\cT$ acts only on the dependence on $x$. Its eigenvalue problem $\cT\varphi=\mu^2\varphi$ is 't~Hooft's equation \eqref{eq:intro-tHooft}, with $\gb^2$ the unit of mass squared (Appendix~\ref{app:conventions}). Nothing in this step knows about the curvature, because $\Htf$ is the flat operator, and $\cK$ is a one-body kinematic operator without infrared subtleties.

The second ingredient is the collinear Casimir on two weight-$\half$ particles. Using \eqref{eq:momentum-rep}, $\cP=k_1+k_2$, $\cD=-\ii\sum_i(k_i\partial_{k_i}+\half)$ and $\cK=-\sum_i\partial_{k_i}k_i\partial_{k_i}$. In Appendix~\ref{app:casimir} we show that on functions of the form $F(P)\varphi(x)$,
\begin{equation}\label{eq:Ccoll-x}
\cC_{\rm coll}\big[F(P)\varphi(x)\big]=F(P)\;\cL\varphi(x),\qquad \cL\equiv-\frac{\dd}{\dd x}\,x(1-x)\,\frac{\dd}{\dd x},
\end{equation}
which is the Legendre operator in $2x-1$, with eigenvalues $\ell(\ell+1)$ on $P_\ell(2x-1)$. It is independent of $F$, as it must be, since $\cC_{\rm coll}$ commutes with $\cP$ and $\cD$.

\subsection{\texorpdfstring{The 't~Hooft equation on AdS$_2$}{The 't Hooft equation on AdS2}}
\label{sec:tHooftAdS}

Combining \eqref{eq:casimir-split}, \eqref{eq:tHooft-operator} and \eqref{eq:Ccoll-x}, the Casimir acts on the internal wavefunction as $\cC=R^2\cT+\cL$, where $\cC$ denotes the action of $\cC_2$ on the internal wavefunction, and the meson dimensions are determined by the following integrodifferential equation
\begin{equation}\label{eq:AdS-tHooft}
\boxed{\;\Delta(\Delta-1)\,\varphi(x)=\big(\ms^2-\gs\big)\Big(\frac1x+\frac1{1-x}\Big)\varphi(x)-\gs\,\PV_0^1\dd x'\,\frac{\varphi(x')}{(x-x')^2}-\frac{\dd}{\dd x}\Big[x(1-x)\,\frac{\dd\varphi}{\dd x}\Big]\;}
\end{equation}
with $\gs=g^2NR^2/2\pi$ and $\ms=mR$. For unequal masses, $(\ms^2-\gs)(\frac1x+\frac1{1-x})\to\frac{\ms_1^2-\gs}{x}+\frac{\ms_2^2-\gs}{1-x}$; in flat space the spectrum of the corresponding two-flavour 't~Hooft equation has been studied analytically with the methods of FLZ in \cite{Artemev:2025cev}. We call \eqref{eq:AdS-tHooft} the AdS$_2$ 't~Hooft equation. Its eigenvalues $c_n$ determine the dimensions through
\begin{equation}
\Delta_n=\tfrac12+\sqrt{\tfrac14+c_n}\, .
\end{equation}

\paragraph{Centre of mass.} The internal wavefunction $\varphi_n(x)$ determines the dimension, but the full state also has a centre-of-mass wavefunction $F(P)$, which is fixed by the requirement that the state be the lowest-weight state of its $\SLtwo$ representation. For an eigenfunction $\varphi_n$, the full lowest-weight state is
\begin{equation}\label{eq:lowest-weight}
\phi(k_1,k_2)=P^{\Delta_n-1}\,e^{-4RP}\,\varphi_n(x).
\end{equation}
On functions $F(P)\varphi_n(x)$ the charges act as $H=\frac1{4R^2}\big[-(P\partial_P^2+2\partial_P)+c_n/P\big]$, $K=4R^2P$ and $D=\ii(P\partial_P+1)$, and the global Hamiltonian in $s=4RP$ is $H_{\rm glob}=\half\big[-(s\partial_s^2+2\partial_s)+c_n/s+s\big]$. The function $s^{a}e^{-s}$ is an eigenfunction with eigenvalue $a+1$ provided $a(a+1)=c_n$. The two-particle measure is $\dd k_1\dd k_2=P\,\dd P\,\dd x$, so normalizability requires $\Delta>0$. For $c_n\ge0$ this selects the root $\Delta_n=\half+\sqrt{\frac14+c_n}$. The centre-of-mass motion is the ground state of conformal quantum mechanics \cite{deAlfaro:1976vlx}, and the descendants $L_-^k|\phi\rangle$ give the towers $\Delta_n+k$. In position space along the null line the centre-of-mass factor is $(4R+\ii\Lambda)^{-\Delta_n}$, with $\Lambda$ the centre-of-mass coordinate. It is analytic in the lower half $\Lambda$-plane and supported on the complete null geodesic, including the part beyond the Poincar\'e horizon.

\subsection{General properties of the equation}
\label{sec:properties}

\paragraph{Positivity.} Using $\PV_0^1\dd x'\,(x-x')^{-2}=-\frac1x-\frac1{1-x}$, one finds that the expectation value of the operator $\cC$ is the quadratic form
\begin{equation}\label{eq:form}
\langle\varphi,\cC\varphi\rangle=\ms^2\!\int_0^1\!\dd x\,\frac{|\varphi|^2}{x(1-x)}+\frac{\gs}{2}\int_0^1\!\!\int_0^1\!\dd x\,\dd x'\,\frac{|\varphi(x)-\varphi(x')|^2}{(x-x')^2}+\int_0^1\!\dd x\,x(1-x)\,|\varphi'|^2 ,
\end{equation}
for wavefunctions vanishing at the endpoints as $x^\beta$, $\beta>0$. All three terms are non-negative. Hence $c_n\ge0$ and $\Delta_n\ge1$, with equality only for $\ms=0$ and $\varphi={\rm const}$. The second term shows the confining nature of the interaction: the kernel is the $x$-space form of $\pi|k|$. The problem is symmetric under charge conjugation corresponding to the map $x\to1-x$, and the eigenfunctions have parity $(-1)^n$. That they are even or odd follows because every term of \eqref{eq:AdS-tHooft} is invariant under $x\to1-x$. That the parity alternates with the level, starting with an even ground state, can be seen in three ways. It holds exactly at $\gs=0$, where the eigenfunctions \eqref{eq:free} are Gegenbauer (symmetric Jacobi) polynomials $P_n^{(2\ms,2\ms)}(2x-1)$ of parity $(-1)^n$. For the ground state it holds at any coupling: in the rapidity-momentum representation of \cref{sec:rapidity} the Birman--Schwinger kernel is pointwise positive, so its leading eigenvector is unique and positive (Perron--Frobenius), hence even. Finally, the parity-resolved semiclassical quantization of Appendix~\ref{sec:integrable} shows that the even and odd levels interlace, and we have confirmed the alternation numerically for all couplings and masses we studied.

\paragraph{Endpoint behaviour.} Near the endpoints the equation is dominated by its most singular terms, and the solutions behave as powers. Inserting $\varphi\sim x^\beta$ as $x\to0$ and using $\PV_0^\infty\dd x'\,x'^\beta/(x-x')^2=-\pi\beta\cot(\pi\beta)\,x^{\beta-1}$, the terms of order $x^{\beta-1}$ cancel provided
\begin{equation}\label{eq:indicial}
\boxed{\;\gs\big(\pi\beta\cot\pi\beta-1\big)+\ms^2-\beta^2=0\;}
\end{equation}
The flat-space condition $\pi\beta\cot\pi\beta=1-m^2/\gb^2$ \cite{tHooft:1974pnl} is modified by the term $-\beta^2$, which comes from the Legendre operator. The function $\pi\beta\cot\pi\beta$ is even, so the roots come in pairs $\pm\beta$. The left-hand side decreases monotonically from $\ms^2$ at $\beta=0$ to $-\infty$ at $\beta\to1^-$, so there is a unique root $\beta_0\in[0,1)$. It interpolates between the free value $\beta_0=\ms$ (as $\gs\to0$, for $\ms<1$) and the flat value as $R\to\infty$ at fixed $m^2/\gb^2$. For small $\ms$,
\begin{equation}\label{eq:beta-small}
\beta_0^2=\frac{\ms^2}{1+\pi^2\gs/3}+O(\ms^4).
\end{equation}
The relation between the root $-\beta_0$ and the quark boundary conditions is discussed in \cref{sec:bc}. For $\ms>1$ and small $\gs$ the smallest positive root is close to $1$ rather than to $\ms$. The free solution $[x(1-x)]^{\ms}$ is then recovered because the coefficient of $x^{\beta_0}$ vanishes as $\gs\to0$, while further roots of \eqref{eq:indicial} near the integers and near $\ms$ describe subleading endpoint terms.

\paragraph{An exact sum rule.} When \eqref{eq:AdS-tHooft} is integrated over $x$, the Legendre term gives a boundary term that vanishes for $\beta_0>0$, and the $\gs$-dependent terms cancel identically, because
\[
\int_0^1\dd x\,\PV_0^1\dd x'\,\frac{\varphi(x')}{(x-x')^2}=-\int_0^1\dd x'\,\varphi(x')\Big(\frac1{x'}+\frac1{1-x'}\Big).
\]
Hence, for every eigenfunction with $\beta_0>0$,
\begin{equation}\label{eq:sumrule}
\Delta_n(\Delta_n-1)\int_0^1\dd x\,\varphi_n(x)=\ms^2\int_0^1\dd x\,\frac{\varphi_n(x)}{x(1-x)} .
\end{equation}
This is the AdS$_2$ version of the identity that underlies the Gell-Mann--Oakes--Renner (GMOR) relation \cite{GellMann:1968rz} in the 't~Hooft model \cite{Zhitnitsky:1985um}.

\subsection{Solvable limits}
\label{sec:limits}

\paragraph{Free quarks, $\gs=0$.} Equation \eqref{eq:AdS-tHooft} reduces to a Sturm--Liouville problem whose eigenfunctions are Jacobi polynomials,
\begin{equation}\label{eq:free}
\varphi_n=\big[x(1-x)\big]^{\ms}P_n^{(2\ms,2\ms)}(2x-1),\qquad \Delta_n=1+2\ms+n=2\Delta_\psi+n .
\end{equation}
These are the two-particle primaries (``double-trace'' operators) of two free AdS$_2$ fermions of dimension $\Delta_\psi=\half+\ms$, as expected. Using alternate quantization for one or both quarks gives $[x(1-x)]^{-\ms}$-type solutions and the corresponding dimensions (discussed in \cref{sec:bc}). This operator, with the free mass terms, is precisely the longitudinal Hamiltonian of basis light-front quantization \cite{Li:2015zda}. Here it arises as the AdS$_2$ collinear Casimir plus the free invariant mass.

\paragraph{Weak coupling.} The $\gs$-dependence of $\cC$ is entirely in the second term of \eqref{eq:form}, so at first order we can use the $\gs = 0$ wavefunctions to compute 
\begin{equation}
\Delta_n=1+2\ms+n+\frac{\gs}{2\Delta_n^{(0)}-1}\,\frac{\frac12\iint\dd x\,\dd x'\,\big(\varphi_n^{(0)}(x)-\varphi_n^{(0)}(x')\big)^2/(x-x')^2}{\int\dd x\,\big(\varphi_n^{(0)}\big)^2}+O(\gs^2).
\end{equation}
For $\ms=\half$ and $n=0$, $\varphi_0^{(0)}=\sqrt{x(1-x)}$, the integrals are elementary giving the result
\begin{equation}\label{eq:weak}
\Delta_0=2+\Big(\frac{\pi^2}{4}-2\Big)\gs+O(\gs^2).
\end{equation}
The positivity of the first-order shift reflects the confining interaction. We have checked \eqref{eq:weak} numerically: $\dd c_0/\dd\gs|_{\gs\to0}=1.402196$, compared with $3\pi^2/4-6=1.402203$.

\begin{figure}[htbp]
\centering
\includegraphics[width=\textwidth]{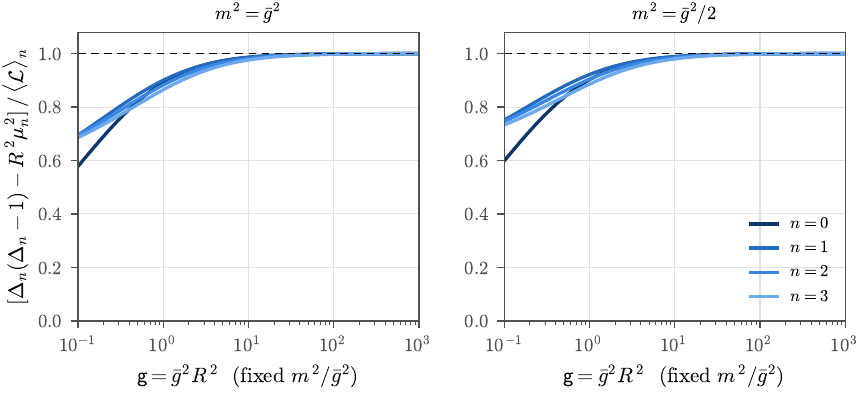}
\caption{Approach to flat space at fixed $m^2/\gb^2$: the curvature correction $\Delta_n(\Delta_n-1)-R^2\mu_n^2$, normalized by the prediction $\vev{\cL}_n$ of \eqref{eq:flat}, for $n=0,\dots,3$. The flat masses $\mu_n^2$ and expectation values $\vev{\cL}_n$ are computed independently from the flat 't~Hooft equation.}
\label{fig:flat}
\end{figure}

\paragraph{Flat-space limit.} As $R\to\infty$ at fixed $m^2/\gb^2$, $\cC=R^2(\cT+\cL/R^2)$, and $\cL/R^2$ can be treated as a perturbation. 
Hence
\begin{equation}\label{eq:flat}
\Delta_n(\Delta_n-1)=R^2\mu_n^2+\vev{\cL}_n+O(R^{-2}),\qquad
\Delta_n=\mu_nR+\frac12+\frac{\frac14+\vev{\cL}_n}{2\mu_nR}+O(R^{-2}),
\end{equation}
where $\mu_n^2$ are 't~Hooft's flat space meson masses and $\vev{\cL}_n=\int x(1-x)\varphi_n'^2/\int\varphi_n^2$ is evaluated in the flat space wavefunctions. The leading term is the standard AdS relation $\Delta(\Delta-1)=\mu^2R^2$ for a bulk particle of mass $\mu$. For $m^2=\gb^2$, for example, $\vev{\cL}_n=0.6802,\,4.7885,\,12.189,\,22.883$ for $n=0,\dots,3$. \Cref{fig:flat} shows the approach to \eqref{eq:flat}.

\begin{figure}[t]
\centering
\begin{minipage}{0.47\textwidth}
\includegraphics[width=\textwidth]{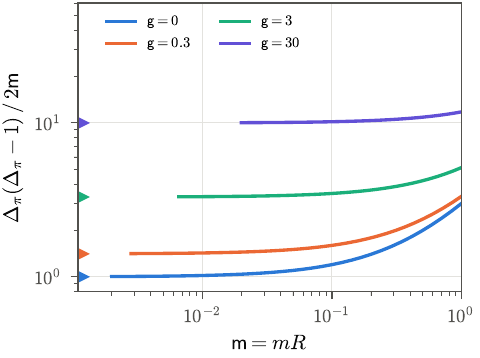}
\end{minipage}\hfill
\begin{minipage}{0.48\textwidth}
\caption{The pion near the chiral limit. As $\ms\to0$, $\Delta_\pi(\Delta_\pi-1)/2\ms$ tends to $\sqrt{1+\pi^2\gs/3}$ (triangles), the AdS$_2$ Gell-Mann--Oakes--Renner relation \eqref{eq:GMOR}. The curve for $\gs=0$ is the exact free result $1+2\ms$. Numerically the ratio to \eqref{eq:GMOR} differs from $1$ by an amount proportional to $\ms$; at $\gs=3$ it is $1.0036$, $1.0011$ and $1.00027$ for $\ms=0.0066$, $0.002$ and $0.0005$.}
\label{fig:pion}
\end{minipage}
\end{figure}

\paragraph{Chiral limit and the pion.} At $\ms=0$ the constant $\varphi=1$ is an exact eigenfunction with $c=0$: $\cL1=0$, and $\cT1=0$ by the identity used above. Hence
\begin{equation}
\Delta_\pi=1\qquad(\ms=0,\ \text{all }\gs).
\end{equation}
This is the counterpart of the massless meson of the flat model. In AdS$_2$ it is a massless bulk scalar with the $\Delta=1$ boundary condition; the other root, $\Delta=0$, is not normalizable. The protected value has a simple explanation. At $N=\infty$ and $m=0$ the axial $U(1)$ is a symmetry of the bulk theory, since its anomaly is suppressed by $1/N$. A reflecting boundary condition for a massless Dirac fermion preserves either the vector or the axial $U(1)$, and gauge invariance requires it to be the vector one, so the AdS boundary conditions break $U(1)_A$ explicitly. A continuous bulk symmetry that is broken only by the boundary condition gives rise to a boundary operator, the so-called ``tilt'' operator\footnote{This is the standard name for the protected boundary (or defect) operator associated with a bulk continuous symmetry that is broken by the boundary \cite{Padayasi:2021sik,Cuomo:2021kfm}. Its dimension equals the dimension of the boundary, here $1$. It should not be confused with the displacement operator, associated with broken translations, or with twist fields, which end monodromy defects.}, which is the boundary value of the bulk symmetry current; its dimension is protected, here $\Delta=1$ because the current is conserved in the bulk. The massless pion is this tilt operator. For small $\ms$, we insert $\varphi\simeq[x(1-x)]^{\beta_0}$ into the sum rule \eqref{eq:sumrule}, which gives $\Delta_\pi(\Delta_\pi-1)\simeq2\ms^2/\beta_0$, and use \eqref{eq:beta-small}:
\begin{equation}\label{eq:GMOR}
\boxed{\;\Delta_\pi(\Delta_\pi-1)=2\ms\sqrt{1+\frac{\pi^2\gs}{3}}+O(\ms^2)\;}
\end{equation}
This AdS$_2$ Gell-Mann--Oakes--Renner relation interpolates between the free value $c=2\ms$, i.e.\ $\Delta_\pi\simeq2\Delta_\psi=1+2\ms$, and 't~Hooft's $\mu_\pi^2=\frac{2\pi}{\sqrt3}\,m\gb$ \cite{tHooft:1974pnl,Zhitnitsky:1985um} as $\gs\to\infty$ with $c=\mu_\pi^2R^2$. It is confirmed numerically in \cref{fig:pion}.

\begin{figure}[htbp]
\centering
\includegraphics[width=\textwidth]{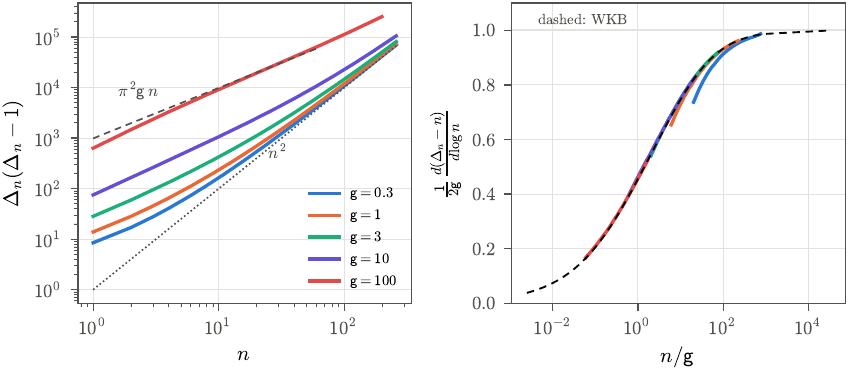}
\caption{Highly excited mesons ($\ms=\half$). Left: $\Delta_n(\Delta_n-1)$ for $\gs=0.3,1,3,10,100$, showing the crossover from 't~Hooft's Regge behaviour $\pi^2\gs n$ (dashed, drawn for $\gs=100$) to $n^2$ (dotted). Right: normalized logarithmic slope of $\Delta_n-n$ against $n/\gs$; the dashed curve is the parameter-free WKB prediction from \eqref{eq:WKB}, which approaches $1$, i.e.\ $\Delta_n\simeq n+2\gs\log n$.}
\label{fig:largen}
\end{figure}

\paragraph{Highly excited states.} For large $n$ the wavefunctions oscillate rapidly, and the operators in \eqref{eq:AdS-tHooft} may be replaced by their symbols, the classical phase-space functions obtained by acting on a local plane wave $\varphi\sim e^{\ii kx}$: the kernel gives $\pi\gs|k|$ and the Legendre term $x(1-x)k^2$. The mass terms affect only the region near the endpoints and contribute $O(1)$ phase shifts. The local dispersion relation $x(1-x)k^2+\pi\gs k=c$, with the Bohr--Sommerfeld condition $\int_0^1k(x)\,\dd x=\pi(n+\delta)$, yields the elementary integral
\begin{equation}
\int_0^1\dd x\,\frac{\sqrt{\pi^2\gs^2+4x(1-x)c}-\pi\gs}{2x(1-x)}=2\pi\gs\big(\vartheta\tan\vartheta+\log\cos\vartheta\big),\qquad \tan\vartheta=\frac{\sqrt c}{\pi\gs}.
\end{equation}
The WKB spectrum is therefore given parametrically by
\begin{equation}\label{eq:WKB}
\boxed{\;\sqrt{c_n}=\pi\gs\tan\vartheta_n,\qquad n+\delta=2\gs\big(\vartheta_n\tan\vartheta_n+\log\cos\vartheta_n\big)\;}
\end{equation}
with an $O(1)$ offset $\delta$. For $n\ll\gs$ ($\vartheta\to0$) it reproduces 't~Hooft's Regge trajectory $c_n\simeq\pi^2\gs\,n$, i.e.\ $\mu_n^2\simeq\pi^2\gb^2n$. For $n\gg\gs$ ($\vartheta\to\pi/2$),
\begin{equation}\label{eq:largen}
\Delta_n=n+2\gs\Big(\log\frac{n}{\pi\gs}+1\Big)+\delta'+o(1),
\end{equation}
with $\delta'$ independent of $n$. This is a logarithmic growth of $\Delta_n-2\Delta_\psi-n$ reminiscent of anomalous dimensions at large spin. The crossover at $n\sim\gs$ occurs when $\Delta\sim\pi\gs$, i.e.\ when the flat-space length of the flux tube, $M/T_s\sim2R$, reaches the curvature scale. Mesons larger than $R$ no longer see a linear potential. \Cref{eq:WKB} also predicts that the local slope $\frac{1}{2\gs}\frac{\dd(\Delta_n-n)}{\dd\log n}=\frac{n}{2\gs}\big(\frac{\pi}{2\vartheta_n}-1\big)$ is a universal function of $n/\gs$. The numerical spectra collapse onto this curve (see \cref{fig:largen}).

\subsection{Special values of the quark mass}
\label{sec:special}

Several values of the quark mass play a special role, in flat space and on AdS$_2$, and it is useful to collect them here together with what changes on a curved background. The most important one is the chiral limit.

\paragraph{Massless quarks: Goldstone bosons, the anomaly and the chiral condensate.} In flat space the chiral limit $m=0$ is special for a reason that goes beyond the form of the equation. The action of massless quarks is invariant under the axial rotations $\psi\to e^{\ii\alpha\gamma^5}\psi$, which rotate the mass operator $\bar\psi\psi$ into $\bar\psi\ii\gamma^5\psi$. For gauge group $SU(N)$ this $U(1)_A$ symmetry is exact in the quantum theory. For $U(N)$ the abelian factor makes it anomalous: the $U(1)$ gauge boson acquires the Schwinger mass $\mu^2=g^2/2\pi=\gb^2/N$, and so does the singlet pseudoscalar, the two-dimensional analogue of the $\eta'$. In the planar limit the two gauge groups are therefore indistinguishable, and one may think of $SU(N)$, or of $U(N)$ at $N=\infty$. At $N=\infty$ the axial symmetry is spontaneously broken. Zhitnitsky \cite{Zhitnitsky:1985um} determined the chiral condensate to be
\begin{equation}\label{eq:condensate-flat}
\vev{\bar\psi\psi}_{\rm flat}=-\frac{N\gb}{\sqrt{12}}\qquad(m\to0),
\end{equation}
and the massless meson with wavefunction $\varphi=1$ is the corresponding Goldstone boson. There is no contradiction with Coleman's theorem \cite{Coleman:1973ci}, which forbids Goldstone bosons in two dimensions: at $N=\infty$ the would-be Goldstone boson is a free field, and at finite $N$ the breaking is only ``almost'' spontaneous, with correlators of $\bar\psi\psi$ that decay as powers of the distance rather than approaching a constant \cite{Witten:1978qu}. The small-mass behaviour $\mu_\pi^2=\frac{2\pi}{\sqrt3}m\gb$ is the Gell-Mann--Oakes--Renner relation $f_\pi^2\mu_\pi^2=-4m\vev{\bar\psi\psi}$ with the pion decay constant $f_\pi^2=N/\pi$: since the chiral-limit wavefunction $\varphi=1$ is also that of the lowest free two-fermion state, the pion couples to the axial current exactly as free fermions do, and $f_\pi$ takes its free-fermion value.

On AdS$_2$ the chiral limit is special in a different way. The massless pion survives, $\Delta_\pi=1$ for all $\gs$, but it is not a Goldstone boson, because the symmetry it would be the Goldstone boson of is broken explicitly: as explained in \cref{sec:limits}, a reflecting boundary condition for a massless fermion preserves either the vector or the axial symmetry but not both, and gauge invariance selects the vector one. The pion is instead the ``tilt'' operator of the broken bulk symmetry, the boundary value of the axial current, and $\Delta_\pi=1$ is protected by the conservation of this current in the bulk. Two consequences follow. First, since the breaking is explicit, a chiral condensate exists at any $N$, not only at $N=\infty$: already free massless fermions on AdS$_2$ have
\begin{equation}\label{eq:condensate-free}
\vev{\bar\psi\psi}_{\rm free}=-\frac{N}{2\pi R},
\end{equation}
a condensate induced by the boundary condition, constant by the isometries.\footnote{By the Weyl map of \cref{sec:action} the massless fermion on AdS$_2$ is a massless fermion on the half-plane $z>0$ with a boundary condition that preserves the vector current, for which the method of images gives $\vev{\bar\psi_{\rm f}\psi_{\rm f}}=-N/2\pi z$, the one-point function of a dimension-one operator in a boundary conformal field theory. Since $\bar\Psi\Psi=\Omega^{-1}\bar\Psi_{\rm f}\Psi_{\rm f}$ with $\Omega=R/z$, the condensate on AdS$_2$ is $-N/2\pi R$. Its phase is fixed by the boundary condition, here the one continuously connected to the standard quantization at $m>0$; an axial rotation of the boundary condition rotates it into $\vev{\bar\psi\ii\gamma^5\psi}$.} Second, the GMOR relation acquires a curvature term. The chiral-limit wavefunction is $\varphi=1$ for all $\gs$, so the pion couples to the axial current as free fermions do also on AdS$_2$, and $f_\pi^2=N/\pi$. Reading \eqref{eq:GMOR} as a GMOR relation, $f_\pi^2\,c_\pi/R^2=-4m\vev{\bar\psi\psi}$, then identifies the chiral condensate on AdS$_2$ as
\begin{equation}\label{eq:condensate}
\boxed{\;\vev{\bar\psi\psi}_{{\rm AdS}_2}=-\frac{N}{2\pi R}\sqrt{1+\frac{\pi^2\gs}{3}}=-N\sqrt{\frac{\gb^2}{12}+\frac{1}{4\pi^2R^2}}\;}
\end{equation}
The curvature adds in quadrature to the flat-space condensate: the boundary condition acts as a second source of chiral symmetry breaking, of strength $1/2\pi R$ per colour, and the two sources combine like orthogonal components. Two checks are exact. At $\gs=0$, \eqref{eq:condensate} reproduces the free-fermion condensate \eqref{eq:condensate-free}, which was computed independently by the method of images; as $R\to\infty$ it reproduces Zhitnitsky's value \eqref{eq:condensate-flat}. In flat space the GMOR relation follows from the axial Ward identity and the dominance of the pion pole. On AdS$_2$ the Ward identity contains a boundary term, since axial charge leaks through the boundary, and we have not derived the AdS$_2$ GMOR relation from it; for intermediate couplings \eqref{eq:condensate} is therefore a conjecture, supported by its agreement at both ends. A direct computation of the condensate at finite $\gs$, which in the light-front formalism involves the zero modes of \cref{sec:vacuum}, would test it. Finally, for gauge group $U(N)$ the anomaly lifts the singlet pion to $\Delta_\pi(\Delta_\pi-1)\simeq\gs/N$, the AdS$_2$ version of the Schwinger mass, which is negligible in the planar limit.

\paragraph{The point $m=\gb$.} At $m^2=\gb^2$ the mass term of the equation cancels: the quark mass squared renormalized by the planar self-energy, $m^2-\gb^2$, vanishes. In flat space this is the point at which 't~Hooft's equation takes its simplest form, with endpoint exponent $\beta_0=\half$; it is the point used in most of the analytic work on the model \cite{Fateev:2009jf}, and the one at which the folded string reproduces 't~Hooft's equation \cite{Bardeen:1975gx,Vegh:2023qsc}. On AdS$_2$ the mass term still cancels at $\ms^2=\gs$, and the equation coincides with Vegh's string equation (\cref{sec:discussion}), but the endpoint exponent is no longer $\half$: by \eqref{eq:indicial} it solves $\pi\gs\cot\pi\beta_0=\beta_0$, which approaches $\half$ only as $\gs\to\infty$. Light quarks, $m<\gb$, have a negative renormalized mass squared, so that the endpoint term of the equation is attractive. In this case a quark interacting with an antiquark placed at the boundary is always pulled back to the boundary by the string connecting them (\cref{sec:anchored}), and on dS$_2$ the same combination sets the height $V_{\rm top}=2\sqrt{m^2-\gb^2+T_s^2R^2}$ of a dissociation barrier (\cref{sec:ds-mesons}).

\paragraph{The point $\ms^2=\gs+\frac14$.} Here the endpoint exponent on AdS$_2$ is $\beta_0=\half$, so the wavefunctions behave as $\sqrt{x(1-x)}$ at the endpoints and the Jacobi basis of Appendix~\ref{sec:numerics} becomes the Chebyshev basis. This is the AdS$_2$ counterpart of the flat point $m=\gb$. It is the edge of the window $\ms^2<\gs+\frac14$ in which the alternate endpoint behaviour $x^{-\beta_0}$ is normalizable (\cref{sec:bc}), the interacting version of the free window $\ms<\half$, and it is also the point at which the alternative ordering of the kinetic term considered in \cite{Forshaw:2025fs} coincides with the gauge-theory equation (\cref{sec:discussion}).

\paragraph{Heavy quarks.} For $\ms\gg1$ the equation reduces to a Schr\"odinger problem for two particles of reduced mass $m/2$ in the AdS$_2$ harmonic trap, bound by the linear potential (\cref{sec:rapidity}). For a quark tied to the boundary the special value is $m^2=\gb^2+T_s^2R^2$: below it the string pulls the quark back to the boundary, above it the AdS$_2$ potential keeps it in the bulk (\cref{sec:anchored}).

\subsection{A collection of numerical results}

We numerically studied eq.~\eqref{eq:AdS-tHooft} by the Rayleigh--Ritz method. More details on the precise numerical implementation can be found in Appendix~\ref{sec:numerics}. Here we list some basic results and properties of the spectrum.

\begin{figure}[htbp]
\centering
\includegraphics[width=\textwidth]{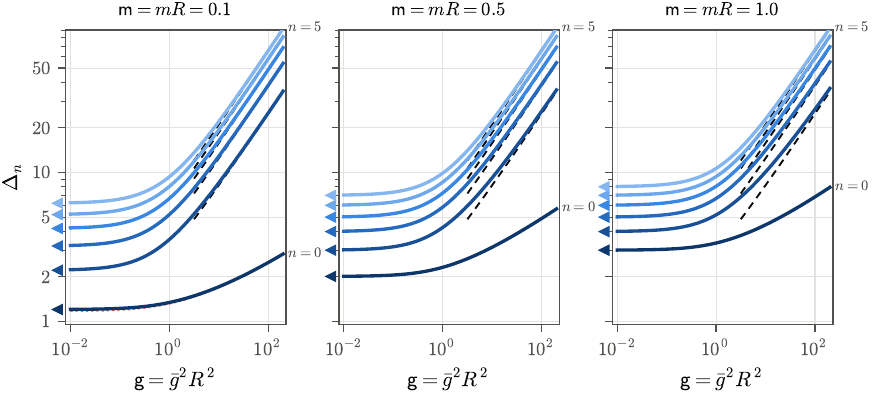}
\caption{Meson dimensions $\Delta_n$, $n=0,\dots,5$ (dark to light), as functions of $\gs=g^2NR^2/2\pi$ for $\ms=mR=0.1,\,0.5,\,1$. Triangles on the left mark the free values $1+2\ms+n$. Dashed lines are the flat-space asymptotics $\half+\sqrt{\frac14+\gs\mu_n^2/\gb^2}$ with the chiral-limit 't~Hooft masses $\mu_n^2/\gb^2=5.882,\,14.14,\,23.08,\,32.30,\,41.68$. In the first panel the dotted curve, which runs close to the $n=0$ curve (at $\ms=0.1$ the numerical $\Delta_0(\Delta_0-1)$ exceeds it by $20\%$ as $\gs\to0$, $9\%$ at $\gs=1$ and $0.7\%$ at $\gs=200$), is the AdS$_2$ GMOR relation \eqref{eq:GMOR}.}
\label{fig:spectrum}
\end{figure}

\Cref{fig:spectrum} shows the six lowest dimensions as functions of $\gs$ for three quark masses, and \cref{tab:spectrum} lists representative values. At weak coupling the dimensions start at the free values $1+2\ms+n$ and grow monotonically with $\gs$. At strong coupling all excited states approach the flat chiral-limit trajectories, $\Delta_n\simeq\half+\sqrt{\frac14+\gs\,\mu_n^2/\gb^2}$. This is expected: $\gs\to\infty$ at fixed $\ms$ is the flat limit at $m^2/\gb^2=\ms^2/\gs\to0$. The ground state instead follows the GMOR curve \eqref{eq:GMOR} for small $\ms$.

\begin{figure}[t]
\centering
\includegraphics[width=\textwidth]{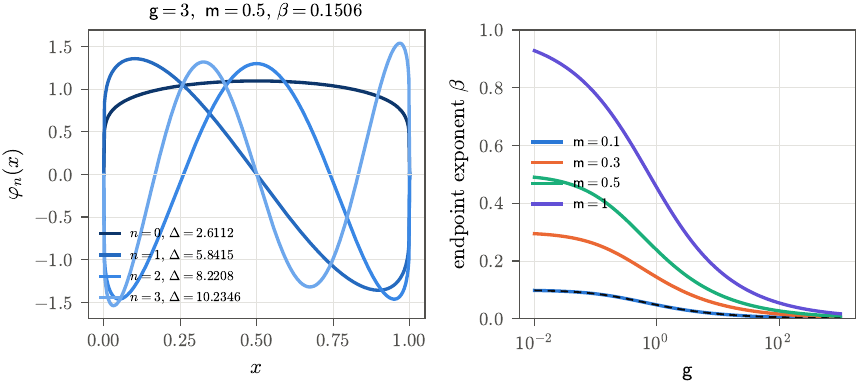}
\caption{Left: normalized wavefunctions of the four lowest mesons at $\gs=3$, $\ms=\half$; the endpoint exponent is $\beta_0=0.1506$. Right: the endpoint exponent $\beta_0$ as a function of $\gs$ for several masses; the dashed curve is the small-mass approximation \eqref{eq:beta-small} for $\ms=0.1$. As $\gs\to0$, $\beta_0\to\ms$ for $\ms<1$, while for $\ms=1$, $\beta_0\simeq1-\sqrt{\gs/2}$.}
\label{fig:wavefunctions}
\end{figure}

\Cref{fig:flat} isolates the first curvature correction in the flat limit. At fixed $m^2/\gb^2$ the quantity $[\Delta_n(\Delta_n-1)-R^2\mu_n^2]/\vev{\cL}_n$ approaches $1$ as $1/\gs$, as predicted by \eqref{eq:flat}. \Cref{fig:largen} shows the spectrum up to $n\simeq250$. The left panel exhibits the crossover from the Regge behaviour $c_n\simeq\pi^2\gs n$ to $c_n\simeq n^2$. In the right panel the logarithmic slope of $\Delta_n-n$, normalized by $2\gs$, collapses for all couplings onto the universal WKB curve \eqref{eq:WKB} as a function of $n/\gs$. \Cref{fig:pion} confirms the AdS$_2$ GMOR relation, and \cref{fig:wavefunctions} shows typical wavefunctions together with the endpoint exponent $\beta_0(\gs,\ms)$.

\begin{table}[t]
\centering\small
\begin{tabular}{rr c ccccc}
\toprule
$\gs$ & $\ms$ & $\beta_0$ & $\Delta_0$ & $\Delta_1$ & $\Delta_2$ & $\Delta_3$ & $\Delta_4$\\
\midrule
0.1 & 0.5 & 0.426172 & 2.043780 & 3.162883 & 4.230717 & 5.281136 & 6.320624\\
1   & 0.5 & 0.237847 & 2.302651 & 4.261117 & 5.836343 & 7.255949 & 8.587429\\
10  & 0.5 & 0.085675 & 3.140729 & 9.180382 & 13.343544 & 16.713016 & 19.633666\\
1   & 0.1 & 0.048253 & 1.337350 & 3.538186 & 5.201859 & 6.666225 & 8.025787\\
10  & 0.1 & 0.017174 & 1.704333 & 8.490494 & 12.803664 & 16.253889 & 19.224401\\
1   & 1   & 0.454288 & 3.351475 & 5.176413 & 6.674908 & 8.053446 & 9.356205\\
10  & 1   & 0.170146 & 4.438249 & 10.044861 & 14.044138 & 17.323342 & 20.185232\\
100 & 1   & 0.054995 & 6.837973 & 26.513220 & 39.500084 & 49.771318 & 58.491776\\
\bottomrule
\end{tabular}
\caption{Lowest meson dimensions. All digits shown are stable under increasing the basis from $N_b=80$ to $120$ (maximal change $4\times10^{-8}$).}
\label{tab:spectrum}
\end{table}

\begin{table}[t]
\centering\small
\begin{tabular}{c cc}
\toprule
$n$ & this work ($R\to\infty$, $N_b=240$) & FLZ \cite{Fateev:2009jf}\\
\midrule
0 & 0.737061746293 & 0.73706174629269\\
1 & 1.753731336917 & 1.7537313369175\\
2 & 2.748160912371 & 2.7481609123706\\
3 & 3.751057581705 & 3.7510575817054\\
4 & 4.749295381038 & 4.7492953810375\\
5 & 5.750492623649 & 5.7504926236487\\
\bottomrule
\end{tabular}
\caption{Flat-space check at $m^2=\gb^2$ ($\alpha=0$ in the conventions of \cite{Fateev:2009jf}): $2\lambda^{\rm FLZ}_n=\mu_n^2/(\pi^2\gb^2)$.}
\label{tab:flat}
\end{table}

\section{Rapidity representation and exact results}
\label{sec:rapidity}

The isometries have been used up in reducing the problem to the single relative coordinate $x$. What remains, \eqref{eq:AdS-tHooft}, is a genuine one-dimensional spectral problem; symmetry alone does not fix its spectrum (the flat mass operator does not commute with $\cK$). In this section we show that in the right variables the problem nevertheless acquires a transparent structure, which yields a number of exact results and makes precise in what sense the AdS$_2$ problem can be thought of as being better behaved than the flat one.

\subsection{Relative rapidity}

We define the relative rapidity $\theta$ of the two lightlike constituents by
\begin{equation}\label{eq:rapidity}
\frac{x}{1-x}=\frac{k_1}{k_2}=e^{2\theta},\qquad x=\tfrac12(1+\tanh\theta),\qquad \theta\in\mathbb R .
\end{equation}
Substituting into \eqref{eq:AdS-tHooft} 
and massaging, the AdS$_2$ 't~Hooft equation becomes exactly
\begin{equation}\label{eq:rapidity-eq}
\boxed{\;-\frac14\,\varphi''(\theta)+\frac{\gs}{2}\int_{\mathbb R}\dd\theta'\,\frac{\varphi(\theta)-\varphi(\theta')}{\sinh^2(\theta-\theta')}+\ms^2\,\varphi(\theta)=\frac{\Delta(\Delta-1)}{4\cosh^2\theta}\,\varphi(\theta)\;}
\end{equation}
In terms of the momentum $p=-\ii\partial_\theta$ conjugate to the rapidity this reads
\begin{equation}\label{eq:Bp}
B(p)\,\varphi=\frac{c}{4}\,{\rm sech}^2\theta\;\varphi,\qquad
B(p)=\frac{p^2}{4}+\gs\,h(p)+\ms^2,\qquad h(p)=\frac{\pi p}{2}\coth\frac{\pi p}{2}-1 ,
\end{equation}
since $\int\dd u\,(1-\cos pu)/\sinh^2u=\pi p\coth(\pi p/2)-2$. The $L^2(\dd x)$ inner product becomes $\int\dd\theta\,\frac{{\rm sech}^2\theta}{2}|\varphi|^2$, and charge conjugation is effected by $\theta\to-\theta$. Several features are immediate:
\begin{enumerate}[label=(\roman*)]
\item \emph{The curvature only adds the term $p^2/4$.} The flat 't~Hooft equation in the same variables is $\big[h(p)+m^2/\gb^2\big]\varphi=\frac{\mu^2}{4\gb^2}{\rm sech}^2\theta\,\varphi$, the form underlying the analytic work of \cite{Fateev:2009jf} and follow-ups. On AdS$_2$ the Legendre (collinear Casimir) term is precisely the free rapidity kinetic energy $p^2/4=-\frac14\partial_\theta^2$: indeed $\cL=-\cosh^2\theta\,\partial_\theta^2$.
\item \emph{At $\gs=0$ the problem is a P\"oschl--Teller well} \cite{Poschl:1933zz}: $-\varphi''-c\,{\rm sech}^2\theta\,\varphi=-4\ms^2\varphi$, with bound states at $c=\ell(\ell+1)$, $\ell=2\ms+n$, and eigenfunctions ${\rm sech}^{2\ms}\theta\,P_n^{(2\ms,2\ms)}(\tanh\theta)$, which is \eqref{eq:free}. The exact solvability of the free problem is the $\mathfrak{so}(2,1)$ potential algebra of the P\"oschl--Teller well, here inherited from the AdS$_2$ isometries.
\item \emph{The 't~Hooft interaction is a relativistic correction to the rapidity kinetic energy.} It is translation invariant in $\theta$, with $\gs h(p)\simeq\frac{\pi\gs}{2}|p|$ at large $|p|$ and $\gs h(p)\simeq\frac{\pi^2\gs}{12}p^2$ at small $p$.
\item \emph{Endpoint exponents are zeros of $B$.} Since $B(2\ii\beta)=\gs(\pi\beta\cot\pi\beta-1)+\ms^2-\beta^2$, the indicial equation \eqref{eq:indicial} is $B(\pm2\ii\beta)=0$. All its roots matter: $\beta_0\in[0,1)$ and one root $\beta_k\in(k,k+1)$ for each $k\ge1$, while $B$ has poles at $p=\pm2\ii k$. Hence
\begin{equation}\label{eq:Bfactor}
\frac{4B(p)}{p^2}=\lim_{K\to\infty}\frac{\prod_{k=0}^{K}\big(1+4\beta_k^2/p^2\big)}{\prod_{k=1}^{K}\big(1+4k^2/p^2\big)} ,
\end{equation}
where both products must be truncated at the same order, since $\beta_k-k\simeq\gs/k$ and neither converges separately.
\item \emph{The sum rule \eqref{eq:sumrule} is the $p=0$ component of \eqref{eq:Bp}}, since $B(0)=\ms^2$.
\end{enumerate}

\subsection{Birman--Schwinger form, trace class and exact sum rules}

With $w(\theta)=\half\,{\rm sech}\,\theta$, \eqref{eq:Bp} is equivalent to $u=c\,\BS u$ for $u=w\varphi$, where
\begin{equation}\label{eq:BS}
\BS(\theta,\theta')=w(\theta)\,G_B(\theta-\theta')\,w(\theta'),\qquad G_B(\theta)=\int\frac{\dd p}{2\pi}\,\frac{e^{\ii p\theta}}{B(p)} .
\end{equation}
This is the Birman--Schwinger operator of the problem \cite{Schwinger:1961zz}: the eigenvalue problem for the ``energy'' $c$ at fixed ``potential'' ${\rm sech}^2\theta$ has been traded for an eigenvalue problem for the coupling, and $\{1/c_n\}$ is the spectrum of $\BS$. For $\ms>0$, $B>0$ and $\BS$ is positive. Because $B(p)\sim p^2/4$, $G_B$ is bounded and $\BS$ is trace class, i.e.\ the sum of its eigenvalues converges absolutely. The spectral (Fredholm) determinant $\det(1-c\BS)=\prod_n(1-c/c_n)$ is then an entire function of $c$ whose zeros are exactly the $c_n$, and its traces give exact sum rules. Using $\int\dd\theta\,e^{\ii q\theta}{\rm sech}^2\theta=\pi q/\sinh(\pi q/2)$ and the parity $(-1)^n$ of the eigenfunctions,
\begin{align}
\sum_{n\ {\rm even}}\frac1{c_n}&=\frac1{4\pi}\int_0^\infty\dd p\;\frac{1+\pi p/\sinh\pi p}{B(p)},\qquad
\sum_{n\ {\rm odd}}\frac1{c_n}=\frac1{4\pi}\int_0^\infty\dd p\;\frac{1-\pi p/\sinh\pi p}{B(p)},\label{eq:sum1}\\
\sum_n\frac1{c_n^2}&=\frac1{4\pi^2}\iint\dd p\,\dd k\;\frac{\widehat W(p-k)^2}{B(p)\,B(k)},\qquad \widehat W(q)=\frac{\pi q}{4\sinh(\pi q/2)} .\label{eq:sum2}
\end{align}
At $\gs=0$ the sum of the two parts of \eqref{eq:sum1} gives $\sum_n1/c_n=1/2\ms$, which is the telescoping sum over $(2\ms+n)(2\ms+n+1)$. In flat space the analogous operator is not trace class, since $h(p)\sim|p|$ makes $\int\dd p/B$ diverge logarithmically, consistently with $\mu_n^2\propto n$; spectral sums there must be regularized \cite{Fateev:2009jf}. On AdS$_2$, $c_n\propto n^2$ and the spectral zeta function $\sum_nc_n^{-s}$ converges for ${\rm Re}\,s>\half$: the curvature regulates the spectral problem as well as the infrared. \Cref{tab:sumrules} compares \eqref{eq:sum1}--\eqref{eq:sum2} with the numerical spectra.

Two corollaries follow.
\begin{itemize}
\item \emph{GMOR from the sum rule.} As $\ms\to0$, $B(p)\simeq\ms^2+\frac14\big(1+\frac{\pi^2\gs}{3}\big)p^2$ near $p=0$. The even sum \eqref{eq:sum1} then diverges as $1/\big(2\ms\sqrt{1+\pi^2\gs/3}\big)$, while the odd sum stays finite. If the excited states remain gapped, the divergence must come from the pion, which gives \eqref{eq:GMOR} independently of the endpoint argument of \cref{sec:limits}.
\item \emph{Chiral limit of the odd sector.} At $\ms=0$,
\begin{equation}
\sum_{n\ \rm odd}\frac1{c_n}\Big|_{\ms=0}=\frac1{4\pi}\int_0^\infty\dd p\;\frac{1-\pi p/\sinh\pi p}{p^2/4+\gs\,h(p)} ,
\end{equation}
which is finite. For example, at $\gs=3$ it equals $0.097219$.
\end{itemize}

\begin{table}[t]
\centering\small
\begin{tabular}{cc ccc}
\toprule
$(\gs,\ms)$ & & $\sum_n(-1)^n/c_n$ & $\sum_n 1/c_n^2$ & $\sum_n 1/c_n$\\
\midrule
$(1,\,0.5)$  & spectrum & 0.28369395 & 0.118714695 & 0.579903 \\
            & integral & 0.28369394 & 0.118714695 & 0.579903 \\
$(3,\,0.3)$  & spectrum & 0.40478634 & 0.190471071 & 0.575404 \\
            & integral & 0.40478633 & 0.190471071 & 0.575405 \\
$(10,\,0.5)$ & spectrum & 0.13923850 & 0.022376108 & 0.20508 \\
            & integral & 0.13923849 & 0.022376108 & 0.20508 \\
\bottomrule
\end{tabular}
\caption{Exact sum rules \eqref{eq:sum1}--\eqref{eq:sum2} against sums over the numerical spectrum (360 levels). The alternating sum is accelerated by averaging the last two partial sums. For $\sum1/c_n$ the tail beyond the computed levels is estimated from \eqref{eq:largen-exact}, which limits the agreement to about $10^{-5}$.}
\label{tab:sumrules}
\end{table}

\subsection{Baxter equation}

Multiplication by $e^{\pm2\theta}$ acts on the Fourier transform $\Phi(p)$ of $\varphi(\theta)$ as a shift by $\mp2\ii$. Writing $4\cosh^2\theta=e^{2\theta}+2+e^{-2\theta}$, \eqref{eq:Bp} becomes a second-order difference equation for $Q(p)\equiv B(p)\Phi(p)$:
\begin{equation}\label{eq:baxter}
\boxed{\;Q(p+2\ii)+2\,Q(p)+Q(p-2\ii)=\frac{c}{B(p)}\,Q(p)\;}
\end{equation}
This is a difference equation of Baxter (TQ) type, of the kind that appears in integrable lattice models and spectral problems; it is the form exploited in flat space by \cite{Fateev:2009jf} and more recently by \cite{Ambrosino:2023dik}. The only change on AdS$_2$ is the deformation $B\to B+p^2/4$ of the coefficient function. At $\gs=0$, $B$ is quadratic and \eqref{eq:baxter} is solved by Gamma functions times polynomials of continuous Hahn/Meixner--Pollaczek type, the Fourier transforms of the Jacobi solutions. The AdS$_2$ problem is therefore a one-parameter deformation of an integrable-in-the-FLZ-sense flat problem, and it has an exactly solvable point with a discrete spectrum, $\gs=0$, which the flat problem lacks. Perturbation theory in $\gs$ is regular: $\gs h(p)$ is relatively bounded by $p^2/4$, so each $\Delta_n(\gs)$ is analytic near $\gs=0$. The integrable structure behind \eqref{eq:baxter}, and the extent to which the FLZ machinery carries over to AdS$_2$, are analysed further in Appendix~\ref{sec:integrable}.

\subsection{Weak coupling in closed form}

The first-order shift is the expectation value of $\gs h(p)$ in the P\"oschl--Teller state, divided by its norm with weight $\frac14{\rm sech}^2\theta$. For the ground state $\varphi_0={\rm sech}^{2\ms}\theta$, whose Fourier transform is proportional to $|\Gamma(\ms+\ii p/2)|^2$,
\begin{equation}\label{eq:weak-general}
\frac{\dd c_0}{\dd\gs}\Big|_{\gs=0}=\frac{4^{2\ms}\,\Gamma(2\ms+\frac32)}{\pi^{3/2}\,\Gamma(2\ms)^2\,\Gamma(2\ms+1)}\int_{-\infty}^{\infty}\!\dd y\;\big(\pi y\coth\pi y-1\big)\,\big|\Gamma(\ms+\ii y)\big|^4 ,
\end{equation}
and $\dd\Delta_0/\dd\gs=(4\ms+1)^{-1}\dd c_0/\dd\gs$. At $\ms=\half$ the integral is elementary and gives $3\pi^2/4-6$, i.e.\ \eqref{eq:weak}. For other masses \eqref{eq:weak-general} gives $0.467364,\,0.890574,\,2.086052,\,3.429493$ at $\ms=0.15,\,0.3,\,0.8,\,1.5$, in agreement with the numerical slopes to the expected $O(\gs)$ accuracy. Since $\cosh(\pi y)=\pi/|\Gamma(\frac12+\ii y)|^2$ and $\pi y/\sinh(\pi y)=|\Gamma(1+\ii y)|^2$, the integrand is a ratio of Gamma functions, and \eqref{eq:weak-general} is a Barnes-type integral.

\subsection{Heavy quarks}

For $\ms\gg1$ the wavefunction is concentrated at $\theta\simeq0$, and the relevant momenta are $p\gtrsim\sqrt{\ms}\gg1$, so that $h(p)\simeq\frac\pi2|p|-1$. We write $\Delta=2\ms+\half+\varepsilon$, so that $c/4=\ms^2+\ms\varepsilon+O(1)$, and expand ${\rm sech}^2\theta\simeq1-\theta^2$. Equation \eqref{eq:rapidity-eq} then becomes, at leading order,
\begin{equation}
\varepsilon+\frac{\gs}{\ms}=\ms\,\theta^2+\frac{p^2}{4\ms}+\frac{\pi\gs}{2\ms}|p| .
\end{equation}
Here the roles of position and momentum are interchanged with respect to the original variables: $\theta$ plays the role of the relative momentum and $p$ of the relative position, $\hat r=r/R=p/\ms$. The reason is the following. For heavy quarks the constituents move slowly in the rest frame of the pair, and the rapidity of a slow particle is its velocity. If the quark carries the relative momentum $+q$ and the antiquark $-q$, their rapidities are $\pm q/m$, so the relative rapidity is $\theta\simeq q/m$, i.e.\ $\theta$ is the relative momentum in units of $m$. The variable $p=-\ii\partial_\theta$ conjugate to $\theta$ is then $m$ times the variable conjugate to $q$, which is the relative position $r$: $p=mr$. The three terms on the right-hand side confirm this. The mass term $\ms\theta^2=Rq^2/m$ is the kinetic energy of the reduced mass $m/2$; the rapidity kinetic term $p^2/4\ms=\ms\hat r^2/4$, which came from the collinear Casimir, is the AdS$_2$ harmonic trap; and the 't~Hooft term $\frac{\pi\gs}{2\ms}|p|=\frac{\pi\gs}{2}|\hat r|$ is the linear potential $T_s|r|$ (times $R$). The last identification holds beyond the heavy-quark limit: the 't~Hooft kernel, which in position space is the linear potential in the affine separation of the constituents, becomes $\gs h(p)\simeq\frac{\pi\gs}{2}|p|$ at large $|p|$, again a linear potential with $|p|$ in the role of the separation. This is why large $|p|$ corresponds to widely separated constituents in general, a fact we shall use in \cref{sec:desitter}. Equation~\eqref{eq:heavy-H} below is thus $R$ times the Hamiltonian of two particles of reduced mass $m/2$ in the AdS$_2$ harmonic trap of frequency $1/R$, bound by the linear potential $T_s|r|$:
\begin{equation}\label{eq:heavy-H}
R\,H_{\rm rel}=-\frac1{\ms}\partial_{\hat r}^2+\frac{\ms}{4}\hat r^2+\frac{\pi\gs}{2}|\hat r|,\qquad
\Delta_n=2mR+\frac12+R\,E^{\rm rel}_n-\frac{\gb^2R}{m}+O(1/\ms).
\end{equation}
The last term is the self-energy shift $m\to m-\gb^2/2m$ of the two quarks. With $\hat r=y\sqrt{2/\ms}$, $RH_{\rm rel}=-\frac12\partial_y^2+\frac12y^2+\eta|y|$ with $\eta=\pi\gs/\sqrt{2\ms}=T_sR^{3/2}/\sqrt{m/2}$. Its levels follow from parabolic-cylinder functions, $D'_\nu(\sqrt2\eta)=0$ (even) and $D_\nu(\sqrt2\eta)=0$ (odd), with $RE^{\rm rel}=\nu+\half-\eta^2/2$. This is also the Hamiltonian one writes down directly for two slow heavy quarks at equal global time. In the static coordinates of AdS$_2$, $\dd s^2=(1+\rho^2/R^2)\dd t^2-\dd\rho^2/(1+\rho^2/R^2)$, a quark of mass $m$ at rest at $\rho$ has energy $m\sqrt{1+\rho^2/R^2}\simeq m+m\rho^2/2R^2$, so two quarks at $\rho=\pm r/2$ feel the harmonic potential $mr^2/4R^2$, while the static Coulomb energy of the flux tube between them is $T_sr$ (\cref{sec:discussion}). Together with the kinetic energy of the reduced mass $m/2$ these are exactly the three terms of \eqref{eq:heavy-H}. \Cref{tab:heavy} shows that the full spectrum approaches \eqref{eq:heavy-H}, with corrections $O(1/\ms)$.

\begin{table}[t]
\centering\small
\begin{tabular}{c cccc}
\toprule
& $n=0$ & $n=1$ & $n=2$ & $n=3$\\
\midrule
$\ms=25$  & 0.76629 & 2.03741 & 3.18112 & 4.31649\\
$\ms=100$ & 0.76452 & 2.03882 & 3.18322 & 4.32124\\
$\ms=400$ & 0.76409 & 2.03922 & 3.18379 & 4.32251\\
\midrule
parabolic cylinder & 0.76395 & 2.03936 & 3.18399 & 4.32295\\
\bottomrule
\end{tabular}
\caption{Heavy-quark limit at fixed $\eta=\pi\gs/\sqrt{2\ms}=\half$: $\Delta_n-2\ms-\half+\gs/\ms$ from the full equation, against the levels $RE_n^{\rm rel}$ of \eqref{eq:heavy-H}. The deviations scale as $1/\ms$.}
\label{tab:heavy}
\end{table}

\subsection{\texorpdfstring{Semiclassics in rapidity and exact large-$n$ asymptotics}{Semiclassics in rapidity and exact large-n asymptotics}}

\begin{figure}[t]
\centering
\includegraphics[width=\textwidth]{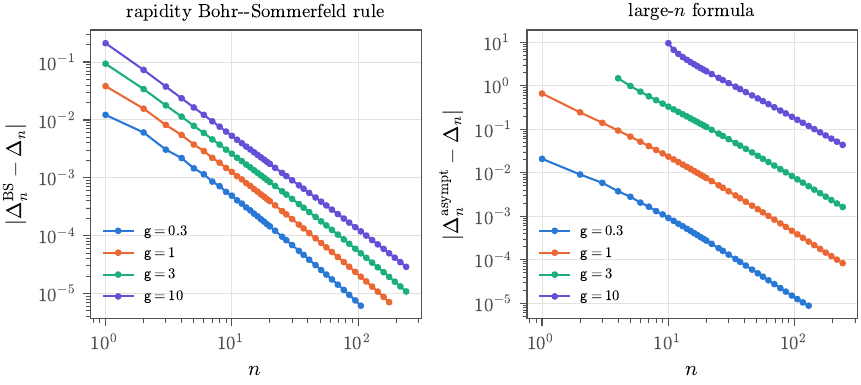}
\caption{Left: error of the rapidity Bohr--Sommerfeld rule \eqref{eq:BS-rule}, whose solution we denote $\Delta_n^{\rm BS}$, for $\ms=\half$; it decreases approximately as $n^{-2}$. Right: error of the closed-form asymptotics \eqref{eq:largen-exact}, decreasing faster than $n^{-1}$, approximately as $n^{-2}\log n$, once $n\gtrsim\gs$.}
\label{fig:rapidity}
\end{figure}

The rapidity form suggests a Bohr--Sommerfeld rule in the phase space $(\theta,p)$. Let
\begin{equation}
\mathcal A(C)=\tfrac12\,{\rm Area}\Big\{(\theta,p):\,B(p)\le\frac C4\,{\rm sech}^2\theta\Big\}=2\int_0^{p_*}\dd p\;{\rm arccosh}\sqrt{\frac{C}{4B(p)}},\qquad B(p_*)=\frac C4 ,
\end{equation}
half the area of the classically allowed region. The standard Bohr--Sommerfeld rule,
\begin{equation}\label{eq:BS-rule}
\mathcal A\big((\Delta_n-\tfrac12)^2\big)=\pi\big(n+\tfrac12\big)
\end{equation}
with $C=(\Delta-\half)^2$ rather than $\Delta(\Delta-1)$ (the analogue of the Langer correction for radial problems), is \emph{exact} at $\gs=0$, where $\mathcal A(C)=\pi(\sqrt C-2\ms)$, and \emph{asymptotically exact} at any coupling. Its error decreases approximately as $n^{-2}$ (\cref{fig:rapidity}), it is below $1\%$ for $n\ge2$ in all cases we examined, and it fails only for the pion at strong coupling. The reason can be seen from the Baxter equation \eqref{eq:baxter}. Near a turning point, $\theta\to\pm\infty$, only one of the exponentials in $4\cosh^2\theta=e^{2\theta}+2+e^{-2\theta}$ survives, and \eqref{eq:baxter} reduces to the first-order difference equation $B(p)\,\Phi(p)=c\,\Phi(p\mp2\ii)$, whose solution is an explicit product of Gamma functions. Its large-$p$ expansion reproduces the WKB phase with the usual $\pi/4$ connection phase, and all corrections are suppressed by powers of $1/p$. There is therefore no additional $O(1)$ phase from the turning points, as there would be for a hard wall or a singular potential. Expanding $\mathcal A$ at large $C$ with the help of \eqref{eq:Bfactor} gives the large-$n$ spectrum in closed form:
\begin{equation}\label{eq:largen-exact}
\boxed{\;\begin{gathered}\Delta_n=n+1+2\beta_0+2\,\Sigma_{\rm reg}+2\gs\log\big(\Delta_n-\tfrac12\big)+O\big(n^{-1}\log n\big),\\
\Sigma_{\rm reg}=\lim_{K\to\infty}\Big[\sum_{k=1}^K(\beta_k-k)-\gs\log K\Big]\end{gathered}\;}
\end{equation}
where the $\beta_k$ are the roots of the indicial equation. Equivalently, $2\beta_0+2\Sigma_{\rm reg}-2\gs\log2=\frac1\pi\mathcal I$ with $\mathcal I=\lim_{\pmax\to\infty}\big[\int_0^{\pmax}\dd p\,\log(4B(p)/p^2)-2\pi\gs\log\pmax\big]$; we have checked this identity numerically. At $\gs=0$, \eqref{eq:largen-exact} reduces to $\Delta_n=1+2\ms+n$. In general it fixes the $O(1)$ term left undetermined in \eqref{eq:largen}. The highly excited mesons behave as two-particle states of effective boundary constituents of dimension $\half+\beta_0+\Sigma_{\rm reg}$, up to the logarithm generated by the confining interaction. \Cref{fig:rapidity} shows that the error of \eqref{eq:largen-exact} decreases faster than $1/n$, approximately as $n^{-2}\log n$, so that the remainder quoted in \eqref{eq:largen-exact} is a conservative estimate.

\section{Boundary conditions}
\label{sec:bc}

In the light-front frame the AdS boundary sits at $|\lambda|\to\infty$, that is, at vanishing light-front momentum. Boundary conditions therefore become statements about the small-momentum (endpoint) behaviour of light-front wavefunctions, and about the flux of the electric field at the two ends of the null line. We should also mention that in global AdS$_2$, since the light front is a complete null geodesic connecting its two boundary components, a light-front wavefunction with support on $k>0$ has the same power-law asymptotics at $\lambda\to+\infty$ and $\lambda\to-\infty$, so the construction imposes the same boundary condition on both components. Theories with different boundary conditions on the two components are $\SLtwo$-invariant, but are not described by a single light-front Fock space of the type we analyse here.

\subsection{Quarks: boundary conditions are endpoint conditions}
\label{sec:bc-quarks}

The single-particle analysis of \cref{sec:oneparticle} showed that the two AdS$_2$ quantizations of a free quark, $\Delta_\psi=\half\pm\ms$, correspond to the behaviour $f(k)\sim k^{\pm\ms}$ as $k\to0$. In a meson, $x\to0$ is the region where the quark carries a vanishing fraction of the light-front momentum, i.e.\ where it is near the boundary; likewise $x\to1$ for the antiquark. This is visible in the free solutions \eqref{eq:free}. In particular the functions $x^{a}(1-x)^{b}P_n^{(2b,2a)}(2x-1)$ with $a,b\in\{\pm\ms\}$ solve the free equation with
\begin{equation}
\Delta_n=1+a+b+n=\Delta_\psi^{(a)}+\Delta_\psi^{(b)}+n,
\end{equation}
covering all four combinations of quark and antiquark boundary conditions. The exponent at each endpoint is the label of the boundary condition of the corresponding constituent.

With interactions the candidate endpoint exponents are the roots $\pm\beta_0$ of \eqref{eq:indicial}. So far we have used $\varphi\sim x^{\beta_0}$, $(1-x)^{\beta_0}$ for our endpoint solutions. For $\ms<1$ this branch is continuously connected to the free standard quantization ($\beta_0\to\ms$ as $\gs\to0$), and it is the one for which the quadratic form \eqref{eq:form} defines the operator; we regard it as the standard boundary condition of the interacting theory. It is natural to associate the root $-\beta_0$ at an endpoint with alternate quantization of the corresponding constituent. Since the quark is not gauge invariant, this identification should be regarded as a conjecture. Square integrability of such a wavefunction requires $\beta_0<\half$, which by the monotonicity of \eqref{eq:indicial} is equivalent to
\begin{equation}
\ms^2<\gs+\tfrac14 .
\end{equation}
At $\gs=0$ this is the free window $\ms<\half$. It is only a necessary condition. With an $x^{-\beta_0}$ falloff every term in \eqref{eq:form} diverges, so the operator must be defined by a self-adjoint extension with boundary terms. Moreover, for eigenvalues $c<0$, as in the free $(-\ms,-\ms)$ case with $\Delta=1-2\ms$, both roots of $\Delta(\Delta-1)=c$ give normalizable centre-of-mass factors in \eqref{eq:lowest-weight}, and an additional condition is needed to select $\Delta$. We leave a complete analysis of these alternative boundary conditions for future work. At $\ms=0$ the two roots merge into $\beta_0=0$ with a logarithmic partner, the analogue of the saturation of a Breitenlohner--Freedman bound \cite{Breitenlohner:1982jf}.

\paragraph{Mixed boundary conditions and the SYK spectrum.} In the window $\ms<\half$ one may also impose mixed (Robin-type) conditions, in which the two endpoint behaviours appear in a fixed ratio,
\begin{equation}\label{eq:mixed}
\varphi(x)\simeq x^{\ms}+\rho\,x^{-\ms}\quad(x\to0),\qquad \varphi(x)\simeq(1-x)^{\ms}+\rho\,(1-x)^{-\ms}\quad(x\to1),
\end{equation}
up to normalization; each $\rho\in\mathbb R$ defines a self-adjoint extension of the free operator. An interesting example arises in Vegh's quantization of the folded string in AdS$_2$ with imaginary radius squared \cite{Vegh:2024fsa,Vegh:2025syk}. Translated to our variables, his Casimir equation [eq.~(27) of \cite{Vegh:2025syk}] is the free equation \eqref{eq:AdS-tHooft} with $\gs=0$ and
\begin{equation}
\ms=\tfrac12-\Delta,\qquad \rho=\frac{\Delta}{1-\Delta}=\frac{1-2\ms}{1+2\ms},
\end{equation}
where $\Delta=1/q$ is the dimension of the SYK fermion with $q$-body interactions, and his boundary condition is \eqref{eq:mixed}, imposed on wavefunctions that are odd under charge conjugation. We have independently verified that the resulting dimensions coincide with the dimensions $h_n$ of the fermion bilinears of the SYK model, the solutions of $k_c(h)=1$ with the kernel eigenvalue $k_c$ of \cite{Maldacena:2016hyu}, to all digits computed: for $q=4$ one finds $h=2,\,3.77354,\,5.67946,\,7.63197,\,9.60396,\dots$, and the agreement persists for $q=6$ and $8$. The 't~Hooft interaction plays no role in this construction. The SYK bilinear spectrum is that of two free AdS$_2$ fermions of mass $\ms=\half-\Delta$, in the window where both quantizations are normalizable, with the particular self-adjoint extension \eqref{eq:mixed}; the value $h=2$ of the first odd level, the stress tensor of SYK, is what fixes $\rho$. Within QCD$_2$ on AdS$_2$ a mixed condition of this type would correspond to a boundary condition on the quark fields that mixes the two quantizations, and the interacting problem with $\gs>0$ and $\rho\neq0$ requires the self-adjoint extension of the full operator mentioned above. We do not pursue this here, but mention it as an interesting future problem.

\subsection{Gauge field: boundary flux and superselection sectors}
\label{sec:flux}

We define the electric flux on the null line, $\mathcal{E}^a(\lambda)=\frac2{g^2}\partial_\lambda A^a_y$, normalized so that Gauss's law reads $\partial_\lambda\mathcal E^a=j^a$. It integrates to
\begin{equation}
\mathcal{E}^a(+\infty)-\mathcal{E}^a(-\infty)=Q^a=\int\dd\lambda\,j^a(\lambda).
\end{equation}
The boundary values $\mathcal{E}^a(\pm\infty)$ are the electric fluxes through the two boundary components. There are two natural classes of boundary conditions for a gauge field in AdS \cite{Marolf:2006nd}, and they define different superselection sectors, i.e.\ sets of states that no local operator connects.
\begin{itemize}
\item \emph{No boundary flux} (Neumann-type for the gauge field: the normal component of the electric field, i.e.\ the normal derivative of the gauge potential, is fixed to zero at the boundary). Gauge transformations need not vanish at the boundary, Gauss's law forces $Q^a=0$, and the Hilbert space consists of colour singlets. This is the boundary condition used in \eqref{eq:coulomb} and throughout most of this work.
\item \emph{Dirichlet-type}. Boundary gauge transformations become a global $SU(N)$ symmetry of the boundary theory, and $\mathcal{E}^a(\pm\infty)$ is dynamical. Sectors with boundary charge exist, but a flux that reaches the boundary costs a Poincar\'e energy $\propto\int\dd\lambda\,\mathcal{E}^2$, which diverges linearly in the cutoff $\lambda_{\max}=2R^2/z_{\min}$. (The proper length to the boundary diverges only logarithmically; the stronger divergence is a redshift effect.) A boundary counterterm is needed, as in the treatment of Maxwell fields in AdS$_2$ \cite{Castro:2008ms}. The zero-flux, singlet sector is common to both classes and is the one solved here.
\end{itemize}

The light-front formalism makes the effect of a background flux transparent. Consider a constant proper electric field $E$ (field strength $F=E\,{\rm vol}$, which is $\SLtwo$-invariant on AdS$_2$) coupled to constituents of charge $q$; this is an abelian background, as in the Schwinger model or for a $U(1)\subset U(N)$ flux. In the gauge \eqref{eq:gauge} it is $A_y=\frac E2\lambda$, which adds $\epsilon\int\dd\lambda\,\lambda\,\psi^\dagger\psi$ to $H$, with $\epsilon=-qE/2$. On a state of net charge $q$ this term is not translation invariant in $\lambda$, but the algebra \eqref{eq:sl2} still closes after the constant shift $D\to D-2\epsilon R^2$, and the Casimir is shifted:
\begin{equation}
\cC_2\to\cC_2-(qER^2)^2\qquad(\text{net charge }q).
\end{equation}
For a single quark this gives $\big(\Delta_\psi-\half\big)^2=\ms^2-(qER^2)^2$, the fermionic analogue of the electric-field shift of the Breitenlohner--Freedman bound \cite{Pioline:2005pf,Faulkner:2009wj}. When $qER^2>\ms$ the background is unstable to pair production, and $\Delta_\psi=\half\pm\ii\sqrt{(qER^2)^2-\ms^2}$ becomes complex: the Casimir then takes a value in the range of the principal series of $\SLtwo$, as for a field below the Breitenlohner--Freedman bound \cite{Breitenlohner:1982jf}. For a neutral pair the dilatation is not shifted; instead the field couples to the dipole and adds the operator $4\epsilon R^2(\lambda_1-\lambda_2)\cP$ to the Casimir. In the 't~Hooft model the flux sourced by a fundamental charge on the boundary has exactly this form, with $qE=T_s=\pi\gb^2/2$. The simplest boundary-charged sector, a quark tied to a boundary antiquark, is solved in \cref{sec:anchored}, including the self-energy of the coloured constituent. At finite $N$ such a flux may be screened near the boundary, an AdS$_2$ analogue of string breaking, but at large $N$ the transitions between screened and unscreened configurations are suppressed by $1/\sqrt N$.

\section{A quark tied to the boundary}
\label{sec:anchored}

So far the antiquark and the quark were both dynamical particles in the interior of AdS$_2$. In this section we pin the antiquark to the boundary and ask what happens to the quark attached to it. Physically, one may picture the quark being sent from the boundary into the interior, while the flux tube (the ``string'') that ties it to the antiquark pulls it back. This is possible with Dirichlet-type boundary conditions for the gauge field (\cref{sec:flux}), which allow electric flux to end on the boundary, so that a colour charge can sit there. We take the antiquark to be static and infinitely heavy, on the boundary component reached by the Poincar\'e patch, and join it to a quark in the bulk by a flux tube. In holography this is a Wilson line on the boundary with a dynamical quark attached at its endpoint.

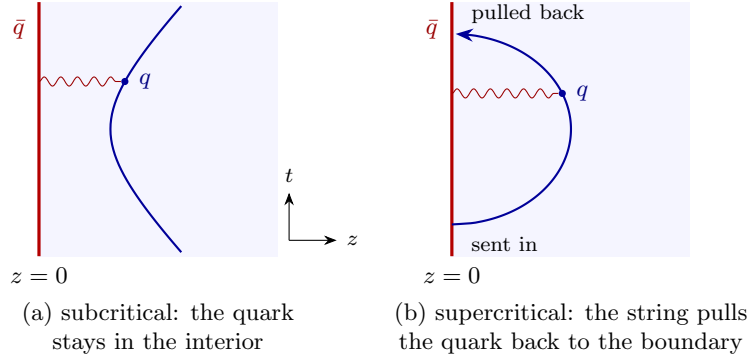
\begin{figure}[t]
\centering
\begin{tikzpicture}[scale=1.05,>=Stealth]
  \begin{scope}
    \fill[blue!4] (0,-1.6) rectangle (3.0,1.6);
    \draw[very thick,red!70!black] (0,-1.6) -- (0,1.6);
    \node[font=\footnotesize,red!60!black,anchor=east] at (-0.05,1.25) {$\bar q$};
    \node[font=\footnotesize,anchor=north] at (0,-1.62) {$z=0$};
    \draw[->] (3.15,-1.4) -- (3.15,-0.8) node[above,font=\scriptsize] {$t$};
    \draw[->] (3.15,-1.4) -- (3.75,-1.4) node[right,font=\scriptsize] {$z$};
    \draw[thick,blue!60!black,domain=-1.55:1.55,smooth,variable=\t] plot ({sqrt(0.9*0.9+\t*\t)},{\t});
    \fill[blue!60!black] ({sqrt(0.81+0.36)},0.6) circle (0.045);
    \node[font=\footnotesize,blue!50!black,anchor=west] at ({sqrt(0.81+0.36)+0.05},0.6) {$q$};
    \draw[decorate,decoration={snake,amplitude=0.6mm,segment length=2.2mm},red!60!black] (0,0.6) -- ({sqrt(0.81+0.36)-0.05},0.6);
    \node[font=\footnotesize,align=center] at (1.5,-2.5) {(a) subcritical: the quark\\stays in the interior};
  \end{scope}
  \begin{scope}[xshift=5.2cm]
    \fill[blue!4] (0,-1.6) rectangle (3.0,1.6);
    \draw[very thick,red!70!black] (0,-1.6) -- (0,1.6);
    \node[font=\footnotesize,red!60!black,anchor=east] at (-0.05,1.25) {$\bar q$};
    \node[font=\footnotesize,anchor=north] at (0,-1.62) {$z=0$};
    \draw[thick,blue!60!black,->,domain=-1.2:1.2,samples=160,smooth,variable=\t] plot ({1.5*sqrt(1-(\t/1.2)^2)},{\t});
    \fill[blue!60!black] ({1.5*sqrt(1-(0.45/1.2)^2)},0.45) circle (0.045);
    \node[font=\footnotesize,blue!50!black,anchor=west] at ({1.5*sqrt(1-(0.45/1.2)^2)+0.05},0.45) {$q$};
    \draw[decorate,decoration={snake,amplitude=0.6mm,segment length=2.2mm},red!60!black] (0,0.45) -- ({1.5*sqrt(1-(0.45/1.2)^2)-0.05},0.45);
    \node[font=\scriptsize,anchor=west] at (0.12,-1.45) {sent in};
    \node[font=\scriptsize,anchor=west] at (0.12,1.45) {pulled back};
    \node[font=\footnotesize,align=center] at (1.5,-2.5) {(b) supercritical: the string pulls\\the quark back to the boundary};
  \end{scope}
\end{tikzpicture}
\caption{A quark (blue) tied by a flux tube (wavy) to a static antiquark on the boundary $z=0$ of the Poincar\'e patch, sketched in Poincar\'e coordinates $(t,z)$. (a) For a weak string the AdS geometry wins: the quark moves in the interior like a free massive particle, coming in from the Poincar\'e horizon $z\to\infty$ and returning to it, with its dimension shifted by the string. (b) For a strong string the quark is pulled back to the boundary. A quark sent in from the boundary returns to it after a finite time, with the scattering amplitude \eqref{eq:anchored-R}.}
\label{fig:anchored-setup}
\end{figure}

The problem turns out to be exactly solvable, and the answer is governed by a competition between two forces, illustrated in \cref{fig:anchored-setup}. The string pulls the quark towards the boundary with the constant force $T_s=g^2C_F/2=\pi\gb^2/2$, the string tension of \cref{sec:action}. The AdS geometry pulls a massive particle back towards the interior, with a force that grows with the distance from the centre and saturates at $m/R$. Since $T_s$ is fixed by the gauge coupling, a ``weak'' or ``strong'' string below means small or large $\gs=\gb^2R^2$ at fixed $\ms=mR$. Our results are as follows:
\begin{itemize}
\item The boundary operator that creates the quark at the end of the Wilson line has dimension
\[
\big(\Delta_{\rm bq}-\tfrac12\big)^2=R^2\big(m^2-\gb^2-T_s^2R^2\big),
\]
where $m^2-\gb^2$ is the quark mass squared renormalized by the planar self-energy.
\item For a weak string, $T_sR<\sqrt{m^2-\gb^2}$ (\emph{subcritical}), the AdS geometry wins and the quark stays in the interior. The string only lowers its dimension and produces a constant scattering phase \eqref{eq:anchored-sub}.
\item For a strong string, $T_sR>\sqrt{m^2-\gb^2}$ (\emph{supercritical}), the string wins and pulls the quark back to the boundary, as in \cref{fig:anchored-setup}(b). The boundary then becomes a region from which the quark can be sent in and to which it returns, and we compute the amplitude for this process in closed form, \eqref{eq:anchored-R}. The quark always returns if its Poincar\'e energy is negative, with a time delay equal to the classical round-trip time; if it is positive, it returns within the Poincar\'e patch only with probability $e^{-4\pi\varkappa}$, with $\varkappa$ given in \eqref{eq:anchored-kappa}.
\end{itemize}

\subsection{The boundary-anchored 't~Hooft equation}

We first derive the light-front Hamiltonian of this system. On every null line $y={\rm const}$ the boundary antiquark sits at $\lambda=+\infty$, the endpoint of the light front on the Poincar\'e boundary (\cref{fig:geometry}). It is static in Poincar\'e time, and in the affine frame it does not move along the null line. By Gauss's law the electric flux equals the quark's colour charge between the quark and the boundary, and vanishes on the other side of the quark, i.e.\ on the part $\lambda'<\lambda$ of the null line, which runs from the quark through the Poincar\'e horizon ($\lambda'=0$) to the second boundary component at $\lambda'\to-\infty$, where no flux enters. For a quark at $\lambda$ the Coulomb energy is therefore
\begin{equation}
\frac{g^2}{4}\int_{\lambda}^{\lambda_{\max}}\dd\lambda'\,\mathcal E^a\mathcal E^a=\frac{g^2C_F}{4}\,(\lambda_{\max}-\lambda)\simeq\frac{T_s}{2}\,(\lambda_{\max}-\lambda),
\end{equation}
a \emph{one-sided} linear potential, in contrast to the two-sided $|\lambda_1-\lambda_2|$ of a meson. (It equals the Poincar\'e energy $T_sR^2(z_{\min}^{-1}-z^{-1})$ of a static string from the quark to a cutoff surface $z=z_{\min}$ near the boundary.) In the language of the meson equation, pinning the antiquark to the boundary sets its momentum fraction $1-x$ to zero, since vanishing light-front momentum is the momentum-space image of $|\lambda|\to\infty$. The two-body problem collapses to a one-body problem for the quark. The principal-value kernel, the Fourier transform of $|\lambda_1-\lambda_2|$, is replaced by the transform of a one-sided potential, and since $\lambda\to-\ii\partial_k$ on positive-momentum wavefunctions \eqref{eq:momentum-rep}, that transform is a first-order derivative.

One point requires care. In the 't~Hooft model the instantaneous Coulomb interaction also acts on each quark separately, through the normal-ordering self-energy of \cref{sec:twobody}, which shifts $m^2\to m^2-\gb^2$. We now show that the anchored quark keeps this shift. We first write the field energy with no flux at $\lambda\to-\infty$ in terms of the colour current,
\begin{equation}
\frac{g^2}{4}\int^{\lambda_{\max}}\!\dd\lambda'\Big(\int_{-\infty}^{\lambda'}\!j\Big)^2=\frac{g^2}{4}\lambda_{\max}Q^aQ^a-\frac{g^2}{4}\,Q^a\!\int\!\dd\lambda\,\lambda\,j^a(\lambda)-\frac{g^2}{8}\iint\dd\lambda\,\dd\lambda'\,j^a(\lambda)\,|\lambda-\lambda'|\,j^a(\lambda').
\end{equation}
The last term is the Coulomb term of \eqref{eq:coulomb}, and its normal ordering produces the same one-body self-energy as for a meson, $m^2\to m^2-\gb^2$. On single-quark states, the middle term is the linear potential projected onto positive momenta. The first term, a constant, is removed together with the remaining constant from normal ordering by the boundary counterterm of \cref{sec:flux}; dilatation covariance fixes the leftover constant to zero. On single-quark wavefunctions $f(k)$ the light-front Hamiltonian can be written as
\begin{equation}\label{eq:anchored-H}
\boxed{\;H=\frac{m^2-\gb^2}{4k}-\frac1{4R^2}\,\partial_k\,k\,\partial_k+\ii\,\frac{\pi\gb^2}{4}\,\partial_k\;}
\end{equation}
This is the ``boundary-anchored'' counterpart of the 't~Hooft equation, and it will be convenient to define
\begin{equation}\label{eq:anchored-nub}
\nu_b^2\equiv\ms^2-\gs-\Big(\frac{\pi\gs}{2}\Big)^2 .
\end{equation}
The first two terms are the renormalized free quark in AdS$_2$ (\cref{sec:oneparticle}), and the last comes from the string. Here $k>0$ is the light-front momentum of the quark, and $H$ is the Poincar\'e Hamiltonian, the generator of translations in Poincar\'e time $t$. Small $k$ corresponds to large $|\lambda|$, i.e.\ to the vicinity of the boundary, and large $k$ to the Poincar\'e horizon $\lambda\to0$.

\paragraph{Two checks of the mass renormalization.} First, the same decomposition applied to a static source \emph{inside} the light front, at $\lambda_0$, reproduces the heavy--light limit of the flat 't~Hooft equation \cite{tHooft:1974pnl}, in which one quark is infinitely heavy. The potential is then two-sided, $|\lambda-\lambda_0|$, and its projection onto positive momenta produces, besides an integral kernel, a one-body term $+\gb^2/4k$ that cancels the self-energy shift; this is why the heavy--light equation can be written with the bare quark mass. For the one-sided potential no such term arises, because multiplication by $\lambda$ maps positive-momentum wavefunctions that vanish at $k=0$ into themselves, so the renormalized mass survives in \eqref{eq:anchored-H}. Second, the same combination appears in the meson equation when the quark and the antiquark are far apart. In the rapidity form of \cref{sec:rapidity}, $B(p)=\frac{p^2}{4}+\gs h(p)+\ms^2$ with $h(p)=\frac{\pi p}{2}\coth\frac{\pi p}{2}-1$ plays the role of the kinetic plus potential energy of the relative motion, and large $|p|$ corresponds to large separation. For $|p|\gtrsim1$, where $h(p)=\frac\pi2|p|-1$ up to exponentially small terms,
\begin{equation}\label{eq:B-large}
B(p)\simeq\frac{p^2}{4}+\frac{\pi\gs}{2}|p|+\ms^2-\gs=\Big(\frac{|p|}{2}+\frac{\pi\gs}{2}\Big)^2+\nu_b^2 ,
\end{equation}
so that $\nu_b^2$ is the constant left over when the linear potential is absorbed into the kinetic term, exactly as the string is absorbed by the transformation that we analyse below.

\subsection{Removing the string gives an effective quark mass}
\label{sec:anchored-twist}

One notices that the isometry algebra still closes. With $K=4R^2\cP=4R^2k$ unchanged, the string term commutes with $K$ up to a constant, and $[H,K]=2\ii D'$ with
\begin{equation}
D'=D+\frac{\pi\gs}{2},
\end{equation}
which also satisfies $[H,D']=\ii H$ and $[K,D']=-\ii K$. This is the constant electric field of \cref{sec:flux} with $qE=T_s=\pi\gb^2/2$, which is $\SLtwo$-invariant on AdS$_2$. One observes the remarkable fact that the string can be removed altogether. Writing $f(k)=k^{\ii\pi\gs/2}\,\tilde f(k)$ one finds
\begin{equation}
4R^2H\,f=k^{\ii\pi\gs/2}\Big[\frac{\nu_b^2}{k}-\partial_kk\,\partial_k\Big]\tilde f,\qquad D'f=k^{\ii\pi\gs/2}D\tilde f,\qquad Kf=k^{\ii\pi\gs/2}K\tilde f,
\end{equation}
so that the quark tied to the boundary is unitarily equivalent to a \emph{free} AdS$_2$ quark of mass $\nu_b/R$. In words, in AdS$_2$ the entire effect of the string is to change the effective mass of the quark. Its dimension therefore follows from the free result $\Delta_\psi=\half+\ms$ of \cref{sec:oneparticle} with $\ms\to\nu_b$:
\begin{equation}\label{eq:anchored-Delta}
\boxed{\;\big(\Delta_{\rm bq}-\tfrac12\big)^2=\nu_b^2=R^2\big(m^2-\gb^2-T_s^2R^2\big)\;}
\end{equation}
Here $\Delta_{\rm bq}$ is the dimension of the boundary operator that creates the quark at the end of the boundary Wilson line. The string lowers the dimension, and the single-particle representation contains one primary: the radial excitations of the quark on its string are the descendants $\Delta_{\rm bq}+n$.

\begin{figure}[t]
\centering
\includegraphics[width=\textwidth]{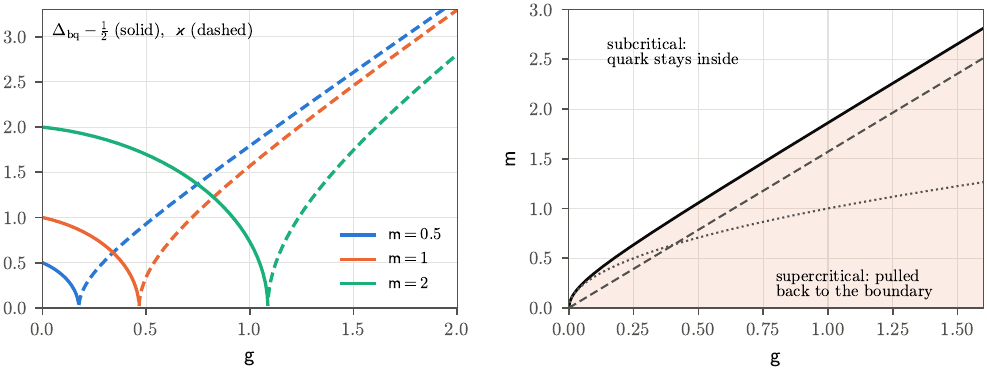}
\caption{A quark tied to the boundary. Left: $\Delta_{\rm bq}-\half$ (solid, subcritical) and $\varkappa=|{\rm Im}\,\Delta_{\rm bq}|$ (dashed, supercritical) as functions of $\gs$ for $\ms=0.5,\,1,\,2$, from \eqref{eq:anchored-Delta}. Right: the two regimes in the $(\gs,\ms)$ plane, separated by $\ms^2=\gs+\pi^2\gs^2/4$ (solid). The dashed line $\ms=\pi\gs/2$ is the classical criterion $T_sR=m$ without the planar mass renormalization, and below the dotted line $\ms^2=\gs$ ($m<\gb$) the string always wins and pulls the quark back to the boundary.}
\label{fig:anchored}
\end{figure}

The result has a transparent classical meaning. A particle of mass $m$ held at a large proper distance from the centre of AdS$_2$ needs a force $m/R$ to keep it from falling back. A string ending on the boundary pulls outward with the constant force $T_s$. The string wins if $T_sR>m$. For $T_sR<m$ the quark has a stable classical equilibrium, at $\sin\sigma=T_sR/m$ in the global coordinates \eqref{eq:global}, with global energy (conjugate to $\tau$) equal to $R\sqrt{m^2-T_s^2R^2}$, the classical value of $\Delta_{\rm bq}-\half$. Equation~\eqref{eq:anchored-Delta} is the quantum version of this criterion, with the planar renormalized mass $m^2-\gb^2$. There are two regimes, \emph{subcritical} and \emph{supercritical}, separated by the critical line $\ms^2=\gs+\pi^2\gs^2/4$ (\cref{fig:anchored}). Since the right-hand side of \eqref{eq:anchored-Delta} is negative whenever $m<\gb$, light quarks are always supercritical.

The same combination also appears later on, in the analysis of the 't Hooft equation in de~Sitter space (\cref{sec:desitter}). In short, the de~Sitter 't Hooft equation is found by sending $R^2\to-R^2$, so that $-\nu_b^2$ becomes $\ms^2-\gs+(\pi\gs/2)^2=(R\,V_{\rm top}/2)^2$ (\cref{sec:ds-mesons}). The stable equilibrium of the anchored quark in AdS$_2$, where the string balances the AdS potential, thus continues to the unstable equilibrium of each quark at the top of the de~Sitter dissociation barrier, where the string balances the cosmological repulsion.

\subsection{Subcritical strings: the quark stays inside the bulk}

Since AdS$_2$ acts as a box, a particle in it never escapes to infinity, and there is no $S$-matrix in the usual sense. A scattering problem can nevertheless be posed in the Poincar\'e patch, because its spectrum of Poincar\'e energies $\omega$ is continuous: a particle comes in from the Poincar\'e horizon, is turned around, and goes back to the horizon. The scattering phase relates the outgoing with the incoming wave.

For $\nu_b^2>0$ the boundary operator has a real dimension, $\Delta_{\rm bq}=\half+\nu_b$ (and also $\half-\nu_b$ for $\nu_b<\half$, as in \cref{sec:oneparticle}). The eigenfunctions of the Poincar\'e Hamiltonian are
\begin{equation}
f_\omega(k)=k^{\ii\pi\gs/2}\,J_{2\nu_b}\big(4R\sqrt{\omega k}\big),\qquad \omega>0 .
\end{equation}
At large $k$, i.e.\ near the Poincar\'e horizon $\lambda\to0$, they are superpositions of an incoming and an outgoing wave with the relative phase $e^{-\ii\pi(2\nu_b+\frac12)}$. A quark that falls in from the Poincar\'e horizon and is turned around by the combined action of the AdS potential and the string acquires, compared with a free quark, the phase shift
\begin{equation}\label{eq:anchored-sub}
e^{2\ii\delta_b}=e^{-2\pi\ii(\Delta_{\rm bq}-\Delta_\psi)},\qquad
\delta_b=\pi\Big[\ms-\sqrt{\ms^2-\gs-\pi^2\gs^2/4}\,\Big]>0 .
\end{equation}
Because of the dilatation symmetry it does not depend on $\omega$. This is an exact analogue of the AdS$_2$ relation between phase shifts and anomalous dimensions \cite{Paulos:2016fap}, here with the anomalous dimension induced by the string and the planar self-energy. In this regime the string does not pull the quark back to the boundary: AdS$_2$ confines it in the interior.

\subsection{Supercritical strings: the quark is pulled back to the boundary}

For $\nu_b^2<0$ we write $\nu_b=\ii\varkappa$ with
\begin{equation}\label{eq:anchored-kappa}
\varkappa=R\sqrt{T_s^2R^2-m^2+\gb^2}=\sqrt{\big(\tfrac{\pi\gs}{2}\big)^2+\gs-\ms^2} ,
\end{equation}
so that $\Delta_{\rm bq}=\half\pm\ii\varkappa$. The Casimir, $\cC_2=\Delta_{\rm bq}(\Delta_{\rm bq}-1)=-\frac14-\varkappa^2$, then takes a value in the range of the principal series of $\SLtwo$, $\Delta=\half+\ii\nu$. A unitary principal-series representation cannot, however, be realized on the single-quark Hilbert space of the light front. There $K=4R^2k>0$, whereas the restriction of a principal-series representation to the subgroup generated by $D$ and $K$ contains both signs of $K$ (\cref{sec:ds-halfspace}). Correspondingly, $H$ has no self-adjoint extension compatible with the dilatations, as we discuss below. This is the same mechanism by which a field below the Breitenlohner--Freedman bound in AdS acquires a dimension on the principal-series line \cite{Breitenlohner:1982jf,Kaplan:2009kr}. This is the regime of the process described at the start of the section. Note that the Poincar\'e energy $\omega$ of the quark can now be negative, because the string releases energy as it shortens. Classically, the Hamiltonian \eqref{eq:anchored-H} is quadratic in $\lambda$, and on a trajectory of Poincar\'e energy $\omega$
\begin{equation}
\lambda k=\frac{\pi\gs}{2}\pm\sqrt{\varkappa^2-4R^2|\omega|\,k}\qquad(\omega<0).
\end{equation}
For $m>\gb$ (i.e.\ $\ms^2>\gs$) both branches reach $\lambda\to+\infty$ as $k\to0$. The quark leaves the boundary on the upper branch, along which $k$ grows, turns at $k_{\max}=\varkappa^2/4R^2|\omega|$, and returns to the boundary on the lower branch after a Poincar\'e time $2\varkappa/|\omega|$. For $\omega>0$ there is no turning point, and the quark escapes through the Poincar\'e horizon. For $m<\gb$, where the renormalized mass squared is negative, the lower branch ends instead at $\lambda\to-\infty$, on the other boundary component, and this simple classical picture no longer applies.

Quantum mechanically, the boundary now plays the role of the region at infinity in an ordinary scattering problem. Near $k=0$ every solution is a superposition of $(k/k_0)^{\ii(\pi\gs/2+\varkappa)}$ (a quark leaving the boundary) and $(k/k_0)^{\ii(\pi\gs/2-\varkappa)}$ (a quark returning to it), where the arbitrary scale $k_0$ fixes the normalization of these boundary waves. The two carry equal and opposite flux, so the boundary now behaves as an asymptotic region. The solution that does not come in from the Poincar\'e horizon is $k^{\ii\pi\gs/2}K_{2\ii\varkappa}(4R\sqrt{|\omega|k})$ for $\omega<0$, and the outgoing Hankel function for $\omega>0$. Their small-$k$ expansions give the amplitude for a quark sent from the boundary to return to it:
\begin{equation}\label{eq:anchored-R}
\boxed{\;\mathcal R(\omega)=\frac{\Gamma(2\ii\varkappa)}{\Gamma(-2\ii\varkappa)}\,\Big[-4R^2k_0\,(\omega+\ii0)\Big]^{-2\ii\varkappa}\;}
\end{equation}
A change of $k_0$ changes $\mathcal R$ by a constant phase. For $\omega<0$, $|\mathcal R|=1$: the string returns the quark with certainty, and the scattering phase is
\begin{equation}\label{eq:anchored-phase}
2\delta_b(\omega)=2\arg\Gamma(2\ii\varkappa)-2\varkappa\log\big(4R^2k_0|\omega|\big).
\end{equation}
For $\omega>0$, $|\mathcal R|=e^{-2\pi\varkappa}$: the quark returns to the boundary inside the Poincar\'e patch only with probability $e^{-4\pi\varkappa}$, and otherwise crosses the Poincar\'e horizon, beyond which it comes back to the boundary outside the patch. We have performed three checks of this picture:
\begin{itemize}
\item For $\omega<0$ the Wigner time delay $-\ii\,\partial_\omega\log\mathcal R=2\varkappa/|\omega|$, the time the quark spends away from the boundary as inferred from the energy dependence of the phase, equals the classical round-trip time.
\item For $\varkappa\gg1$, Stirling's formula turns \eqref{eq:anchored-phase} into $2\varkappa\log\big(\varkappa^2/e^2R^2k_0|\omega|\big)-\frac\pi2+O(1/\varkappa)$. This is the classical action $\oint\lambda\,\dd k$ of the excursion, with a Maslov phase $-\pi/2$ from the turning point.
\item Direct numerical integration of the eigenvalue equation of \eqref{eq:anchored-H} reproduces \eqref{eq:anchored-R} for both signs of $\omega$, to $10^{-9}$ or better in most cases and to better than $10^{-6}$ in all cases we examined.
\end{itemize}
Formally, up to the factor $(4R^2k_0)^{-2\ii\varkappa}/\Gamma(-2\ii\varkappa)$, \eqref{eq:anchored-R} is the frequency-space boundary two-point function $\propto\Gamma(1-2\Delta)\big[-(\omega+\ii0)\big]^{2\Delta-1}$ of an operator of dimension $\Delta=\half-\ii\varkappa$. The scattering phase of the quark on its string is the boundary correlator of the operator that creates it, continued to complex dimension.

Two qualifications are in order. First, small $k$ describes the vicinity of \emph{both} boundary components, as discussed in \cref{sec:bc}. A wave $k^{\ii\alpha}$ is distributed between $\lambda\to+\infty$ and $\lambda\to-\infty$ with probabilities in the ratio $e^{2\pi\alpha}$. The statement that the returning quark arrives on the boundary component carrying the antiquark is therefore semiclassical, valid for $\pi\gs/2-\varkappa\gg1/2\pi$, which requires $m>\gb$. In general $\mathcal R$ is the amplitude between the two channels at $k\to0$. Second, \eqref{eq:anchored-R} does not depend on what happens to the quark when it reaches the boundary. A complete description of the dynamics would also require specifying this, e.g.\ by a reflection phase $e^{\ii\theta_b}$ fixed by the quark's boundary condition and the boundary counterterm. This choice inevitably breaks the dilatations, because the relative phase of $k^{\ii\varkappa}$ and $k^{-\ii\varkappa}$ is not scale invariant. The quark would then bounce between the boundary and the turning point. Bound states occur when the total phase of a round trip is a multiple of $2\pi$, $\mathcal R(\omega_n)e^{\ii\theta_b}=1$, which gives a geometric tower of Poincar\'e bound states, as in the Efimov effect,
\begin{equation}
\omega_n=-\frac{1}{4R^2k_0}\exp\Big[\frac{2\arg\Gamma(2\ii\varkappa)+\theta_b}{2\varkappa}\Big]\,e^{-\pi n/\varkappa},\qquad n\in\mathbb Z .
\end{equation}
The tower is unbounded below: the string pulls the quark into the boundary and releases an unbounded Poincar\'e energy. This resembles the fate of an operator with dimension $\Delta$ that collides with its ``shadow'' operator $1-\Delta$ (the other root of $\Delta(\Delta-1)=\nu_b^2-\frac14$) at $\Delta=\half$, so that both move into the complex plane~\cite{Kaplan:2009kr}. It is also the counterpart for a confining string of the instability of a charged particle in a strong AdS$_2$ electric field, whose effective mass then falls below the Breitenlohner--Freedman bound $\Delta=\half$ \cite{Pioline:2005pf,Faulkner:2009wj}. At $N=\infty$ the string cannot break, and for $m>\gb$ the instability is the collapse of the quark onto the boundary charge. At finite $N$ it would be cut off by screening effects, which we have not analysed.

\section{\texorpdfstring{QCD$_2$ on de~Sitter space}{QCD2 on de Sitter space}}
\label{sec:desitter}

De~Sitter space is the other maximally symmetric curved spacetime. It describes a universe expanding at a constant Hubble rate $1/R$, and every observer in it is surrounded by a cosmological horizon at distance $R$, which radiates at the Gibbons--Hawking temperature $T_{\rm dS}=1/2\pi R$ \cite{Gibbons:1977mu}. In this section we ask what happens to the mesons of the 't~Hooft model when the background expands.

The construction of \cref{sec:gauge,sec:symmetry} used three facts: a null gauge invariant under the isometries, affine null coordinates in which the Yang--Mills sector is exactly flat, and the observation that the curvature then enters only through the kinematic operator $\cK$. All three hold on two-dimensional de~Sitter space as well. For the light-front Hamiltonian, the Casimir and the meson equation, the passage from AdS$_2$ to dS$_2$ is induced by simply sending $R^2\to-R^2$. This is not an analytic continuation of the physics, but an exact statement about the two Lorentzian geometries: in affine null coordinates the dS$_2$ metric, $\dd y\,\dd\lambda-\lambda^2\dd y^2/4R^2$, is the Lorentzian AdS$_2$ metric \eqref{eq:affine-metric} with $R^2\to-R^2$, and in \cref{sec:ds-geometry} we derive all operator statements directly in Lorentzian dS$_2$. (It is reminiscent of, but different from, the continuation that relates de~Sitter correlators to those of Euclidean AdS \cite{Sleight:2020obc,Hogervorst:2021uvp,DiPietro:2021sjt}.) What the substitution does not determine are the Hilbert space and the conditions at the endpoints, and it is there that the physics changes substantially. On AdS$_2$ the collinear Casimir acts as a confining harmonic trap. On dS$_2$ it appears with the opposite sign and acts as the repulsive potential of the cosmological horizon, and at large separation it always wins over the linear potential between the quarks. The meson equation turns from a bound-state problem into a scattering problem.

Our main findings are the following:
\begin{itemize}
\item The dimensions of the mesons obey the dS$_2$ 't~Hooft equation \eqref{eq:ds-tHooft}. In it, $\Delta(1-\Delta)=\mu^2R^2$, the de~Sitter version of the mass, equals $R^2$ times 't~Hooft's mass operator \emph{minus} the collinear Casimir, which on the quark--antiquark wavefunction is the Legendre operator $-\frac{\dd}{\dd x}x(1-x)\frac{\dd}{\dd x}$ of \eqref{eq:AdS-tHooft}.
\item Even at $N=\infty$, where mesons cannot decay into other mesons, every meson is unstable. The de~Sitter expansion pulls the quark and the antiquark out of causal contact and stretches the flux tube across the horizon (even if it cannot break at $N=\infty$). We call this effect \emph{cosmological dissociation}.
\item The mesons are therefore resonances, with complex $\Delta_n$. We can define them by a certain analytic continuation of the AdS$_2$ spectrum in $R^2$, equivalent to requiring purely outgoing waves at the endpoints $x\to0,1$ of the momentum fraction. These endpoints are the configurations in which one constituent carries no light-front momentum, i.e.\ sits at future or past infinity $\mathcal I^\pm$, and ``outgoing'' means that constituents only escape through $\mathcal I^+$, flying apart across the horizon, while none arrive from $\mathcal I^-$ (Gamow states, \cref{sec:ds-resonances}). We compute them in \cref{sec:ds-mesons} (see \cref{tab:ds}).
\item The widths are thermal. Up to a prefactor, $\Gamma_n\propto\exp[-(V_{\rm top}-\mu_n)/T_{\rm dS}]$, where $V_{\rm top}$ is the top of the energy barrier that a quark--antiquark pair joined by a string must cross in the static patch, evaluated with the planar renormalized quark mass. About $\gb^2R^2$ mesons, those smaller than the horizon, are narrow and long-lived.
\item In the chiral limit the pion remains exactly massless. For small non-zero quark mass, a light bound pion exists only above a critical string tension $T_s=3/2\pi R^2$, of order the Hubble rate squared. Below the same threshold there is no barrier at all, and every meson is broad.
\end{itemize}
This section is organized as follows. In \Cref{sec:ds-geometry} we derive the light-front Hamiltonian, in \cref{sec:ds-equation} the meson equation, and in \cref{sec:ds-halfspace} we explain why the light-front Hilbert space no longer fixes the conditions at the endpoints. In \Cref{sec:ds-dissociation} we identify the physics behind this, and in \cref{sec:ds-resonances} we study the de~Sitter meson resonances. We present some further results in \cref{sec:ds-mesons,sec:ds-pion}.

\subsection{Geometry and the light-front Hamiltonian}
\label{sec:ds-geometry}

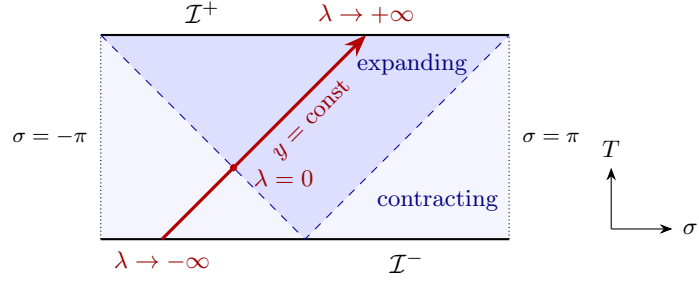
\begin{figure}[t]
\centering
\begin{tikzpicture}[scale=1.35,>=Stealth]
  \fill[blue!4] (-2,-1) rectangle (2,1);
  \fill[blue!13] (0,-1) -- (2,1) -- (-2,1) -- cycle;
  \draw[thick] (-2,1) -- (2,1);
  \draw[thick] (-2,-1) -- (2,-1);
  \node[font=\small,anchor=south] at (-1.0,1.02) {$\mathcal I^+$};
  \node[font=\small,anchor=north] at (1.0,-1.02) {$\mathcal I^-$};
  \draw[densely dotted] (-2,-1) -- (-2,1); \draw[densely dotted] (2,-1) -- (2,1);
  \node[font=\scriptsize,anchor=east] at (-2.03,0) {$\sigma=-\pi$};
  \node[font=\scriptsize,anchor=west] at (2.03,0) {$\sigma=\pi$};
  \draw[dashed,blue!60!black] (-2,1) -- (0,-1) -- (2,1);
  \node[font=\footnotesize,blue!50!black] at (1.05,0.72) {expanding};
  \node[font=\footnotesize,blue!50!black] at (1.3,-0.62) {contracting};
  \draw[very thick,red!70!black,->] (-1.4,-1) -- node[pos=0.66,below,sloped,font=\footnotesize,red!60!black] {$y={\rm const}$} (0.6,1);
  \fill[red!70!black] (-0.7,-0.3) circle (0.035);
  \node[font=\footnotesize,red!60!black,anchor=west] at (-0.6,-0.4) {$\lambda=0$};
  \node[font=\footnotesize,red!60!black,anchor=south] at (0.6,1.02) {$\lambda\to+\infty$};
  \node[font=\footnotesize,red!60!black,anchor=north] at (-1.4,-1.02) {$\lambda\to-\infty$};
  \draw[->] (3.0,-0.9) -- (3.0,-0.3) node[above,font=\footnotesize] {$T$};
  \draw[->] (3.0,-0.9) -- (3.6,-0.9) node[right,font=\footnotesize] {$\sigma$};
\end{tikzpicture}
\caption{Global dS$_2$ (a cylinder of finite height; the dotted edges are identified). The expanding Poincar\'e patch (dark) is bounded by the cosmological horizon (dashed), the future light cone of a point on $\mathcal I^-$; its complement is the contracting patch. The light front is a complete null geodesic $y={\rm const}$ from $\mathcal I^-$ to $\mathcal I^+$, with affine parameter $\lambda\in\mathbb R$; it crosses the horizon at $\lambda=0$, and its endpoints are antipodal. Unlike in AdS$_2$, where every massive particle bounces between the boundaries and crosses the light front repeatedly, here the light front is not a Cauchy surface: under light-front evolution particles enter at $\mathcal I^-$ ($\lambda\to-\infty$) and leave through $\mathcal I^+$ ($\lambda\to+\infty$).}
\label{fig:ds-geometry}
\end{figure}

De~Sitter space of radius $R$ is the hyperboloid $-X_0^2+X_1^2+X_2^2=R^2$. In global coordinates,
\begin{equation}\label{eq:ds-global}
\dd s^2=\frac{R^2}{\cos^2T}\big(\dd T^2-\dd\sigma^2\big),\qquad -\tfrac\pi2<T<\tfrac\pi2,\quad \sigma\sim\sigma+2\pi ,
\end{equation}
it is a cylinder of finite height, bounded by past and future infinity $\mathcal I^\mp$ at $T=\mp\pi/2$. Comparing with \eqref{eq:global}, dS$_2$ is AdS$_2$ with the roles of time and space exchanged (and the overall sign of the metric reversed). The expanding Poincar\'e patch is
\begin{equation}\label{eq:ds-poincare}
\dd s^2=\frac{R^2}{\tau^2}\big(\dd\tau^2-\dd\xi^2\big)=\Omega^2\,\dd x^+\dd x^-,\qquad x^\pm=\tau\pm\xi,\qquad \Omega=\frac{R}{|\tau|}=-\frac{2R}{x^++x^-},
\end{equation}
with conformal time $\tau<0$ and $\mathcal I^+$ at $\tau\to0^-$. In terms of cosmic time $t_c=-R\log(-\tau/R)$ the metric is $\dd t_c^2-e^{2t_c/R}\dd\xi^2$, a two-dimensional universe expanding exponentially at the Hubble rate $1/R$. The patch is the causal future of a point on $\mathcal I^-$, bounded by a cosmological horizon, and its complement in global dS$_2$ is the contracting patch $\tau>0$ (\cref{fig:ds-geometry}). The isometries act as $x^+\to f(x^+)$, $x^-\to-f(-x^-)$ with $f$ a M\"obius map, and the Killing vectors are
\begin{equation}\label{eq:ds-killing}
L_{-1}=\partial_+-\partial_-,\qquad L_0=x^+\partial_++x^-\partial_-,\qquad L_1=(x^+)^2\partial_+-(x^-)^2\partial_- ,
\end{equation}
i.e.\ $L_{-1}=\partial_\xi$ and $L_0=\tau\partial_\tau+\xi\partial_\xi$. Each is again of the form $a(x^+)\partial_++b(x^-)\partial_-$, which is all that was used in \cref{sec:gauge} to show that the null gauge $A_-=0$ is invariant under the isometries.

The affine parameter of the null lines $x^+=y$ is fixed by $\dd\lambda=\Omega^2\dd x^-$,
\begin{equation}\label{eq:ds-lambda}
\lambda=-\frac{4R^2}{x^++x^-}=-\frac{2R^2}{\tau},\qquad \Omega=\frac{|\lambda|}{2R},\qquad
\dd s^2=\dd y\,\dd\lambda-\frac{\lambda^2}{4R^2}\,\dd y^2,\qquad \sqrt{-g}=\tfrac12 .
\end{equation}
The only change with respect to \eqref{eq:affine-metric} is the sign of $g_{yy}$. Inside the expanding patch $\lambda\in(0,\infty)$, with $\lambda\to\infty$ on $\mathcal I^+$ and $\lambda=0$ on the cosmological horizon. Continued through the horizon, $\lambda<0$ covers the contracting patch, and $\lambda\to-\infty$ on $\mathcal I^-$. The light front is the complete null geodesic $\lambda\in\mathbb R$, which runs from $\mathcal I^-$ to $\mathcal I^+$ and whose two endpoints are antipodal points of dS$_2$. In these coordinates
\begin{equation}\label{eq:ds-killing-affine}
L_{-1}=\partial_y,\qquad L_0=y\,\partial_y-\lambda\,\partial_\lambda,\qquad L_1=y^2\partial_y-\big(4R^2+2y\lambda\big)\partial_\lambda .
\end{equation}
Two differences from AdS$_2$ are worth stressing. First, $L_1$ has $-4R^2$ in place of $4R^2$. Second, $\partial_y$ at fixed $\lambda$ is now spacelike ($g_{yy}<0$): it is the Poincar\'e translation $\partial_\xi$. Light-front evolution in de~Sitter space is generated by a conserved but indefinite charge, as it must be, since de~Sitter space has no positive conserved energy. The generator that is timelike inside the static patch centred at $\xi=0$ is the dilatation $L_0$, which acts kinematically on the null line.

The derivation of \cref{sec:gauge} now goes through line by line. The Yang--Mills action depends only on the area form, which is flat in $(y,\lambda)$, so the gauge sector and the Coulomb term \eqref{eq:coulomb} are unchanged, and so is the mass term. The only geometric input is again the symplectic term, now with $\partial_{x^+}|_{x^-}=\partial_y+\frac{\lambda^2}{4R^2}\partial_\lambda$. The charges are
\begin{equation}\label{eq:ds-charges}
\boxed{\;H=\Htf-\frac{\cK}{4R^2},\qquad D=-\cD,\qquad K=-4R^2\,\cP\;}
\end{equation}
which satisfy the algebra \eqref{eq:sl2}, and the Casimir \eqref{eq:casimir} becomes
\begin{equation}\label{eq:ds-casimir}
\cC_2=-4R^2\,\cP\,\Htf+\Big[\tfrac12\big(\cK\cP+\cP\cK\big)-\cD^2\Big]=-R^2\,\hat M^2_{\rm flat}+\cC_{\rm coll}.
\end{equation}
The algebraic content of \cref{sec:gauge,sec:symmetry,sec:equation} (the charges, their algebra, the Casimir decomposition and the reduction to the internal wavefunction) thus holds on dS$_2$ after replacing $R^2\to-R^2$. Statements that rely on the Hilbert space, such as the positivity of $c_n$, the lowest-weight centre-of-mass factor \eqref{eq:lowest-weight} and the vacuum argument of \cref{sec:vacuum}, do not carry over though, for reasons that we explain in \cref{sec:ds-halfspace}.

\subsection{\texorpdfstring{The dS$_2$ 't~Hooft equation}{The dS2 't Hooft equation}}
\label{sec:ds-equation}

A unitary irreducible representation of the de~Sitter group $SO(1,2)$ is labelled by $\Delta$, with $\cC_2=\Delta(\Delta-1)$, and a free particle of mass $\mu$ carries
\begin{equation}\label{eq:ds-mass}
\Delta(1-\Delta)=\mu^2R^2 .
\end{equation}
De~Sitter space has no boundary at spatial infinity, and $\Delta$ means something different than in AdS. A field of mass $\mu$ decays at late times as $(-\tau)^{\Delta}$ and $(-\tau)^{1-\Delta}$, and \eqref{eq:ds-mass} is its wave equation evaluated on these falloffs. Heavy particles have complex $\Delta$ with real part $\half$, so their late-time falloff oscillates. In the language of representation theory, the principal series $\Delta=\half+\ii\nu$ describes particles with $\mu R>\half$, the complementary series $0<\Delta<1$ light particles, and the discrete series $\Delta=j\in\{1,2,\dots\}$ fields with $\mu^2R^2=j(1-j)\le0$, such as the massless scalar ($j=1$) \cite{Anninos:2023lin,Sun:2021thf}. The same $\Delta$ (together with $1-\Delta$) are the scaling dimensions of the boundary operators that a bulk field induces at late times on $\mathcal I^+$, and they determine the quasinormal frequencies $\omega=-\ii(\Delta+k)/R$ of the static patch \cite{Jafferis:2013qia,Loparco:2023rug}. They are the de~Sitter counterparts of the AdS$_2$ boundary conformal dimensions.

On a single quark, $H=\frac{m^2}{4k}+\frac1{4R^2}\partial_kk\,\partial_k$ and $K=-4R^2k$, and the Casimir is the number $-(\ms^2+\frac14)$. Hence $\Delta_\psi=\half\pm\ii\ms$: a Dirac fermion of any mass is in the principal series (of the double cover of $\SLtwo$) \cite{Pethybridge:2021rwf,Letsios:2025pqo}. On the colour-singlet quark--antiquark sector the reduction of \cref{sec:equation} goes through unchanged, and the internal wavefunction of a state of definite Casimir obeys
\begin{equation}\label{eq:ds-tHooft}
\Delta(1-\Delta)\,\varphi(x)=\big(\ms^2-\gs\big)\Big(\frac1x+\frac1{1-x}\Big)\varphi(x)-\gs\,\PV_0^1\dd x'\,\frac{\varphi(x')}{(x-x')^2}+\frac{\dd}{\dd x}\Big[x(1-x)\,\frac{\dd\varphi}{\dd x}\Big]
\end{equation}
with $\gs=\gb^2R^2$ and $\ms=mR$ in terms of the de~Sitter radius. This is the dS$_2$ 't~Hooft equation: $\Delta(1-\Delta)$ equals $R^2$ times 't~Hooft's mass operator \emph{minus} the collinear Casimir. In the rapidity variables of \cref{sec:rapidity} it reads
\begin{equation}\label{eq:ds-Bp}
B_{\rm dS}(p)\,\varphi=\frac{E}{4}\,{\rm sech}^2\theta\,\varphi,\qquad B_{\rm dS}(p)=-\frac{p^2}{4}+\gs\,h(p)+\ms^2,\qquad E\equiv\Delta(1-\Delta).
\end{equation}
Here, as in \cref{sec:rapidity}, $x=\half(1+\tanh\theta)$, $\theta$ is the relative rapidity of the constituents and $p=-\ii\partial_\theta$ its conjugate momentum, and $h(p)=\frac{\pi p}{2}\coth\frac{\pi p}{2}-1$ combines the 't~Hooft kernel with the self-energy. The only change with respect to \eqref{eq:Bp} is the sign of the rapidity kinetic energy $p^2/4$ that is flipped.

\subsection{De Sitter light-front and boundary conditions}
\label{sec:ds-halfspace}

In short, we found that on AdS$_2$ the boundary reflects every particle back into the interior, and the light front sees the whole Hilbert space. On dS$_2$ particles leave through future infinity, so the light-front description is incomplete, and the meson equation must be supplemented by a prescription at the endpoints of the momentum fraction. The rest of this subsection explains this in more detail and may be skipped by readers interested only in the results.

Since de~Sitter space has no positive conserved energy, it is worth stating clearly what kind of equation \eqref{eq:ds-tHooft} is. It is not an energy eigenvalue problem. The light-front Hamiltonian $H$ generates the spacelike translation $\partial_\xi$, and \eqref{eq:ds-tHooft} is the eigenvalue equation of the quadratic Casimir \eqref{eq:ds-casimir}, built from the three conserved charges on the null line, exactly as in AdS$_2$. The Casimir is the de~Sitter notion of mass, $\Delta(1-\Delta)=\mu^2R^2$ for a free particle, and it reduces to an equation for the internal wavefunction because it commutes with the kinematic generators $K=-4R^2\cP$ and $D$. Energy conservation is thus replaced by the isometry algebra, and the only dynamical input is the flat 't~Hooft operator $\hat M^2_{\rm flat}$. Nor is it a calculation in the interior of a static patch. The light front is a complete null geodesic through a bifurcation point: it is one generator of the bifurcate Killing horizon shared by a static patch and its antipodal one, and the static-patch time translation, the boost $L_0$, acts on it kinematically, as the dilatation $\lambda\to e^s\lambda$ with fixed point $\lambda=0$ (\cref{sec:ds-geometry}). The static-patch Hamiltonian is therefore never diagonalized (its spectrum is continuous, see \cref{sec:ds-mesons}); what is quantized is the field on the horizon. One could instead use Eddington--Finkelstein coordinates adapted to global dS$_2$, as in \cref{sec:affine} for AdS$_2$: with $u=T+\sigma$ and $\lambda_g=2R^2\tan T$ the metric is $\dd s^2=\dd u\,\dd\lambda_g-(R^2+\lambda_g^2/4R^2)\dd u^2$, and light-front evolution is the rotation $\partial_\sigma$ of the spatial circle. The light-front time $u$ is then periodic, and a shift $u\to u+2\pi$ must act trivially on physical states (up to the fermion sign), so the light-front Hamiltonian of that frame, the compact generator of $SO(1,2)$, must have integer spectrum. This is nothing but the statement that physical states form representations of the de~Sitter group, a condition that the Poincar\'e-type frame used here does not impose. On a single null line, where $K\le0$, only discrete-series states can satisfy it (\cref{sec:ds-remarks}); the resonances with complex $\Delta$ found below do not, which is another way of seeing that they are late-time data rather than states of a Hilbert space. Restricted to either of the half-lines $\lambda\gtrless0$, each of which is a horizon of one of the two patches, the light-front vacuum is a thermal state with respect to this boost at inverse temperature $2\pi$ \cite{Kay:1988mu}, i.e.\ at the Gibbons--Hawking temperature $T_{\rm dS}=1/2\pi R$ in static time \cite{Gibbons:1977mu}. This is the origin of the Boltzmann factors in the widths of \cref{sec:ds-mesons}, and the resonances $\Delta_n$ found below are static-patch observables: they are its quasinormal frequencies, $\omega=-\ii(\Delta_n+k)/R$.

On AdS$_2$ the light-front Hilbert space determined everything: $K=4R^2\cP\ge0$, every state belongs to a positive-energy (lowest-weight) discrete-series representation $D^+_\Delta$, i.e.\ one in which $H_{\rm glob}$ is bounded from below (see the footnote in \cref{sec:solve}), and normalizability, together with the choice of quark quantization where both are allowed, selected the endpoint behaviour of the wavefunctions and hence the spectrum. On dS$_2$ this fails, for a reason that is both group-theoretical and geometrical.

The group-theoretical reason is that $K=-4R^2\cP\le0$ on the light-front Fock space. The operators $D$ and $K$ generate the two-dimensional ``$ax+b$'' subgroup of $\SLtwo$, whose only infinite-dimensional unitary irreducible representations are the two in which $K$ is positive or negative. Restricted to this subgroup, the discrete series $D^\pm$ of $\SLtwo$ contain only one of them, but the principal and complementary series contain both. A Hilbert space on which $K\le0$ can therefore carry a unitary representation of $\SLtwo$ only if it consists of discrete-series representations of one sign ($D^-$ here) and trivial ones, whereas a massive particle in de~Sitter space is in the principal series. (The mixed solutions with $\Delta=n+1$ found below are precisely such a $D^-$ content.) Concretely, the dynamical generator $H$ is not essentially self-adjoint on the light-front Fock space, and none of its self-adjoint extensions is compatible with the dilatations. In the single-particle sector, every solution of $Hf=\ii f$ or $Hf=-\ii f$ behaves near $k=0$ as a combination of $k^{\ii\ms}$ and $k^{-\ii\ms}$, both of which are square integrable, so the deficiency indices are $(1,1)$ and probability can flow through $k=0$. A self-adjoint extension would fix the relative phase of $k^{\ii\ms}$ and $k^{-\ii\ms}$, which the dilatations $k\to e^sk$ rotate. On AdS$_2$ the corresponding behaviours were the real powers $k^{\pm\ms}$; for $\ms\ge\half$ normalizability selected one of them, and for $\ms<\half$ there was a freedom in the quantization of the quark, i.e.\ standard, alternate or mixed (\cref{sec:oneparticle,sec:bc}).

The geometrical reason is that small light-front momentum means $|\lambda|\to\infty$, i.e.\ $\mathcal I^\pm$. On AdS$_2$ a massive particle bounces between the reflecting boundaries and crosses the light front again and again. On dS$_2$ it enters the light front from $\mathcal I^-$ and leaves it through $\mathcal I^+$ (\cref{fig:ds-geometry}), and nothing returns. The probability flux through $k=0$ describes precisely these particles entering the light front from $\mathcal I^-$ and leaving it through $\mathcal I^+$.

What survives, though, is the algebraic structure. The kinematic generators act as before, and the Casimir \eqref{eq:ds-casimir} is an exact operator identity, so \eqref{eq:ds-tHooft} holds for the internal wavefunction of any quark--antiquark state of definite Casimir. For free fields the light-front Fock vacuum is the restriction of the Bunch--Davies state to the algebra of the null geodesic. The null geodesic is one generator of the bifurcate Killing horizon of a static patch, and the Bunch--Davies state is invariant under its boost, so this follows from the uniqueness theorem of Kay and Wald \cite{Kay:1988mu}, which is formulated for scalar fields but applies to the Dirac field in the same way. For the interacting theory we have not shown that $\cK$ creates no pairs from the Fock vacuum: the argument of \cref{sec:vacuum} used $\Delta>1$, which fails for the complex dimensions found below, and in our computations we assume the light-front vacuum throughout. A partial analysis of this assumption, which shows that the evolution cannot create pairs with non-zero light-front momentum but leaves the zero-mode sector open, is given in Appendix~\ref{app:ds-vacuum}. What the light front no longer supplies are the conditions at the endpoints $x=0,1$, which encode what happens to a constituent that reaches $\mathcal I^\pm$. On AdS$_2$ these were boundary conditions. On dS$_2$ they are a prescription for the flux, and as we now show, all of the new physics is hidden in these conditions.

\subsection{Endpoint scattering and cosmological dissociation}
\label{sec:ds-dissociation}

Recall from \cref{sec:rapidity} that the rapidity momentum $p$ is conjugate to the relative rapidity of the constituents, and that large $|p|$ corresponds to widely separated constituents; in the heavy-quark limit $p$ is literally correlated with their separation (we provide more details below). The function $B_{\rm dS}$ is positive at $p=0$, where it equals $\ms^2$, and decreases as $-p^2/4$ at large $p$. It therefore has a pair of real zeros $\pm p_*$ for every $\gs\ge0$ and $\ms>0$. Equivalently, the indicial equation, obtained from \eqref{eq:indicial} by $R^2\to-R^2$,
\begin{equation}\label{eq:ds-indicial}
\gs\big(\pi\beta\cot\pi\beta-1\big)+\ms^2+\beta^2=0 ,
\end{equation}
always has a pair of purely imaginary roots $\beta=\pm\ii y_*$, with $y_*=p_*/2$ the positive solution of
\begin{equation}\label{eq:ds-ystar}
y_*^2=\gs\big(\pi y_*\coth\pi y_*-1\big)+\ms^2 ,
\end{equation}
in addition to real roots. Near the endpoints the wavefunction therefore contains oscillating terms $x^{\pm\ii y_*}$ and $(1-x)^{\pm\ii y_*}$, which are bounded but carry a probability flux into or out of the endpoint. Their meaning is clear from the preceding subsection: $x\to0$ is the configuration in which the quark carries no light-front momentum, i.e.\ sits on $\mathcal I^\pm$. The oscillating terms describe a constituent that escapes to future infinity (or arrives from past infinity). For free quarks $y_*=\ms$, and $x^{\pm\ii\ms}$ is the principal-series label $\Delta_\psi-\half=\pm\ii\ms$ of the quark itself. With interactions, $y_*$ is the label of a constituent that reaches $\mathcal I^\pm$ dragging its flux tube; at strong coupling $y_*\simeq\pi\gs$.

The origin of the real zeros is transparent in the heavy-quark limit, where the rapidity momentum $p$ becomes the separation of the quarks, $p=\ms\,r/R$ (see \cref{sec:rapidity}). Repeating the expansion that led to \eqref{eq:heavy-H}, with $\Delta=\half+\ii(2\ms+\varepsilon)$, one finds
\begin{equation}\label{eq:ds-heavy}
R\,H_{\rm rel}=-\frac1{\ms}\,\partial_{\hat r}^2-\frac{\ms}{4}\,\hat r^2+\frac{\pi\gs}{2}\,|\hat r|,\qquad \varepsilon=R\,E^{\rm rel}-\frac{\gs}{\ms}+O(1/\ms),
\end{equation}
with $\hat r=r/R$. It will be convenient to define $\kappa_h\equiv\pi\gs/\sqrt{2\ms}$ (the parameter called $\eta$ in \eqref{eq:heavy-H}). Equation~\eqref{eq:ds-heavy} is $R$ times the Hamiltonian of two quarks of reduced mass $m/2$ in the static patch, bound by the linear potential $T_s r$ and pushed apart by the inverted harmonic potential $-\frac{m}{4R^2}r^2$ due to the cosmological horizon. More generally, consider a static quark--antiquark pair at $\rho=\pm r/2$ in the static patch $\dd s^2=(1-\rho^2/R^2)\dd t^2-\dd\rho^2/(1-\rho^2/R^2)$, joined by a straight flux tube, with $r$ the separation in the static coordinate $\rho$. Its energy is
\begin{equation}\label{eq:ds-static}
V(r)=2m\sqrt{1-\frac{r^2}{4R^2}}+T_s\,r,\qquad V_{\max}=\max_rV=2\sqrt{m^2+T_s^2R^2},
\end{equation}
since the flux tube contributes $T_s\int\sqrt{-g_{tt}g_{\rho\rho}}\,\dd\rho=T_s\,r$ while the rest energy of the quarks is redshifted. The potential rises from $2m$ to a maximum at $r_*=2T_sR^2/\sqrt{m^2+T_s^2R^2}<2R$ and then falls. Every meson is therefore metastable. At $N=\infty$ the string cannot break, since pair creation is suppressed by $1/N$, but nothing prevents the pair from tunnelling through the barrier, after which the quarks fall through the horizon. The quark and the antiquark are then carried out of causal contact, connected by a flux tube that stretches across the horizon. This \emph{cosmological dissociation} is a leading-order planar effect of the expansion, not suppressed by $1/N$. In the rapidity language it is the statement that the collinear Casimir, a harmonic trap with coefficient $+\frac14$ on AdS$_2$, is an inverted one with coefficient $-\frac14$ on dS$_2$, and always overwhelms the confining term $\gs h(p)\simeq\frac{\pi\gs}{2}|p|$ at large separation.

\subsection{Resonances as the continuation of the AdS spectrum}
\label{sec:ds-resonances}

\begin{figure}[t]
\centering
\includegraphics[width=\textwidth]{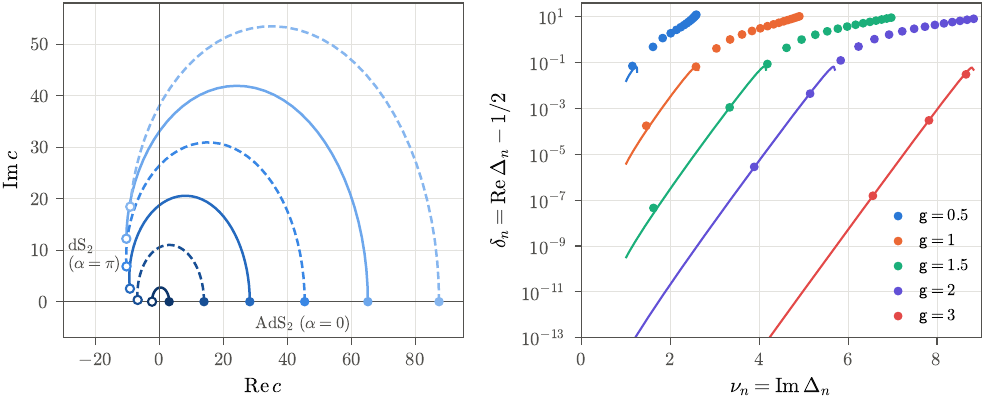}
\caption{Left: the spectral flow of the lowest eigenvalues $c_n(\alpha)$ of \eqref{eq:ds-continuation} from AdS$_2$ ($\alpha=0$, filled circles on the positive real axis) to dS$_2$ ($\alpha=\pi$, open circles at $c=-E_n$), for $\gs=1$, $\ms=\half$; even (solid) and odd (dashed) levels. Right: the widths $\delta_n={\rm Re}\,\Delta_n-\half$ of the dS$_2$ resonances against their masses $\nu_n={\rm Im}\,\Delta_n$ at $\ms=\half$ and $\gs=0.5,\,1,\,1.5,\,2,\,3$ (points), compared with the Gamow formula \eqref{eq:ds-gamow} (lines). The lines end at the top of the barrier, $\nu=2\sqrt{\max B_{\rm dS}}$; above it the levels become broad.}
\label{fig:ds-flow}
\end{figure}

In quantum mechanics, a metastable state, such as a particle trapped behind a barrier through which it can tunnel, has a complex energy $E-\ii\Gamma/2$, where $\Gamma$ is the decay rate. Its wavefunction is purely outgoing far away; it corresponds to a Gamow state. Under an appropriate complex scaling, or rotation of a coordinate or momentum into the complex plane, we can turn outgoing waves into decaying ones. Resonances then become ordinary eigenvalues of a non-hermitian operator (a standard prescription of computing them \cite{Aguilar:1971ve,Balslev:1971vb}). Since the mesons decay, the natural spectral data are the positions of the resonances: the complex dimensions $\Delta_n$ of states whose wavefunctions are purely \emph{outgoing} at both endpoints, i.e.\ Gamow states, with the quarks escaping through $\mathcal I^+$ and none arriving from $\mathcal I^-$. Here it is convenient to define them by analytic continuation from AdS$_2$. To do this, we replace $R^2\to e^{\ii\alpha}R^2$ in the AdS$_2$ equation \eqref{eq:Bp}, i.e.\ $\gs\to e^{\ii\alpha}\gs$ and $\ms^2\to e^{\ii\alpha}\ms^2$, with the eigenvalue $c=\Delta(\Delta-1)$ not rescaled:
\begin{equation}\label{eq:ds-continuation}
B_\alpha(p)\,\varphi=\frac c4\,{\rm sech}^2\theta\,\varphi,\qquad B_\alpha(p)=\frac{p^2}{4}+e^{\ii\alpha}\big(\gs\,h(p)+\ms^2\big),\qquad 0\le\alpha\le\pi .
\end{equation}
For $0\le\alpha<\pi$, $B_\alpha$ has no zeros on the real axis. The Birman--Schwinger operator \eqref{eq:BS} then remains compact, now complex symmetric rather than hermitian, and its eigenvalues $c_n(\alpha)$ depend analytically on $\alpha$. At $\alpha=\pi$, $B_\pi=-B_{\rm dS}$ and $c=-E$. As $\alpha\to\pi$ the zeros of $B_\alpha$ approach $+p_*$ from below and $-p_*$ from above, so the limiting Green's function behaves as $e^{-\ii p_*|\theta|}$, which is outgoing at both endpoints. Equivalently, after the complex rotation $p\to p\,e^{\ii\phi}$ with $\phi>0$ the operator is compact also at $\alpha=\pi$, and its eigenvalues are independent of $\phi$, as expected for resonances. We define the de~Sitter spectrum by
\begin{equation}\label{eq:ds-def}
E_n\equiv\Delta_n(1-\Delta_n)=-c_n(\alpha\to\pi),
\end{equation}
and the opposite continuation, $\alpha\to-\pi$, gives the complex-conjugate (incoming) set. This is the precise sense in which $R^2\to-R^2$ maps the AdS$_2$ boundary spectrum to dS$_2$. \Cref{fig:ds-flow} shows the flow of the lowest levels from the AdS$_2$ values (real, positive $c_n$) to the dS$_2$ resonances.

We write $\Delta_n=\half+\delta_n+\ii\nu_n$, so that $E_n=\frac14+\nu_n^2-\delta_n^2-2\ii\nu_n\delta_n$. The two real numbers have a simple meaning: $\nu_n\simeq\mu_nR$ measures the mass and $\delta_n$ the width. A stable particle has $\delta_n=0$: a stable massive particle with $\mu R>\half$ carries a principal-series representation, $\Delta=\half+\ii\nu$ with $\nu=\sqrt{\mu^2R^2-\frac14}$, and the resonances are displaced off the principal-series line ${\rm Re}\,\Delta=\half$ by $\delta_n$. For $\delta_n>0$ the state is a resonance of mass $\mu_n\simeq\nu_n/R$: in the flat-space limit $E_n/R^2\to M^2-\ii M\Gamma$ with
\begin{equation}\label{eq:ds-width}
\Gamma_n=\frac{2\delta_n}{R}.
\end{equation}
Of the four dimensions with $\Delta(1-\Delta)\in\{E_n,E_n^*\}$, the decaying set is $\{\Delta_n,\Delta_n^*\}$, and we expect the corresponding quasinormal modes of the static patch to decay as $e^{-(\frac12+\delta_n+k)t/R}$ rather than $e^{-(\frac12+k)t/R}$, the additional rate being related to the dissociation process itself.

\paragraph{Free quarks.} At $\gs=0$, \eqref{eq:ds-tHooft} is again a hypergeometric problem, now with imaginary exponents: $\varphi=x^{a}(1-x)^{b}P_n^{(2b,2a)}(2x-1)$ with $a,b\in\{\pm\ii\ms\}$ and $\Delta=1+a+b+n$. The outgoing solutions, $a=b=\ii\ms$, give
\begin{equation}\label{eq:ds-free}
\Delta_n=1+2\ii\ms+n=2\Delta_\psi+n ,
\end{equation}
the late-time ``double-trace'' dimensions of two free fermions, each in the principal series $\Delta_\psi=\half+\ii\ms$. These are falloffs (poles) rather than unitary representations: the two-particle Hilbert space of two principal-series particles decomposes into a continuum of principal-series representations together with discrete-series ones \cite{Repka:1978,Penedones:2023uqc}. 
The mixed solutions, $a=-b=\ii\ms$, give real $\Delta=n+1$, i.e.\ $E=-n(n+1)$. These are the discrete-series representations $D^-_{n+1}$, which are known to appear in the tensor product of two principal-series representations of $\SLtwo$ \cite{Repka:1978,Penedones:2023uqc}. Their wavefunctions, $e^{2\ii\ms\theta}$ times a polynomial in $\tanh\theta$, are reflectionless plane waves in the rapidity. The light-front equation thus captures both the late-time poles and the discrete-series content of the free two-particle problem.

\subsection{The meson resonances}
\label{sec:ds-mesons}

We solve \eqref{eq:ds-continuation} at $\alpha=\pi$ by the momentum-space Nystr\"om method of Appendix~\ref{app:numerics}, on the rotated contour $p=s\,e^{\ii\phi}$, $s\in\mathbb R$. The eigenvalues are found to be independent of $\phi$ (we use $\phi=0.3$ and $0.45$) and of the grid (up to the accuracy of our method). At $\alpha=0$ the same code reproduces the AdS$_2$ spectrum, and tracking the eigenvalues in $\alpha$ connects the two (\cref{fig:ds-flow}). \Cref{tab:ds} lists the lowest resonances for $\ms=\half$. Three regimes are visible. The lowest states are very narrow: at $\gs=2$ the width of the ground state is below the $10^{-9}$ accuracy of the numerics (the semiclassical estimate below gives $\delta_0\approx4\times10^{-12}$), and $\delta_n$ grows by several orders of magnitude per level. Near $\nu\approx2\sqrt{\max_pB_{\rm dS}}$, the top of the barrier, $\delta_n$ reaches $10^{-2}$--$10^{-1}$, and above it the levels are broad, with $\delta_n=O(1)$, and no longer meson-like. (In \cref{tab:ds} the levels are labelled by increasing $\nu_n$, with parity $(-1)^n$.) For $\gs<3/\pi^2$, $B_{\rm dS}$ has no maximum at $p\neq0$, i.e.\ no barrier, and every level is broad. This is the same critical coupling $\gs_c$ that controls the pion (\cref{sec:ds-pion}): from $B_{\rm dS}\simeq\ms^2+\frac14\big(\frac{\pi^2\gs}{3}-1\big)p^2$ at small $p$, the barrier exists only when $B_{\rm dS}$ initially rises.

\begin{table}[t]
\centering\footnotesize
\begin{tabular}{cc c cc cc}
\toprule
$\gs$ & $n$ & ${\rm Re}\,E_n$ & $-{\rm Im}\,E_n$ & $-{\rm Im}\,E_n$ (Gamow) & $\nu_n$ & $\delta_n$\\
\midrule
$1$ & 0 & 2.35822 & $5.1\times10^{-4}$ & $3.2\times10^{-4}$ & 1.4520 & $1.8\times10^{-4}$\\
 & 1 & 6.85623 & $0.3411$ & $0.3599$ & 2.5711 & $0.06634$\\
 & 2 & 9.26790 & $2.530$ & -- & 3.0318 & $0.4173$\\
 & 3 & 10.34589 & $6.850$ & -- & 3.3389 & $1.026$\\
 & 4 & 10.39515 & $12.2$ & -- & 3.6081 & $1.695$\\
 & 5 & 9.15023 & $18.4$ & -- & 3.8329 & $2.406$\\
\midrule
$2$ & 0 & 3.24194 & $<10^{-9}$ & $1.3\times10^{-11}$ & 1.7297 & $<10^{-9}$\\
 & 1 & 15.32733 & $2.2\times10^{-5}$ & $2.1\times10^{-5}$ & 3.8830 & $2.8\times10^{-6}$\\
 & 2 & 26.68248 & $0.04565$ & $0.04807$ & 5.1413 & $4.4\times10^{-3}$\\
 & 3 & 34.22662 & $1.456$ & -- & 5.8303 & $0.1249$\\
 & 4 & 38.83214 & $6.283$ & -- & 6.2319 & $0.5041$\\
 & 5 & 42.55735 & $14.0$ & -- & 6.5904 & $1.061$\\
\midrule
$3$ & 0 & 3.86500 & $<10^{-9}$ & $1.5\times10^{-19}$ & 1.9013 & $<10^{-9}$\\
 & 1 & 22.76422 & $<10^{-9}$ & $2.4\times10^{-11}$ & 4.7449 & $<10^{-9}$\\
 & 2 & 43.21733 & $2.0\times10^{-6}$ & $2.0\times10^{-6}$ & 6.5549 & $1.6\times10^{-7}$\\
 & 3 & 61.37144 & $4.7\times10^{-3}$ & $4.8\times10^{-3}$ & 7.8180 & $3.0\times10^{-4}$\\
 & 4 & 75.10547 & $0.5326$ & $0.6223$ & 8.6520 & $0.03078$\\
 & 5 & 83.92857 & $4.414$ & -- & 9.1508 & $0.2412$\\
\bottomrule
\end{tabular}
\caption{Lowest dS$_2$ meson resonances for $\ms=\half$ from the complex-scaled Nystr\"om method, with $E_n=\Delta_n(1-\Delta_n)$ and $\Delta_n=\half+\delta_n+\ii\nu_n$. The column ``Gamow'' is the semiclassical width \eqref{eq:ds-gamow} evaluated at the exact ${\rm Re}\,E_n$; it is omitted for broad levels. The real parts are stable to all digits shown under changes of the grid and of the rotation angle; imaginary parts below $10^{-9}$ are beyond the accuracy of the method, and for these levels the Gamow column gives the width.}
\label{tab:ds}
\end{table}

\paragraph{Semiclassics.} The structure is captured quantitatively by the rapidity semiclassics of \cref{sec:rapidity}, applied to the symbol $4\cosh^2\theta\,B_{\rm dS}(p)$ at the level $\nu^2=E-\frac14=-(\Delta-\half)^2$. (This is the dS$_2$ counterpart of the variable $(\Delta-\half)^2$ in \eqref{eq:BS-rule}, the analogue of the Langer correction.) The classically allowed region, $4\cosh^2\theta\,B_{\rm dS}(p)\le\nu^2$, consists of an island around the origin, inside the barrier, together with the outer region beyond it. The island carries the Bohr--Sommerfeld quantization $\mathcal A(\nu^2)=\pi(n+\half)$, with $\mathcal A=2\int_0^{p_1}\dd p\,{\rm arccosh}\sqrt{\nu^2/4B_{\rm dS}}$. Tunnelling to the outer region occurs at imaginary rapidity, through the forbidden interval $p_1<p<p_2$ in which $B_{\rm dS}>\nu^2/4$, and gives the Gamow width
\begin{equation}\label{eq:ds-gamow}
{\rm Im}\,E_n\simeq-\frac{e^{-2S(\nu_n^2)}}{T_{\rm cl}(\nu_n^2)},\qquad S(\nu^2)=\int_{p_1}^{p_2}\dd p\;\arccos\sqrt{\frac{\nu^2}{4B_{\rm dS}(p)}},\qquad T_{\rm cl}=2\,\frac{\dd\mathcal A}{\dd\nu^2},
\end{equation}
where $T_{\rm cl}$ is the classical period of the motion in the island. 
Evaluated at the numerical ${\rm Re}\,E_n$, \eqref{eq:ds-gamow} agrees with the exact widths to a few per cent for the narrow excited levels, deteriorating only close to the top of the barrier (\cref{tab:ds}), and within a factor of two for the ground state, for which the island is too small for semiclassics.

\paragraph{The widths are thermal.} The exponent $S$ can be evaluated in closed form, and the result has a simple physical interpretation. For $|p|\gtrsim1$ one has $h(p)=\frac\pi2|p|-1$ up to exponentially small terms, and with this form of $B_{\rm dS}$ the integral $S$ is elementary. The forbidden interval is bounded by the roots of a quadratic, and $\dd(2S)/\dd\nu=-2\int_{p_1}^{p_2}\dd p/\sqrt{(p-p_1)(p_2-p)}=-2\pi$ exactly. Hence $2S=2\pi(\nu_{\rm top}-\nu)$, where $\nu_{\rm top}$ is the value at which the barrier disappears, and the widths take the form
\begin{gather}
2S=2\pi\Big[\sqrt{\pi^2\gs^2+4(\ms^2-\gs)}-\nu\Big],\label{eq:ds-2S}\\
\Gamma_n\propto\exp\Big[-\frac{V_{\rm top}-\mu_n}{T_{\rm dS}}\Big],\qquad V_{\rm top}=2\sqrt{m^2-\gb^2+T_s^2R^2} ,\label{eq:ds-thermal}
\end{gather}
with $T_{\rm dS}=1/2\pi R$ the Gibbons--Hawking temperature \cite{Gibbons:1977mu} and $T_s=\pi\gb^2/2$. This is the Boltzmann factor for thermal activation of a meson of mass $\mu_n=\nu_n/R$ over the static-patch barrier \eqref{eq:ds-static}: $V_{\rm top}$ is $V_{\max}$ with the quark mass replaced by 't~Hooft's renormalized mass, $m^2\to m^2-\gb^2$. The light-front tunnelling exponent thus reproduces, exactly in the linear approximation to $h$, the static-patch energetics at the de~Sitter temperature. We have checked \eqref{eq:ds-2S} against the full action $S$, with $h$ not approximated: the two agree up to corrections of order $e^{-\pi p_1}$ from the inner turning point, e.g.\ to $3\times10^{-4}$ (absolute, in $2S$) at $p_1\simeq3$ and to $10^{-9}$ at $p_1\simeq7$. For light quarks at strong coupling ($T_sR\gg m,\gb$), $V_{\rm top}\simeq2T_sR$ is the energy of a flux tube spanning the static patch from horizon to horizon. For heavy quarks, $V_{\rm top}-2m\simeq T_s^2R^2/m$, and with $\nu=2\ms+\varepsilon$, \eqref{eq:ds-2S} becomes $2\pi(\kappa_h^2/2-RE^{\rm rel})$, the exact tunnelling exponent of the inverted-oscillator problem \eqref{eq:ds-heavy} at energy $E^{\rm rel}$. Decay rates of particles in de~Sitter space have been discussed in other contexts in \cite{Bros:2006gs,Volovik:2009qt}; here the decaying object is a bound state, and the rate is fixed by an effective potential barrier rather than by the masses of the decay products.

\paragraph{How many mesons fit in de~Sitter space.} The narrow resonances are those with $\nu_n<\nu_{\rm top}\simeq\pi\gs$. Along the Regge trajectory, $\nu_n^2\simeq\pi^2\gs\,n$, so their number is
\begin{equation}
n_{\rm narrow}\simeq\gs=\gb^2R^2 .
\end{equation}
These are the mesons whose flat-space size $\mu_n/T_s$ is smaller than the diameter $2R$ of the static patch. The same scale marks the crossover of the AdS$_2$ spectrum at $n\sim\gs$ (\cref{sec:limits}). In AdS$_2$, mesons larger than the curvature radius are squeezed by the AdS potential; in dS$_2$ they are torn apart.

\paragraph{A maximal static-patch energy for confined mesons.} One may ask whether the static-patch Hamiltonian, the generator $L_0/R$ that is timelike inside the static patch, has a maximal energy. As an operator it does not. In the light-front variables it acts kinematically, as the dilatation $D=-\cD$ of the null line, which is a generator of the horizon of the static patch, with $\lambda=0$ the fixed point of the boost. On a single quark $D=\ii(k\partial_k+\half)$ on $L^2(\mathbb R_+,\dd k)$, whose spectrum is the whole real line. More generally, the generator of a hyperbolic one-parameter subgroup has continuous spectrum covering $\mathbb R$ in every non-trivial unitary representation of $\SLtwo$, including the principal series of a massive particle. A maximal energy can therefore only be a dynamical statement, and for the confined sector it is one. A quark--antiquark pair joined by a string can remain inside the static patch only if its energy lies below the maximum $V_{\max}$ of the static potential \eqref{eq:ds-static}; above it the pair crosses the barrier and the quarks leave through the horizon. For light quarks $V_{\max}\to2T_sR$, the energy of a string stretched from horizon to horizon. Equivalently, a yo-yo string with massless endpoints has $E=T_sL_{\max}$, where $L_{\max}$ is its maximal length in static coordinates (since $\sqrt{-g_{tt}g_{\rho\rho}}=1$), and $L_{\max}<2R$ for a string that stays inside the patch. The light-front equation reproduces this, with the planar mass $m^2\to m^2-\gb^2$: by \eqref{eq:ds-2S} the tunnelling exponent is $2\pi(RV_{\rm top}-\nu_n)$, so only the $n_{\rm narrow}\simeq\gs$ mesons with $\mu_n<V_{\rm top}$ are long-lived, and every level above $V_{\rm top}$ is broad. In static-patch language, a meson at rest at the centre has static energy $\mu_n$, and the corresponding quasinormal modes, which oscillate with frequency ${\rm Re}\,\omega\simeq\nu_n/R$, are long-lived only for $\nu_n/R\lesssim V_{\rm top}$. The bound is soft rather than sharp: states above $V_{\rm top}$ exist as broad resonances and as the continuum of dissociated pairs, and the states below it decay by thermal activation. A bounded static-patch spectrum is also a feature of double-scaled SYK, which has been proposed to describe the de~Sitter static patch \cite{Sekino:2025bsc,Miyashita:2025rpt}. Here an analogous bound arises without gravity, from confinement, for the states that the static patch can hold.

\paragraph{Flat-space limit.} As $R\to\infty$ at fixed $m/\gb$ the widths vanish, at fixed $n$, as $e^{-4\pi T_sR^2}$, and the real parts follow from \eqref{eq:flat} by $R^2\to-R^2$:
\begin{equation}\label{eq:ds-flat}
\Delta_n(1-\Delta_n)=R^2\mu_n^2-\vev{\cL}_n+O(R^{-2}).
\end{equation}
The collinear Casimir that raises the AdS$_2$ dimensions lowers the dS$_2$ masses by the same amount. For $m^2=\gb^2$ and $\gs=10,\,20,\,40$ we find $R^2\mu_0^2-{\rm Re}\,E_0=0.6902,\,0.6851,\,0.6826$ and $R^2\mu_1^2-{\rm Re}\,E_1=4.8664,\,4.8261,\,4.8070$, approaching $\vev{\cL}_{0,1}=0.6802,\,4.7885$ (\cref{sec:limits}). For comparison, the AdS$_2$ values of $c_{0,1}-R^2\mu_{0,1}^2$ are $0.6709,\,0.6755,\,0.6778$ and $4.7195,\,4.7530,\,4.7704$. At leading order the $O(1/\gs)$ corrections are equal and opposite to those of AdS$_2$, as \eqref{eq:ds-continuation} implies.

\paragraph{Heavy quarks.} With $u=\hat r\sqrt{\ms/2}$, \eqref{eq:ds-heavy} becomes $-\frac12\partial_u^2-\frac12u^2+\kappa_h|u|$, as in AdS$_2$ but with an inverted oscillator. Its outgoing solution on $u>0$ is the parabolic-cylinder function $D_s\big(\sqrt2e^{-\ii\pi/4}(u-\kappa_h)\big)$ with $s=\ii(E^{\rm rel}-\kappa_h^2/2)-\half$, and the resonances follow from $D'_s(-\sqrt2e^{-\ii\pi/4}\kappa_h)=0$ (even) and $D_s(-\sqrt2e^{-\ii\pi/4}\kappa_h)=0$ (odd). At $\kappa_h=2$ the lowest even and odd resonances are $RE^{\rm rel}=1.1045-0.0010\,\ii$ and $2.1216-0.1324\,\ii$. The full equation gives $\varepsilon+\gs/\ms=1.1121-0.0013\,\ii$ and $2.1050-0.1418\,\ii$ at $\ms=25$, and $1.1063-0.0011\,\ii$ and $2.1168-0.1350\,\ii$ at $\ms=100$, with deviations governed by a $1/\ms$ expansion.

\subsection{The pion and a critical string tension}
\label{sec:ds-pion}

\begin{figure}[t]
\centering
\includegraphics[width=0.62\textwidth]{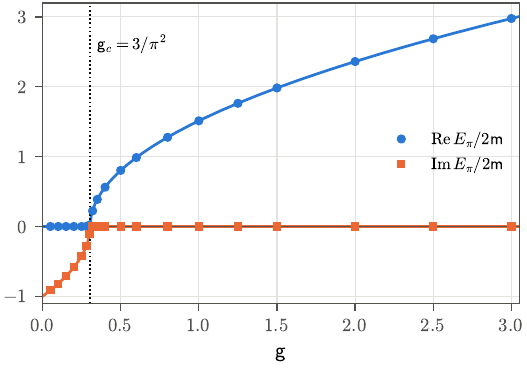}
\caption{The pion on dS$_2$: $E_\pi/2\ms$ with $E_\pi=\Delta_\pi(1-\Delta_\pi)$, extrapolated to $\ms\to0$ from $\ms=0.01,\,0.02$ (points), against \eqref{eq:ds-GMOR} (lines): real part (blue) and imaginary part (orange). At $\gs_c=3/\pi^2$ (dotted) the light pion turns into a broad resonance. Close to $\gs_c$ the limit $\ms\to0$ is not uniform, which explains the deviations there.}
\label{fig:ds-pion}
\end{figure}

At $\ms=0$ the constant $\varphi=1$ solves \eqref{eq:ds-tHooft} with $E=0$, exactly as in AdS$_2$. The massless pion survives on dS$_2$ at every coupling, with $\Delta_\pi=1$: a massless scalar, on the boundary between the complementary and the discrete series. For small $\ms$ the AdS$_2$ argument of \cref{sec:limits}, continued by $R^2\to-R^2$, gives
\begin{equation}\label{eq:ds-GMOR}
\boxed{\;\Delta_\pi(1-\Delta_\pi)=2\ms\sqrt{\frac{\pi^2\gs}{3}-1}+O(\ms^2)\;}
\end{equation}
The square root changes character at
\begin{equation}\label{eq:ds-gc}
\gs_c=\frac{3}{\pi^2},\qquad\text{i.e.}\qquad T_s=\frac{3}{2\pi R^2},
\end{equation}
the same value below which $B_{\rm dS}$ has no barrier (\cref{sec:ds-mesons}). For $\gs>\gs_c$ the endpoint exponent $\beta_0^2\simeq\ms^2/(\pi^2\gs/3-1)$ is real. The pion is then a light, narrow particle in the complementary series, with $1-\Delta_\pi\simeq2\ms\sqrt{\pi^2\gs/3-1}$ (here $\Delta_\pi$ is the root continuously connected to $\Delta_\pi=1$ at $\ms=0$; $\Delta$ and $1-\Delta$ label the same representation), and as $\gs\to\infty$ \eqref{eq:ds-GMOR} reproduces 't~Hooft's $\mu_\pi^2=\frac{2\pi}{\sqrt3}m\gb$. For $\gs<\gs_c$ the right-hand side is imaginary, $E_\pi\simeq-2\ii\ms\sqrt{1-\pi^2\gs/3}$, and there is no light bound state: the lowest level is a broad resonance, continuously connected to the free two-quark pole. The transition is controlled by the curvature of $B_{\rm dS}$ at the origin, $B_{\rm dS}\simeq\ms^2+\frac14\big(\frac{\pi^2\gs}{3}-1\big)p^2$. For $\gs<\gs_c$ the expansion beats confinement already for the lightest quark--antiquark configuration. \Cref{fig:ds-pion} confirms \eqref{eq:ds-GMOR} on both sides of $\gs_c$. In terms of the Hubble rate $1/R$, \eqref{eq:ds-gc} says that the string tension must exceed a number of order the Hubble rate squared. A condition of this form, a confining scale that must exceed the Hubble scale, has been discussed in the context of the ``Festina Lente'' bound \cite{Montero:2019ekk,Mishra:2022fic}; here it arises dynamically in a solvable model.

\paragraph{The chiral condensate on dS$_2$.} The de~Sitter GMOR relation \eqref{eq:ds-GMOR} can be read in the same way as its AdS$_2$ counterpart in \cref{sec:special}. With the de~Sitter mass $\mu_\pi^2=E_\pi/R^2$ and the free-fermion value $f_\pi^2=N/\pi$ of the pion decay constant, which is again fixed by the chiral-limit wavefunction $\varphi=1$, the relation $f_\pi^2\mu_\pi^2=-4m\vev{\bar\psi\psi}$ identifies the chiral condensate on dS$_2$, for $\gs>\gs_c$, as
\begin{equation}\label{eq:ds-condensate}
\vev{\bar\psi\psi}_{{\rm dS}_2}=-\frac{N}{2\pi R}\sqrt{\frac{\pi^2\gs}{3}-1}=-N\sqrt{\frac{\gb^2}{12}-T_{\rm dS}^2}\,,\qquad T_{\rm dS}=\frac1{2\pi R}\,.
\end{equation}
This is the AdS$_2$ result \eqref{eq:condensate} with $R^2\to-R^2$: the curvature now subtracts in quadrature from the flat-space condensate, which is reduced by the expansion rather than enhanced by a boundary. The condensate vanishes at $\gs_c$, with the mean-field behaviour $\vev{\bar\psi\psi}\propto(\gs-\gs_c)^{1/2}$, and the critical point admits a suggestive thermal reading: it is where the de~Sitter temperature equals the flat-space condensate per colour, $T_{\rm dS}=\gb/\sqrt{12}=|\vev{\bar\psi\psi}_{\rm flat}|/N$. For $\gs<\gs_c$ there is no light pion, the right-hand side of \eqref{eq:ds-GMOR} is imaginary, and no condensate is defined in this way: the chiral symmetry is effectively restored, because the quark--antiquark pairs that would form the condensate are torn apart by the expansion before they bind. For free quarks this is precisely as it should be, since the Bunch--Davies state of massless fermions on dS$_2$ is chirally symmetric and there is no boundary to induce a condensate. As on AdS$_2$, \eqref{eq:ds-condensate} assumes the GMOR relation and we can independently check only the flat space limit. In the chiral limit itself the pion is the massless scalar with $\Delta_\pi=1$, which carries the discrete series $D_1$ (\cref{sec:ds-remarks}); being the phase of the condensate it is a compact scalar, so the zero-mode ambiguity of a non-compact massless scalar on dS$_2$ \cite{Anninos:2023lin} does not arise. For gauge group $U(N)$ the anomaly would lift it to $\Delta_\pi(1-\Delta_\pi)\simeq\gs/N$, negligible in the planar limit.

\subsection{Further remarks}
\label{sec:ds-remarks}

We found that the dS$_2$ problem differs from the AdS$_2$ one, in various regards. The equation \eqref{eq:ds-tHooft}, the Casimir decomposition, the free solutions, the chiral-limit pion and the heavy-quark limit are exact. The definition of the spectrum as resonances is a choice of spectral data, the de~Sitter analogue of the boundary spectrum, and it is the one that analytic continuation in $R^2$ produces\footnote{It would of course be interesting to devise other physical setups, where different choices might be relevant.}. The full Hilbert-space content of the planar quark--antiquark sector is a different question. It is expected to consist of the continuum of dissociated configurations, with its decomposition into principal-series (and possibly discrete-series) representations, in which the mesons appear as peaks of the K\"all\'en--Lehmann density of a bilinear operator \cite{Loparco:2023rug,Penedones:2023uqc}. Properly constructing it requires both signs of $K$, i.e.\ more than one null line or an in-in formulation, and we leave this more involved construction for future work. At order $1/N$ mesons can also decay into pairs of mesons. In de~Sitter space such decays are allowed for any masses, although they are exponentially suppressed when they are kinematically forbidden in flat space \cite{Bros:2006gs}. Which mechanism dominates depends on $N$ and on the couplings; at fixed couplings and $N\to\infty$ the dissociation width \eqref{eq:ds-thermal} is the leading one. Finally, double-scaled SYK at infinite temperature has been argued to describe a strongly coupled 't~Hooft model in the flat-space limit of the de~Sitter static patch \cite{Sekino:2025bsc,Miyashita:2025rpt}. The finite-$R$ effects found here, in particular the thermal widths, could provide a useful point of comparison and a refinement of that proposal.

\paragraph{Late-time correlators and discrete series.} The resonance dimensions are also the natural data for late-time correlators. The late-time expansion of a two-point function in de~Sitter space is controlled by the poles of its K\"all\'en--Lehmann density \cite{Loparco:2023rug}. For a meson operator $O$ we therefore expect the falloffs $|\tau\tau'|^{\Delta_n}$ and their complex conjugates, i.e.\ damped oscillations whose extra decay $\delta_n$ is the dissociation rate, together with the complementary-series exponent of the light pion for $\gs>\gs_c$, which dominates at late times. The coefficients, and the density itself, require the full Hilbert space and are left for future work. The discrete-series content, on the other hand, is visible on a single null line, which carries representations of type $D^-$ (\cref{sec:ds-halfspace}). At $E=0$ the term proportional to ${\rm sech}^2\theta$ in \eqref{eq:ds-Bp} drops out, and the plane waves at the real zeros $\pm p_*=\pm2y_*$ of $B_{\rm dS}$ solve the equation exactly, for all $\gs$ and $\ms$:
\begin{equation}\label{eq:ds-D1}
\varphi_\pm(x)=\Big(\frac{x}{1-x}\Big)^{\pm\ii y_*},\qquad \Delta=1 .
\end{equation}
These normalizable states carry $D^-_1$. They describe a constituent that leaves through $\mathcal I^+$ while the other arrives from $\mathcal I^-$, with no net flux, so they belong to the continuum of dissociated configurations rather than to the mesons. They overlap with the light-front charge density $\psi^\dagger\psi$, the good component of the conserved vector current (and of the dual axial current), with $\int\dd x\,\varphi_\pm=\pi y_*/\sinh(\pi y_*)$, but not with $\bar\psi\psi$, whose overlap $\propto\int\dd x\,\varphi_\pm/x(1-x)$ vanishes. Their weight is exponentially small at strong coupling and vanishes in the flat-space limit. In the chiral limit the massless pion, with $\Delta_\pi=1$, is a $D_1$ state as well. The higher discrete series $D^-_{n+1}$, $n\ge1$, of free quarks (\cref{sec:ds-resonances}) rely instead on the transparency of the free problem, the vanishing of its reflection amplitude, at the integer values $E=-n(n+1)$. We have computed the reflection amplitude of the interacting rapidity problem and find that its zeros move continuously with the coupling: at $\ms=\half$ the zero that starts at $E=-2$ for $\gs=0$ lies at $E\simeq-1.75$, $-1.08$ and $-0.42$ for $\gs=0.05$, $0.2$ and $0.5$. For generic $\gs>0$ these states are therefore absent, and only the $D_1$ states \eqref{eq:ds-D1}, presumably protected by the conserved $U(1)$ current, survive in the planar quark--antiquark sector.

\section{Discussion}
\label{sec:discussion}

The main observation behind this work is that the light-cone solution of two-dimensional QCD does not need a symmetry that translates along the light front: an isometry-invariant null gauge and a light-front frame in which the gauge sector is flat are enough, and on AdS$_2$ and dS$_2$ the geometry supplies both. The curvature then reduces to the single kinematic operator $\pm\cK/4R^2$, and the resulting meson equation is as tractable as 't~Hooft's, with exactly solvable limits, an underlying integrable structure inherited from flat space, and a spectrum that can be computed to high precision.

\paragraph{Three faces of confinement on curved space.} The same operator, the collinear Casimir with coefficient $\pm1/R^2$, controls how confinement is modified by the curvature, and our results show three distinct behaviours.
\begin{itemize}
\item \emph{AdS$_2$ squeezes.} Low-lying mesons follow the flat Regge trajectory, but beyond $n\sim\gs$, when the flux tube becomes longer than the curvature radius, the AdS box takes over and the dimensions grow as $\Delta_n\simeq n+2\gs\log n$. The remnant of the string tension is a logarithmic ``anomalous dimension''. The boundary condition breaks the axial symmetry explicitly, and the condensate it induces adds in quadrature to the dynamical one, $\vev{\bar\psi\psi}=-N\sqrt{\gb^2/12+1/4\pi^2R^2}$.
\item \emph{The boundary drags.} A quark tied by a string to a charge on the boundary feels two competing forces. The AdS potential wins for $T_sR<\sqrt{m^2-\gb^2}$, and the string only lowers the dimension. Beyond this point the dimension becomes complex, $\Delta_{\rm bq}=\half\pm\ii\varkappa$, with a Casimir in the range of the principal series, the quark is pulled back to the boundary, and the problem becomes a scattering problem with the amplitude \eqref{eq:anchored-R}.
\item \emph{dS$_2$ tears.} The cosmological horizon pushes the quarks apart, and every meson becomes a resonance even at $N=\infty$. The widths are Boltzmann factors at the Gibbons--Hawking temperature, $\Gamma_n\propto\exp[-(V_{\rm top}-\mu_n)/T_{\rm dS}]$, about $\gs$ mesons are narrow, and for $m>0$ below the critical tension $T_s=3/2\pi R^2$ not even the pion is bound. The chiral condensate is reduced to $-N\sqrt{\gb^2/12-T_{\rm dS}^2}$ and vanishes at the same point, where the de~Sitter temperature equals the flat-space condensate per colour.
\end{itemize}
Technically the 't Hooft equations in AdS$_2$ and dS$_2$ are related by $R^2\to-R^2$, and the continuation also relates the second behaviour to the third: the dimension of the quark tied to the AdS$_2$ boundary continues to the top of the de~Sitter barrier, $-\nu_b^2\to(RV_{\rm top}/2)^2$.

\paragraph{Relation to folded strings, light-front holography and SYK.} At $m^2=\gb^2$ the mass terms in \eqref{eq:AdS-tHooft} cancel, and the equation becomes identical to the generalized 't~Hooft equation that Vegh obtained from the light-cone quantization of a folded string in AdS$_2$, realized as a degenerate four-segment string in AdS$_3$ [eq.~(80) of \cite{Vegh:2023qsc}]. The identifications are $M^2\leftrightarrow\Delta(\Delta-1)$ and $4g/\pi\leftrightarrow\gs$, where $g=L^2/2\pi\alpha'$ is the string tension in units of the AdS radius. In that work it was noted that the flat-space folded string reproduces 't~Hooft's equation at this mass, a connection that goes back to \cite{Bardeen:1975gx}, and the spectrum was computed numerically. Our derivation shows that his equation is the exact planar meson equation of QCD$_2$ on AdS$_2$, with the string tension $T_s=\pi\gb^2/2$ of the flux tube in the role of the total tension $1/\pi\alpha'$ of the folded string (two strands of tension $1/2\pi\alpha'$), and hence provides its microscopic interpretation (extending it also for a general mass dependence). 

The general-mass equation was proposed independently by Forshaw and Sandapen \cite{Forshaw:2025pion,Forshaw:2025fs} as a phenomenological model of the longitudinal dynamics of the pion in light-front holography. Their three parameters $(m_q,g,\sigma)$ correspond to $(m,\gb,1/R)$. They also observed that at $m_q^2=g^2+\sigma^2/4$ their equation takes the form of Vegh's string equation with the alternative ordering $\sqrt{x(1-x)}\,s^2\sqrt{x(1-x)}$ of the kinetic term, $s=-\ii\,\dd/\dd x$, instead of $s\,x(1-x)\,s$ \cite{Vegh:2023qsc}. Indeed, the two orderings differ by $+\frac14\big(\frac1x+\frac1{1-x}\big)$, so the string equation with the alternative ordering is our equation at $\ms^2=\gs+\frac14$, i.e.\ precisely at their constraint, at which the endpoint exponent is $\beta_0=\half$. In the gauge theory there is no ordering ambiguity, and the quark mass is a free parameter. Forshaw and Sandapen found that one of their models accommodates this constraint when $\sigma\gg g$, which they describe as weak string coupling; in our variables this is $\gs\ll1$, the regime in which the Legendre term dominates. The Legendre operator is also the ``longitudinal confinement'' term of basis light-front quantization \cite{Li:2015zda}, whose eigenfunctions are the Jacobi functions \eqref{eq:free}, and it is closely related to the de~Alfaro--Fubini--Furlan algebra that underlies light-front holographic QCD \cite{deAlfaro:1976vlx,Brodsky:2014yha}, where 't~Hooft-type longitudinal dynamics has also been used \cite{deTeramond:2021yyi,Ahmady:2021lsh,Katz:2007br}. Here this term has a precise origin: it is the collinear conformal Casimir that appears when a confining theory is placed on AdS$_2$.

In \cite{Vegh:2024fsa,Vegh:2025syk} the folded string in AdS$_2$ was quantized by a different route. For imaginary $R^2$ and a specific boundary condition, the spectrum reproduces the dimensions of the fermion bilinears of the SYK model. As shown in \cref{sec:bc}, in our language this is the free AdS$_2$ equation ($\gs=0$) with quark mass $\ms=\half-\Delta_{\rm SYK}$ and the mixed endpoint condition \eqref{eq:mixed}. It would be interesting to understand whether an interacting version, with $\gs>0$, has a role to play in the SYK context.

\paragraph{Relation to equal-time quantization.} The model can also be quantized at equal global time in the axial gauge $A_\sigma=0$. The Coulomb energy of a static singlet pair is then $\frac{g^2C_FR^2}{2}|\tan\sigma_1-\tan\sigma_2|$, or $\frac{g^2C_FR^2}{2}|z_1^{-1}-z_2^{-1}|$ in Poincar\'e time. Both are linear in the affine parameter of the null geodesics and coincide with our light-front potential $\frac{g^2C_F}{4}|\lambda_1-\lambda_2|$. In equal-time quantization, however, the planar vacuum is non-trivial, and the meson problem requires a curved-space gap equation together with forward and backward amplitudes, as in flat space \cite{Bars:1977ud,RajeevTurgut:1997}. One might object that a light-cone reduction cannot work on AdS$_2$, because $\partial_{x^+}$ at fixed $x^-$ is not a Killing vector, the null line ends at a time-dependent boundary point, and no light-front momentum is conserved. The affine frame removes all three obstructions: $\partial_y$ at fixed $\lambda$ is the Killing vector $\partial_t$, the boundary sits at $\lambda=\infty$ for all $y$, and the conserved charge $K=4R^2\cP$ plays the role of the light-front momentum. Where the two approaches can be compared (static potentials, first-order perturbation theory and the heavy-quark limit \eqref{eq:heavy-H}) they agree. In fact, in Eddington--Finkelstein coordinates adapted to global AdS$_2$ the light-front Hamiltonian is the global Hamiltonian itself (\cref{sec:affine}), so the light-front and the equal-time descriptions diagonalize the same operator; they differ in the quantization surface, a null line versus a spacelike slice, and hence in the structure of the vacuum. This of course deserves further study.

\paragraph{Open problems.} Let us now close with a list of open problems that can be studied in the future, from technical to conceptual:
\begin{itemize}
\item \emph{Integrability.} The Baxter equation \eqref{eq:TQrel} should admit an FLZ-type large-$c$ analysis based on the product $\prod_j\Gamma(s\pm\beta_j)/\prod_k\Gamma(s\pm k)$ over the indicial roots. This would turn \eqref{eq:largen-exact} and \eqref{eq:ratio-asy} into complete asymptotic expansions.
\item \emph{Boundary conformal data and $1/N$.} The light-front wavefunctions determine the couplings of mesons to boundary operators, and the flat light-front overlap formulas \cite{Einhorn:1976uz} give the planar three-meson couplings. Together with the $1/N$ expansion, which carries over verbatim using the affine coordinate $\lambda$, they would give the boundary four-point functions and, in the flat-space limit, the meson $S$-matrix \cite{Paulos:2016fap}.
\item \emph{Zero modes and the chiral condensate.} A complete treatment of the light-front zero modes on AdS$_2$ is still missing. It should clarify how the chiral condensate $\vev{\bar\psi\psi}$, which is allowed by the isometries, is encoded in the zero modes and in the endpoint behaviour of the wavefunctions, and how this interplays with the explicit breaking of the axial symmetry by the quark boundary conditions (\cref{sec:vacuum,sec:limits}). A direct computation of the condensate at finite coupling would test the GMOR-based formulas \eqref{eq:condensate} and \eqref{eq:ds-condensate}. On dS$_2$ the zero-mode sector is also the only place where the light-front vacuum could differ from the de~Sitter-invariant one, and it is the light-front image of the infrared region of de~Sitter space (Appendix~\ref{app:ds-vacuum}).
\item \emph{Boundary-charged sectors.} Natural extensions of \cref{sec:anchored} are mesons in the presence of a boundary charge, two quarks tied to the boundary, and the boundary condition that completes the supercritical dynamics, i.e.\ the reflection phase of the quark at the boundary, which fixes the scale of the Efimov-like tower.
\item \emph{Finite-$N$ effects.} All our results hold at $N=\infty$. The $1/N$ expansion carries over to the affine frame, since the gauge sector is flat, and the first corrections are of three kinds. Mesons interact, decay and mix with multi-meson states, which on AdS$_2$ shifts the boundary dimensions at order $1/N$ and on dS$_2$ adds the $1/N$ decays into pairs of mesons to the planar dissociation widths (\cref{sec:ds-remarks}). Colour flux can be screened by pair creation, with an amplitude of order $1/\sqrt N$ per pair: the string of the boundary quark of \cref{sec:anchored} can break, which cuts off the supercritical collapse and the unbounded Efimov-like tower, and it remains to be understood whether the Dirichlet sectors of \cref{sec:flux}, whose flux costs an energy linear in the cutoff, survive at finite $N$ or are always screened. Finally, for gauge group $U(N)$ the anomaly lifts the singlet pion at order $1/N$ (\cref{sec:special}), and at finite $N$ Coleman's theorem turns the flat-space condensate into power-law correlations \cite{Witten:1978qu}; on AdS$_2$ the explicit breaking by the boundary condition protects the condensate, whereas on dS$_2$ its fate at finite $N$ is an open question.
\item \emph{De~Sitter space.} The full Hilbert space of the planar quark--antiquark sector on dS$_2$, including the continuum of dissociated states and its representation content, remains to be constructed, and with it the K\"all\'en--Lehmann densities of meson operators \cite{Loparco:2023rug,Penedones:2023uqc} and the competition between dissociation and $1/N$ decays. It would also be interesting to test whether the thermal form \eqref{eq:ds-thermal} of the dissociation rate holds for confining strings in higher-dimensional de~Sitter space, where it would be relevant for confining sectors during inflation.
\item \emph{Other matter and backgrounds.} The reduction of \cref{sec:gauge} uses only the null gauge and the area-preserving property of affine null coordinates. Adjoint QCD$_2$, the Schwinger model and its multi-flavour generalizations (already in flat space the two-flavour model has a rich dynamics, recently clarified in \cite{Cuomo:2026kmy}), and supersymmetric versions should be tractable in the same way. So should scalar QCD$_2$, with charged scalars in place of the quarks, whose planar meson equation in flat space is known \cite{Shei:1977ci,Tomaras:1978nt} and has recently been analysed with the methods of FLZ \cite{Meshcheriakov:2025lsx}: the ``seagull'' coupling vanishes in the null gauge, but scalar constituents carry collinear weight $0$ instead of $\half$, so the collinear Casimir, and with it the curvature term and the endpoint behaviour, differ from the fermionic case. On any two-dimensional background affine null coordinates bring the metric to the form $\dd y\,\dd\lambda+h\,\dd y^2$, so the gauge sector is again flat and the curvature enters only through the coupling of $h$ to $T_{\lambda\lambda}$. (A)dS$_2$ is singled out because $h=\pm\lambda^2/4R^2$ makes this coupling kinematic; black-hole backgrounds lead to genuinely $\lambda$-dependent one-body potentials. Coupling to JT gravity \cite{Iliesiu:2019lfc} would make the geometry dynamical, through the boundary (Schwarzian) mode of AdS$_2$, increasing the complexity of the model in an interesting way.
\end{itemize}

\section*{Acknowledgments}

We wish to thank Dio Anninos and David Vegh for useful discussions and correspondence. We are especially grateful to Juan Maldacena for important discussions and suggestions that have greatly benefited this work. We also thank A.~Antunes, F.~Ambrosino, P.~Meshcheriakov and K.~Siampos for their comments on the first version. P.B.\ would also like to thank the Institute for Advanced Study for hospitality during the course of this work. This research was supported in part by the Research Foundation -- Flanders (FWO) through a Grant for a Long Stay Abroad (grant number V431626N). P.B.\ acknowledges financial support from the European Research Council (grant BHHQG-101040024), funded by the European Union. Views and opinions expressed are those of the author only and do not necessarily reflect those of the European Union or the European Research Council. Neither the European Union nor the granting authority can be held responsible for them. We acknowledge the use of Claude as a research aid to help identify errors, work through derivations and assist with coding and numerics. All AI-assisted calculations and derivations were independently checked by the author.

\clearpage
\appendix
\section{Conventions and normalization of the coupling}
\label{app:conventions}

\paragraph{Light-cone conventions.} We use $x^\pm=t\pm z$, so that $\dd s^2_{\rm flat}=\dd x^+\dd x^-$, $\dd^2x=\half\dd x^+\dd x^-$, $\partial_\pm=\half(\partial_t\pm\partial_z)$, $A_\pm=\half(A_t\pm A_z)$ and $F_{\mu\nu}F^{\mu\nu}=-8F_{+-}^2$. A plane wave $e^{-\ii(p_+x^++p_-x^-)}$ has $p_\pm=\half(E\mp p^z)$, so that $M^2=4p_+p_-$. With $\gamma^0=\sigma^1$, $\gamma^1=\ii\sigma^2$ one has $\gamma^0+\gamma^1=2\sigma^+$ and $\gamma^0-\gamma^1=2\sigma^-$, and $\bar\psi\gamma^\mu\partial_\mu\psi=2\chi^\dagger\partial_+\chi+2\eta^\dagger\partial_-\eta$ for $\psi=(\eta,\chi)^T$.

\paragraph{The coupling.} With the action \eqref{eq:action}, the Coulomb term \eqref{eq:coulomb} gives, for a colour-singlet quark--antiquark pair (on which $T^a_q T^a_{\bar q}=-C_F$), the light-front potential $V=\frac{g^2C_F}{4}|\lambda_1-\lambda_2|$, with $C_F\to N/2$ at large $N$. Using $|\lambda|=\int\frac{\dd q}{2\pi}\,e^{\ii q\lambda}\big(-\frac2{q^2}\big)$, the potential acts on $\phi(k_1,k_2)$ as
\begin{equation}
(V\phi)(k_1,k_2)=-\frac{g^2N}{8\pi}\int\dd q\,\frac{\phi(k_1+q,k_2-q)}{q^2}=-\frac{g^2N}{8\pi P}\int_0^1\dd x'\,\frac{\varphi(x')}{(x-x')^2},
\end{equation}
so that $4PV$ contributes $-\frac{g^2N}{2\pi}\PV\frac{\varphi(x')}{(x-x')^2}$ to $\hat M^2$. This fixes $\gb^2=g^2N/2\pi$ in \eqref{eq:tHooft-operator}. Three cross-checks:
\begin{enumerate}[label=(\roman*)]
\item In 't~Hooft's normalization, with $\frac14\tr G_{\mu\nu}G^{\mu\nu}$ and $A$ a matrix field \cite{tHooft:1974pnl}, $g_{\rm tH}^2=g^2/2$, so $\gb^2=g^2_{\rm tH}N/\pi$, his unit of mass squared.
\item In the conventions of \cite{Fateev:2009jf,Litvinov:2024riz}, $2\pi^2\lambda^{\rm FLZ}\varphi=\big[\frac{\alpha}{x}+\frac{\alpha}{1-x}\big]\varphi-\PV\frac{\varphi(x')}{(x'-x)^2}$ with $\alpha=m^2/\gb^2-1$ and $\mu^2=2\pi^2\gb^2\lambda^{\rm FLZ}$.
\item The Regge asymptotics $\mu_n^2\simeq\pi^2\gb^2 n$ agrees with the Bohr--Sommerfeld spectrum $M^2=2\pi T_sn$ of two massless particles joined by a string of tension $T_s=g^2C_F/2=\pi\gb^2/2$.
\end{enumerate}
In the affine frame the equal-$y$ light-front potential has slope $\frac{\pi\gb^2}{4}=\frac{T_s}2$ in $\lambda$.

\section{The two-particle collinear Casimir}
\label{app:casimir}

On a single weight-$\half$ particle in the momentum representation \eqref{eq:momentum-rep},
\begin{equation}
\cP=k,\qquad \cD=-\ii\big(k\partial_k+\tfrac12\big),\qquad \cK=-\partial_k\,k\,\partial_k ,
\end{equation}
which are hermitian on $L^2(\mathbb R_+,\dd k)$ and satisfy $\half(\cK\cP+\cP\cK)-\cD^2=-\frac14$. For two particles we use $P=k_1+k_2$ and $x=k_1/P$, so that
\begin{equation}
\partial_{k_1}=\partial_P+\frac{1-x}{P}\partial_x,\qquad \partial_{k_2}=\partial_P-\frac xP\partial_x,\qquad \sum_ik_i\partial_{k_i}=P\partial_P .
\end{equation}
A short computation gives, for $\Phi=F(P)\varphi(x)$,
\begin{equation}
\sum_i\partial_{k_i}k_i\partial_{k_i}\,\Phi=\big(PF''+2F'\big)\varphi-\frac FP\,\cL\varphi,\qquad \cL=-\partial_x\,x(1-x)\,\partial_x .
\end{equation}
Hence $\cK\Phi=-(PF''+2F')\varphi+\frac FP\cL\varphi$, and $\cD\Phi=-\ii(PF'+F)\varphi$. Then
\begin{align}
\tfrac12\big(\cK\cP+\cP\cK\big)\Phi&=-\big(P^2F''+3PF'+F\big)\varphi+F\,\cL\varphi,\nonumber\\
-\cD^2\Phi&=\big(P^2F''+3PF'+F\big)\varphi ,
\end{align}
and the $F$-dependent terms cancel:
\begin{equation}
\cC_{\rm coll}\,\Phi=F(P)\,\cL\varphi(x).
\end{equation}
The spectrum of $\cL$ on $L^2([0,1],\dd x)$ with regular endpoint behaviour is $\ell(\ell+1)$, $\ell=0,1,2,\dots$, with eigenfunctions $P_\ell(2x-1)$. This reproduces the decomposition of the product of two lowest-weight-$\half$ representations into representations of lowest weight $\ell+1=1,2,\dots$, with the Gegenbauer polynomials $C^{(1/2)}_\ell=P_\ell$ familiar from the conformal partial-wave expansion of light-ray operators \cite{Braun:2003rp}. We have also verified the identity by finite differences.

The same computation determines the action of $H$ on $F(P)\varphi_n(x)$ quoted in \eqref{eq:lowest-weight}. Since $\Htf\Phi=\frac{F}{4P}\cT\varphi$, one has
\[
H\,\Phi=\frac1{4R^2}\Big[-(PF''+2F')+\frac{c_n}{P}F\Big]\varphi_n
\]
whenever $(R^2\cT+\cL)\varphi_n=c_n\varphi_n$.

\section{Integrable structure and exact spectral sums}
\label{sec:integrable}

In flat space, Fateev, Lukyanov and Zamolodchikov (FLZ) showed that the rapidity form of 't~Hooft's equation carries an integrable structure \cite{Fateev:2009jf}. It consists of a Baxter (TQ) equation for suitably relaxed solutions, a quantum Wronskian, and relations between the spectral determinants and special values of the $Q$-functions. Together these give a systematic large-$n$ expansion and closed forms for the spectral sums, such as $\sum_{n\,\rm even}\lambda_n^{-2}=7\zeta(3)$ at $m^2=\gb^2$, where $\lambda_n=\mu_n^2/2\pi^2\gb^2$ is the eigenvalue in FLZ's normalization. The method was extended to arbitrary quark masses, parametrized by $\alpha=m^2/\gb^2-1$, by Litvinov and Meshcheriakov (LM) \cite{Litvinov:2024riz}, to two flavours with unequal masses \cite{Artemev:2025cev} and to scalar QCD$_2$ \cite{Meshcheriakov:2025lsx}, and Ambrosino and Komatsu related it to an inhomogeneous Fredholm problem \cite{Ambrosino:2023dik}. A notable outcome of the large-$n$ expansion is that it contains explicitly alternating terms, the first of which is of order $(-1)^n/n^2$ in $\lambda_n$ \cite{Fateev:2009jf,Litvinov:2024riz}. The spectrum therefore does not lie on a single Regge trajectory but on precisely two, one for even and one for odd $n$, i.e.\ one for each charge-conjugation parity, and the two approach each other only as $n\to\infty$. Alternating terms of this kind are also present, in modified form, for unequal quark masses \cite{Artemev:2025cev} and in scalar QCD$_2$ \cite{Meshcheriakov:2025lsx}. On AdS$_2$ the even and odd levels are likewise governed by distinct quantization conditions, see item~(iii) below. In this section we ask which parts of this structure survive the AdS$_2$ deformation $B\to B+p^2/4$ of \eqref{eq:baxter}. The answer has five parts.
\begin{enumerate}[label=(\roman*)]
\item The Birman--Schwinger kernel is \emph{integrable} in the sense of Its, Izergin, Korepin and Slavnov (IIKS) \cite{Its:1990}, with the same rank-two structure as in flat space.
\item The relaxed $Q$-functions obey a homogeneous Baxter equation and a \emph{quantum Wronskian} identity. They are built from solutions of the eigenvalue equation with a source term added, \eqref{eq:relaxed}. Such solutions exist for every value of $c$ and not only at the eigenvalues, i.e.\ the eigenvalue condition has been ``relaxed'', hence the name used by FLZ \cite{Fateev:2009jf}.
\item The \emph{ratio of the even and odd spectral determinants} is a special value of a $Q$-function. This follows exactly from the rank-two structure and parity, and the argument also proves the corresponding flat-space relations of FLZ and LM. Semiclassically, the poles and zeros of $D_-/D_+$ are governed by the rapidity Bohr--Sommerfeld phase of \cref{sec:rapidity}.
\item The FLZ \emph{derivative relation}, which would express the individual determinants through point values of $Q$, does \emph{not} survive in its flat-space form for $\gs>0$.
\item In its place, the partial-fraction expansion of $1/B$ over the indicial roots turns the AdS$_2$ kernel into a positive superposition of solvable P\"oschl--Teller kernels. The spectral sums then follow in closed form as \emph{sums over the roots of the indicial equation}, and they reduce exactly to the $\zeta(3)$-type values of FLZ in the flat limit.
\end{enumerate}

\subsection{\texorpdfstring{An integrable kernel and relaxed $Q$-functions}{An integrable kernel and relaxed Q-functions}}

In momentum space the Birman--Schwinger operator $\BS$ of \eqref{eq:BS} has the same nonzero spectrum as
\begin{equation}
\widetilde{\BS}(p,k)=\frac1{2\pi}\,\frac{\widehat W(p-k)}{\sqrt{B(p)B(k)}},\qquad \widehat W(q)=\frac{\pi q}{4\sinh(\pi q/2)} .
\end{equation}
Since $\sinh\frac{\pi(p-k)}2=\cosh\frac{\pi p}2\cosh\frac{\pi k}2\,\big(\tanh\frac{\pi p}2-\tanh\frac{\pi k}2\big)$, it obeys
\begin{equation}\label{eq:IIKS}
\big(M(p)-M(k)\big)\,\widetilde{\BS}(p,k)=\frac18\Big[f_1(p)f_2(k)-f_2(p)f_1(k)\Big],
\end{equation}
with
\begin{equation}
M(p)=\tanh\frac{\pi p}{2},\qquad f_1(p)=\frac{p}{\cosh\frac{\pi p}{2}\,\sqrt{B(p)}},\qquad f_2(p)=\frac{1}{\cosh\frac{\pi p}{2}\,\sqrt{B(p)}} .
\end{equation}
The commutator of $\widetilde{\BS}$ with multiplication by $\tanh(\pi p/2)$ therefore has rank two, so $\widetilde{\BS}$ is an integrable operator \cite{Its:1990}. The structure comes entirely from the ${\rm sech}^2\theta$ weight, the same in flat space; $B$ only enters the vectors $f_j$. Consequently the resolvent $\mathcal R=\widetilde{\BS}(1-c\widetilde{\BS})^{-1}$ is integrable as well, $\big(M(p)-M(k)\big)\mathcal R(p,k)=\frac18\big[F_1(p)F_2(k)-F_2(p)F_1(k)\big]$, with
\begin{equation}
F_j=(1-c\widetilde{\BS})^{-1}f_j .
\end{equation}

In rapidity space the $F_j$ are the solutions of the \emph{relaxed} problems
\begin{equation}\label{eq:relaxed}
B(p)\,\varphi_j-\frac c4\,{\rm sech}^2\theta\;\varphi_j=\sigma_j,\qquad \sigma_1=\frac{\ii}{\pi}\,{\rm sech}\,\theta\tanh\theta,\qquad \sigma_2=\frac1\pi\,{\rm sech}\,\theta ,
\end{equation}
that is \eqref{eq:Bp} with sources proportional to $w'$ and $w$. In $x$-space these are $(\cC-c)\varphi_2=\frac2\pi[x(1-x)]^{-1/2}$ and $(\cC-c)\varphi_1=\frac{2\ii}{\pi}(2x-1)[x(1-x)]^{-1/2}$. They are solvable for every $c$ outside the spectrum, and $\varphi_1$ ($\varphi_2$) has poles in $c$ only at the odd (even) levels. The source terms relax the eigenvalue problem, which has solutions only at $c=c_n$, to a problem that can be solved at any $c$; the spectrum is then recovered from the conditions that the source decouples, see below. With $\Phi_j$ the Fourier transform of $\varphi_j$, we define
\begin{equation}\label{eq:Qdef}
\mathcal Q_j(p)\equiv\cosh\frac{\pi p}{2}\;B(p)\,\Phi_j(p)=\{p,\,1\}+\frac c4\cosh\frac{\pi p}2\int\dd\theta\;e^{-\ii p\theta}\,{\rm sech}^2\theta\;\varphi_j(\theta) .
\end{equation}
These functions have four properties.
\begin{enumerate}[label=(\alph*)]
\item $\mathcal Q_1$ is odd and $\mathcal Q_2$ is even.
\item They are analytic in the strip $|{\rm Im}\,p|<2+2\beta_*$, with $\beta_*=\min(\beta_0,\half)$, and polynomially bounded on horizontal lines.
\item Since $\cosh(\pm\ii\pi/2)=0$, they are normalized by $\mathcal Q_1(\pm\ii)=\pm\ii$ and $\mathcal Q_2(\pm\ii)=1$.
\item Because $\cosh\frac{\pi(p\pm2\ii)}{2}=-\cosh\frac{\pi p}2$, and $p$ and $1$ are annihilated by the second difference, they obey the homogeneous Baxter equation
\begin{equation}\label{eq:TQrel}
\boxed{\;\mathcal Q(p+2\ii)+\mathcal Q(p-2\ii)-2\,\mathcal Q(p)=-\frac{c}{B(p)}\,\mathcal Q(p)\;}
\end{equation}
\end{enumerate}
For an eigenfunction, $\sigma=0$ and $\mathcal Q=\cosh\frac{\pi p}2\,Q$ with $Q=B\Phi$ the function of \eqref{eq:baxter}, so \eqref{eq:TQrel} is \eqref{eq:baxter}. The eigenvalue problem is thus the TQ equation \eqref{eq:TQrel} supplemented by the analyticity of (b) and the vanishing of $\mathcal Q$ at $p=\pm\ii$. The relaxed solutions drop the last condition. FLZ relax with sources $\{p,1\}/\sinh\frac{\pi p}2$ and normalize at $p=0,\pm2\ii$. As we now show, both choices span the same two-dimensional space of solutions.

\paragraph{Quantum Wronskian.} Two solutions of a second-order difference equation have a ``quantum Wronskian'', the analogue of the Wronskian of a second-order differential equation. By \eqref{eq:TQrel} the combination
\begin{equation}\label{eq:qW}
W(p)=\mathcal Q_1(p+\ii)\,\mathcal Q_2(p-\ii)-\mathcal Q_1(p-\ii)\,\mathcal Q_2(p+\ii)
\end{equation}
is $2\ii$-periodic. It is analytic in $|{\rm Im}\,p|<1+2\beta_*$, a strip wider than the period, so it is entire. Being polynomially bounded, it is a constant: as a function of $e^{\pi p}$ it is analytic on $\mathbb C^*$ with at most logarithmic growth at $0$ and $\infty$. Its value at $p=0$ follows from the normalization, so $W\equiv2\ii$. Consequently every solution with the properties (b) is a linear combination of $\mathcal Q_1$ and $\mathcal Q_2$. Evaluating $W$ at $p=\ii$, where $\mathcal Q_1(0)=0$, gives
\begin{equation}\label{eq:Q1Q2}
\mathcal Q_1(2\ii\,|\,c)\;\mathcal Q_2(0\,|\,c)=2\ii\qquad\text{for all }c .
\end{equation}

\subsection{Spectral determinants}

Let $D(c)=\det(1-c\BS)=D_+(c)D_-(c)$, with $D_\pm=\prod_{n\ \rm even/odd}(1-c/c_n)$. Since $c_n\simeq n^2$, these are entire functions of order $\half$.

\paragraph{Trace identities.} The diagonal and anti-diagonal of the integrable resolvent are $\mathcal R(p,p)=\frac18(F_1'F_2-F_2'F_1)/M'$ and $\mathcal R(p,-p)=\frac18F_1F_2/M$. Using $F_j=\mathcal Q_j/(\cosh\frac{\pi p}2\sqrt B)$, we obtain
\begin{equation}\label{eq:traceid}
-\partial_c\log\big(D_+D_-\big)=\frac1{4\pi}\int_{\mathbb R}\dd p\;\frac{\mathcal Q_1'\mathcal Q_2-\mathcal Q_2'\mathcal Q_1}{B(p)},\qquad
-\partial_c\log\frac{D_+}{D_-}=\frac14\int_{\mathbb R}\dd p\;\frac{\mathcal Q_1\mathcal Q_2}{\sinh(\pi p)\,B(p)} .
\end{equation}
These are the AdS$_2$ counterparts of the flat-space identities of \cite{Fateev:2009jf,Litvinov:2024riz}. We have also checked that they hold numerically to $10^{-12}$.

\paragraph{The ratio theorem.} For all $c$,
\begin{equation}\label{eq:ratio}
\boxed{\;\begin{gathered}\frac{D_-(c)}{D_+(c)}=\mathcal Q_2(0\,|\,c)=\ms^2\!\int_{\mathbb R}\dd\theta\;\varphi_2(\theta\,|\,c),\\[2pt]
\frac{D_+(c)}{D_-(c)}=\frac{\mathcal Q_1(2\ii\,|\,c)}{2\ii}=1+\frac{\ii c}{4}\int_{\mathbb R}\dd\theta\,\tanh\theta\;\varphi_1(\theta\,|\,c)\end{gathered}\;}
\end{equation}
\emph{Proof.} Let $\widetilde{\BS}_\pm$ be the restrictions of $\widetilde{\BS}$ to even and odd functions, so that $D_\pm(c)=\det(1-c\widetilde{\BS}_\pm)$. For an even function $u$ one has $\langle f_1,u\rangle=0$, and \eqref{eq:IIKS} gives $M\widetilde{\BS}u-\widetilde{\BS}Mu=\frac18f_1\langle f_2,u\rangle$. Since multiplication by $M$ maps even functions to odd ones,
\begin{equation}
\widetilde{\BS}_-\,M=M\,\widehat{\BS},\qquad \widehat{\BS}=\widetilde{\BS}_+-\frac18\,|g\rangle\langle f_2|,\qquad g=\frac{f_1}{M}=\frac{p}{\sinh\frac{\pi p}2\,\sqrt{B}} .
\end{equation}
Multiplication by $M$ is injective, and every odd eigenvector $v$ of $\widetilde{\BS}_-$ satisfies $v/M\in L^2$, being smooth and vanishing at $p=0$. Hence $\widehat{\BS}$ and $\widetilde{\BS}_-$ have the same nonzero eigenvalues, with multiplicities, and $\det(1-c\widetilde{\BS}_-)=\det(1-c\widehat{\BS})$. For the rank-one perturbation the matrix determinant lemma gives
\begin{equation}
\frac{D_-}{D_+}=1+\frac c8\,\big\langle g,(1-c\widetilde{\BS}_+)^{-1}f_2\big\rangle=1+\frac c8\int\dd p\;\frac{p\,\Phi_2(p)}{\sinh\frac{\pi p}2} ,
\end{equation}
which is $\mathcal Q_2(0|c)$ by \eqref{eq:Qdef}. The second identity follows from \eqref{eq:Q1Q2}. $\square$

\begin{figure}[t]
\centering
\includegraphics[width=\textwidth]{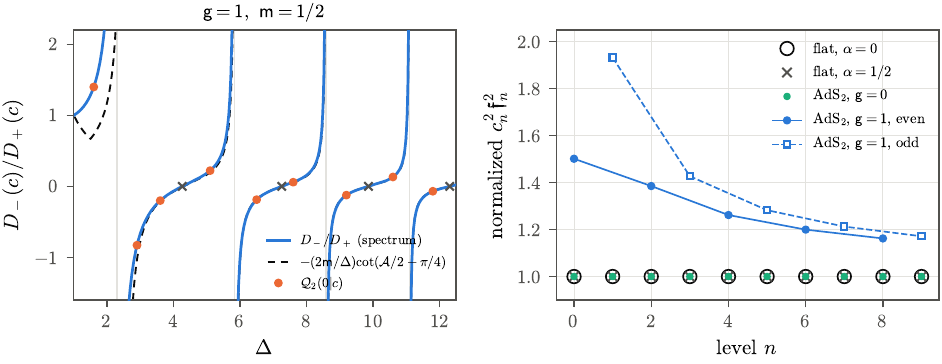}
\caption{Left: $D_-/D_+$ at $\gs=1$, $\ms=\half$, from the spectrum (solid), from the relaxed solution, $\mathcal Q_2(0|c)$ (dots), and the asymptotic form \eqref{eq:ratio-asy} (dashed). Vertical lines mark the even levels (poles) and crosses the odd levels (zeros). Right: the overlaps $\mathfrak f_n^2$ that encode the FLZ derivative relation, normalized so that they equal $1$ whenever that relation holds: $c_n(c_n-4\alpha)\mathfrak f_n^2/16\pi$ (even) and $c_n^2\mathfrak f_n^2/16\pi$ (odd) in flat space, and $\pi c_n^2\mathfrak f_n^2/16(2\Delta_n-1)$ on AdS$_2$. The relation holds in flat space and at $\gs=0$, $\ms=\half$, but not at $\gs=1$.}
\label{fig:integrable}
\end{figure}

The argument is algebraic, so \eqref{eq:ratio} also holds exactly for any parity-symmetric Nystr\"om discretization of $\widetilde{\BS}$ (\cref{tab:ratio}). The free-point formula below provides a test in the continuum. Three remarks are in order.
\begin{itemize}
\item \emph{Relation to flat space.} By two-dimensionality, the FLZ relaxed solutions are $Q_-=\mathcal Q_2/\mathcal Q_2(0)$ and $Q_+=2\ii\,\mathcal Q_1/\mathcal Q_1(2\ii)$. In this language \eqref{eq:ratio} reads $Q_-(\ii)/Q_-(0)=D_+/D_-$ and $Q_+(\ii)/Q_+(2\ii)=D_-/2D_+$, which are exactly the ratio relations of \cite{Fateev:2009jf,Litvinov:2024riz}; LM regard them as conjectural for $\alpha\neq0$. The proof uses only the rank-two structure and parity, and it goes through in flat space, where $\widetilde{\BS}$ is Hilbert--Schmidt but not trace class. With Carleman--Fredholm determinants $D^{(2)}_\pm=\prod(1-c/c_n)e^{c/c_n}$ it gives $D^{(2)}_-/D^{(2)}_+=\mathcal Q_2(0|c)\exp\big(-c\sum_n(-1)^n/c_n\big)$. This establishes the flat-space ratio relations for all $\alpha$, up to the convention-dependent exponential normalization of $D_\pm$. On AdS$_2$ the determinants need no regularization and no exponential appears.
\item \emph{Free point.} At $\gs=0$ the P\"oschl--Teller spectrum gives
\begin{equation}\label{eq:ratio-g0}
\begin{aligned}\frac{D_-}{D_+}\Big|_{\gs=0}&=\ms\,\frac{\Gamma(\ms+\frac12-\frac\Delta2)\,\Gamma(\ms+\frac\Delta2)}{\Gamma(\ms+1-\frac\Delta2)\,\Gamma(\ms+\frac12+\frac\Delta2)}\\
&=\ms\tan\!\Big[\pi\Big(\frac\Delta2-\ms\Big)\Big]\frac{\Gamma(\frac\Delta2-\ms)\,\Gamma(\frac\Delta2+\ms)}{\Gamma(\frac\Delta2-\ms+\frac12)\,\Gamma(\frac\Delta2+\ms+\frac12)} ,\end{aligned}
\end{equation}
which the numerical $\mathcal Q_2(0|c)$ reproduces to $10^{-13}$.
\item \emph{Large $c$.} For large positive $c$, away from the spectrum,
\begin{equation}\label{eq:ratio-asy}
\frac{D_-(c)}{D_+(c)}=-\frac{2\ms}{\Delta}\,\cot\Big(\frac{\mathcal A\big((\Delta-\frac12)^2\big)}{2}-\frac\pi4\Big)\Big[1+O(\Delta^{-1})\Big],
\end{equation}
with $\mathcal A$ the rapidity area of \cref{sec:rapidity}. At $\gs=0$ the phase is exact by \eqref{eq:ratio-g0}, while the prefactor receives $O(1/\Delta)$ corrections from the ratio of Gamma functions, $1.2\%$ at $\Delta\simeq42$. At $\gs=1$ and $3$ the deviations of the prefactor from $2\ms/\Delta$ are of the same size, $1$--$2\%$ at $\Delta\simeq40$--$50$. Its poles and zeros are the even and odd solutions of the Bohr--Sommerfeld rule \eqref{eq:BS-rule}: the semiclassical quantization is parity-resolved, and the two sectors interlace (\cref{fig:integrable}, left).
\end{itemize}

\begin{table}[t]
\centering\small
\begin{tabular}{cc cc}
\toprule
$(\gs,\ms)$ & $c$ & $\mathcal Q_2(0|c)$ & $D_-/D_+$\\
\midrule
$(1,\,0.5)$ & $-5$  & $0.460828062724$ & $0.460828062724$\\
            & $1.3$ & $1.648128762580$ & $1.648128762580$\\
            & $7$   & $-0.447149556507$ & $-0.447149556507$\\
$(3,\,0.3)$ & $-40$ & $0.102508361325$ & $0.102508361325$\\
            & $1.3$ & $2.205672300584$ & $2.205672300584$\\
$(0.5,\,1)$ & $7$   & $86.0912074$ & $86.0912074$\\
\bottomrule
\end{tabular}
\caption{The ratio theorem \eqref{eq:ratio}: $\mathcal Q_2(0|c)=\ms^2\int\varphi_2$ from the relaxed problem \eqref{eq:relaxed}, against the ratio of spectral determinants from the spectrum. The values are converged in the grid ($\pmax=80$, $h=0.02$). The two columns agree to machine precision on any parity-symmetric grid, as the proof shows.}
\label{tab:ratio}
\end{table}

\paragraph{The derivative relation does not survive in its flat-space form.} FLZ also conjectured that $\partial_\lambda\log D_\pm$ equals $2\ii\,\partial_\nu\log Q_\mp$ at $\nu=\ii$, where $\lambda$ and $\nu$ are their eigenvalue and momentum variables ($\lambda=c/2\pi^2$ and $\nu=p$ in our notation), and LM a modified form for $\alpha\neq0$ \cite{Fateev:2009jf,Litvinov:2024riz}. These relations express the \emph{individual} determinants through point values, and they are what produces closed forms for the individual sums $G_\pm^{(s)}$. In the present language $2\ii\,\mathcal Q_2'(\ii)$ equals $-\frac{\pi c}8\langle f_2,(1-c\widetilde{\BS})^{-1}f_2\rangle$. Matching residues, any relation of this type is equivalent to the statement that the overlaps $\mathfrak f_n^2=\langle f_j,v_n\rangle^2$ of the normalized eigenvectors with the sources ($j=2$ for even and $j=1$ for odd $n$) are fixed functions of $c_n$.

In flat space ($\gb=1$, $c=\mu^2/\gb^2=2\pi^2\lambda$) we find numerically $c_n(c_n-4\alpha)\mathfrak f_n^2=16\pi$ for even $n$ and $c_n^2\mathfrak f_n^2=16\pi$ for odd $n$, to ten digits at all levels computed. This is precisely the residue content of the FLZ relation and of its LM modification. On AdS$_2$ at $\gs=0$ and $\ms=\half$ one finds instead $c_n^2\mathfrak f_n^2=\frac{16}\pi(2\Delta_n-1)$ exactly. For $\gs>0$, however, no rational function of $(c_n,\Delta_n)$ reproduces the overlaps: fits of increasing order converge only algebraically, and the overlaps carry the logarithms generated by the confining interaction (\cref{fig:integrable}, right). On AdS$_2$, therefore, the \emph{ratio} of the determinants is a point value, while no derivative relation with rational coefficients exists for the individual determinants; they are given by the trace identity \eqref{eq:traceid}. The AdS$_2$ deformation preserves the TQ equation and its quantum Wronskian but not the additional structure behind the FLZ derivative relation, which is special to flat space and to the free point at $\ms=\half$.

\subsection{Spectral sums as sums over indicial roots}

Write $I(s)\equiv B(2\ii s)=\ms^2-s^2+\gs(\pi s\cot\pi s-1)$ for the indicial function, whose positive zeros are $\beta_0<\beta_1<\beta_2<\cdots$ (\cref{sec:rapidity}). Since $1/B$ is meromorphic with simple poles at $p=\pm2\ii\beta_j$ and decays as $4/p^2$, it has the Mittag-Leffler expansion
\begin{equation}\label{eq:ML}
\frac1{B(p)}=\sum_{j\ge0}\frac{4\beta_j\,r_j}{p^2+4\beta_j^2},\qquad r_j=\frac{2}{|I'(\beta_j)|}\simeq\frac{2\gs}{j^3},\qquad
G_B(\theta)=\sum_{j\ge0}r_j\,e^{-2\beta_j|\theta|} ,
\end{equation}
with the sum rules $\sum_jr_j\beta_j=1$ (from $p\to\infty$) and $\sum_jr_j/\beta_j=1/\ms^2$ (from $p=0$). The Green function $G_B$ is a positive superposition of exponentials. The Birman--Schwinger operator therefore decomposes as
\begin{equation}\label{eq:Kroots}
\BS=\sum_{j\ge0}r_j\,E_{\beta_j},\qquad E_\beta(\theta,\theta')=w(\theta)\,e^{-2\beta|\theta-\theta'|}\,w(\theta') ,
\end{equation}
where $E_\beta/\beta$ is the Birman--Schwinger operator of the free (P\"oschl--Teller) problem with $\ms\to\beta$, with spectrum $1/(2\beta+n)(2\beta+n+1)$. The 't~Hooft interaction thus replaces the single free endpoint exponent $\ms$ by the tower of indicial roots, with positive weights $r_j$; the same data fix the large-$n$ spectrum through $\Sigma_{\rm reg}$ in \eqref{eq:largen-exact}. Spectral sums become sums over roots of traces of products of P\"oschl--Teller kernels, and these are elementary. From the free spectrum, $\operatorname{Tr}E_\beta=\half$ and $\operatorname{Tr}\Par E_\beta=\beta[\psi(\beta+\half)-\psi(\beta)]-\half$, with $\Par$ the parity operator $\theta\to-\theta$ and $\psi$ here the digamma function. Since the trace of $E_aE_b$ involves only $e^{-2(a+b)|\theta-\theta'|}$, $\operatorname{Tr}E_aE_b=\operatorname{Tr}E^2_{(a+b)/2}$. The twisted trace follows by expanding ${\rm sech}^2$ in exponentials. The result is
\begin{equation}\label{eq:rootsums}
\boxed{\;\begin{aligned}
\sum_{n\ \rm even}\frac1{c_n}&=\frac12\sum_jr_j\beta_j\Big[\psi\big(\beta_j+\tfrac12\big)-\psi(\beta_j)\Big],\\
\sum_{n\ \rm odd}\frac1{c_n}&=\frac12\sum_jr_j\Big\{1-\beta_j\Big[\psi\big(\beta_j+\tfrac12\big)-\psi(\beta_j)\Big]\Big\},\\
\sum_n\frac1{c_n^2}&=\sum_{j,k}r_jr_k\,J(\beta_j+\beta_k),\qquad
\sum_n\frac{(-1)^n}{c_n^2}=\sum_{j,k}r_jr_k\,L(\beta_j,\beta_k),
\end{aligned}\;}
\end{equation}
with
\begin{equation}
\begin{gathered}
J(x)=\frac{x^2}{2}\,\psi'(x)-\frac x2-\frac14,\qquad L(a,b)=F(a,b)+F(b,a),\\
F(a,b)=\sum_{n\ge1}\frac{(-1)^n\,n}{n+a+b}\Big[\frac12-(2b+n)\,\tilde\beta(2b+n)\Big],
\end{gathered}
\end{equation}
and $\tilde\beta(x)=\half[\psi(\frac{x+1}2)-\psi(\frac x2)]$. At $\gs=0$ only $\beta_0=\ms$ survives, with $r_0=1/\ms$, and \eqref{eq:rootsums} reduces to the free sums. Since $r_j=O(j^{-3})$ the root sums converge rapidly, and they reproduce the direct evaluations to better than $10^{-10}$ (\cref{tab:roots}). Higher sums $\sum_n c_n^{-s}$ are $s$-fold root sums of traces of products of $s$ P\"oschl--Teller kernels.

\paragraph{Flat-space limit and FLZ.} The same formulas hold in flat space with the flat indicial roots. At $m^2=\gb^2$ the flat symbol is $\frac{\pi p}2\coth\frac{\pi p}2$, whose roots are the half-integers $\beta_j=j+\half$, with $r_j=4/\pi^2(2j+1)$. Then \eqref{eq:rootsums} gives
\begin{equation}
\sum_n\frac1{c_n^2}=\frac{16}{\pi^4}\sum_{N\ge1}\frac{J(N)}{N}\sum_{k=1}^N\frac1{2k-1} ,
\end{equation}
which we find equal to $(7\zeta(3)+2)/4\pi^4$ to $5\times10^{-13}$, i.e.\ $G_+^{(2)}+G_-^{(2)}=7\zeta(3)+2$ in FLZ normalization ($c=2\pi^2\lambda$). The parity-twisted sum reproduces $G_+^{(2)}-G_-^{(2)}=7\zeta(3)-2$ to better than $10^{-12}$ once the double sum is evaluated with convergence acceleration. The first alternating sum gives $G_+^{(1)}-G_-^{(1)}=2$, as in \cite{Fateev:2009jf}. The individual first sums diverge logarithmically in flat space and are regularization dependent.

The AdS$_2$ closed forms are therefore a deformation of FLZ's. The half-integers are replaced by the AdS$_2$ indicial roots, and the $\zeta$-values by the corresponding root sums. For $\alpha\neq0$ the flat roots are already transcendental, and the root sums play the role of the single integrals $\mathbf u_{2k-1}(\alpha)$ of LM \cite{Litvinov:2024riz}.

\begin{table}[t]
\centering\footnotesize
\begin{tabular}{cc cccc}
\toprule
$(\gs,\ms)$ & & $\sum1/c_n$ & $\sum(-1)^n/c_n$ & $\sum1/c_n^2$ & $\sum(-1)^n/c_n^2$\\
\midrule
$(1,\,0.5)$   & roots  & 0.5799030973 & 0.2836939429 & 0.1187146952 & 0.1068888562\\
              & direct & 0.5799030973 & 0.2836939429 & 0.1187146952 & 0.1068888562\\
$(3,\,0.3)$   & roots  & 0.5754052635 & 0.4047863304 & 0.1904710709 & 0.1867857279\\
              & direct & 0.5754052635 & 0.4047863304 & 0.1904710709 & 0.1867857279\\
$(0.3,\,0.8)$ & roots  & 0.5186305764 & 0.1582270223 & 0.0558442060 & 0.0397310794\\
              & direct & 0.5186305764 & 0.1582270223 & 0.0558442060 & 0.0397310794\\
\midrule
flat, $\alpha=0$ & roots &  & & 0.0267285071 & 0.0164625248\\
                 & FLZ   &  & & 0.0267285071 & 0.0164625248\\
\bottomrule
\end{tabular}
\caption{Spectral sums from the indicial roots \eqref{eq:rootsums} against direct evaluation by the integrals \eqref{eq:sum1}--\eqref{eq:sum2} and their parity-twisted analogue, including the large-$p$ tails. The indicial roots are, e.g., $\beta_0=0.237847$, $\beta_1=1.327423$ with $r_0=0.922849$, $r_1=0.107426$ at $(\gs,\ms)=(1,\half)$. The last two rows are the flat-space check against FLZ \cite{Fateev:2009jf}, $(7\zeta(3)\pm2)/4\pi^4$.}
\label{tab:roots}
\end{table}

\subsection{\texorpdfstring{Is the AdS$_2$ problem easier than the flat one?}{Is the AdS2 problem easier than the flat one?}}

We can now give a precise answer. The $\SLtwo$ symmetry reduces the problem to one variable and makes its free part exactly solvable, but it does not determine the interacting spectrum. In rapidity variables the interaction is translation invariant, while the curvature enters only through the $p^2/4$ term, which together with the ${\rm sech}^2\theta$ weight forms the P\"oschl--Teller problem of the free theory; no further symmetry relates the two. AdS$_2$ is not more integrable than flat space.

The integrable structure that makes the flat problem tractable does, however, survive the deformation largely intact: the IIKS form of the kernel, the TQ equation and its quantum Wronskian, the trace identities, and the ratio relation for the spectral determinants. The only property that fails is the derivative relation. In several respects the AdS$_2$ version is better behaved.
\begin{itemize}
\item It is trace class, so determinants and spectral zeta functions need no regularization.
\item Its determinants are ordinary Fredholm determinants, so the ratio relation holds without exponential factors.
\item There is an exactly solvable point, $\gs=0$, around which perturbation theory is regular.
\item The spectral sums are closed-form sums over indicial roots that interpolate smoothly between Hurwitz-type values at $\gs=0$ and FLZ's $\zeta$-values in the flat limit.
\item The pion, heavy-quark and large-$n$ limits are controlled analytically, and the semiclassics in rapidity is uniformly accurate and parity-resolved.
\end{itemize}
What the AdS$_2$ problem lacks is a point-value formula for the individual determinants. Whether a modified derivative relation exists, with transcendental rather than rational coefficients, is an open question.

\section{Numerical solution and method}
\label{sec:numerics}

\subsection{Method and checks}

We solve \eqref{eq:AdS-tHooft} by the Rayleigh--Ritz method, i.e.\ by expanding $\varphi$ in a finite basis with the exact endpoint behaviour and diagonalizing the matrix of the quadratic form \eqref{eq:form}:
\begin{equation}
\varphi(x)=\sum_{k=0}^{N_b-1}a_k\,\big[4x(1-x)\big]^{\beta_0}\,p_k(2x-1),
\end{equation}
where $\beta_0$ is the root of \eqref{eq:indicial} and $p_k$ are orthonormal with respect to the weight $(1-t^2)^{\beta_0}$ (Gegenbauer polynomials). The Legendre and mass terms are integrated exactly with Gauss--Jacobi quadrature. The singular kernel is handled through the finite Hilbert transform: its action on $(1-t^2)^{\beta_0}p_k(t)$ splits into an explicit non-analytic piece $\propto\cot(\pi\beta_0)$ and an analytic remainder. The remainder obeys the three-term recurrence of the $p_k$, initialized by a ${}_2F_1$ function. The details are in Appendix~\ref{app:numerics}. The resulting symmetric generalized eigenvalue problem converges rapidly. Already for $N_b=40$ the ground state is converged to $10^{-10}$ and the next few levels to $10^{-7}$, and the results are insensitive to the quadrature order once a Golub--Welsch rule is used (\cref{tab:conv}).

We have checked the solver in five independent ways.
\begin{enumerate}[label=(\alph*)]
\item At $\gs=0$ it reproduces the free spectrum $\Delta_n=1+2\ms+n$ to $10^{-11}$.
\item Dropping $\cL$ (the flat limit) and setting $m^2=\gb^2$, it reproduces the high-precision values of $2\lambda^{\rm FLZ}_n=\mu_n^2/\pi^2\gb^2$ from Fateev, Lukyanov and Zamolodchikov \cite{Fateev:2009jf} to better than $6\times10^{-13}$ for $n\le5$ with $N_b=240$ (\cref{tab:flat}). It also agrees with the five-digit tables of Litvinov and Meshcheriakov \cite{Litvinov:2024riz} at $m^2/\gb^2=\half$ and $\frac32$, to within one unit in the last quoted digit, and in the chiral limit it gives $\mu_n^2/\gb^2=0,\,5.8816,\,14.143,\,23.084,\,32.304$, consistent with the values $5.88$, $14.1$, $23.1$, $32.3$ in the literature \cite{Armoni:2000uw,Jia:2017uul}.
\item For $\beta_0=\half$, i.e.\ $\ms^2=\gs+\frac14$, the basis coincides with the Chebyshev basis $\sqrt{1-t^2}\,U_{k}(t)$, in which the hypersingular operator is known in closed form. The kernel matrices computed in the two ways agree to machine precision.
\item The weak-coupling result \eqref{eq:weak}, the flat-limit correction \eqref{eq:flat}, the GMOR relation \eqref{eq:GMOR} and the WKB formula \eqref{eq:WKB} are all reproduced (\cref{fig:flat,fig:pion,fig:largen}).
\item A completely independent solver, a Nystr\"om discretization (replacement of the integral by a quadrature sum) of the Birman--Schwinger operator \eqref{eq:BS} in rapidity-momentum space (Appendix~\ref{app:numerics}), reproduces the spectra to relative accuracy $10^{-10}$ (for example $c_0=2.999550854$ and $c_5=87.42464756$ at $(\gs,\ms)=(1,\half)$ in both methods). The exact sum rules of \cref{sec:rapidity} are satisfied (\cref{tab:sumrules}). The relaxed solutions of Appendix~\ref{sec:integrable} obey the quantum Wronskian to $10^{-14}$ and the trace identities to $10^{-12}$; the ratio theorem holds exactly even for the discretized operator, and the root sums \eqref{eq:rootsums} agree with direct evaluations (\cref{tab:ratio,tab:roots}).
\end{enumerate}

\subsection{Further details on the numerical method}
\label{app:numerics}

We work in $t=2x-1\in[-1,1]$ and use the basis $\phi_k(t)=w(t)\,p_k(t)$, $w=(1-t^2)^{\beta}$ with $\beta=\beta_0(\gs,\ms)$ the root of \eqref{eq:indicial}, and $p_k$ orthonormal with respect to $w$ (in this appendix $w$, $\phi_k$ and $M$ denote the weight, the basis functions and the mass matrix, unrelated to the symbols of \cref{sec:rapidity} and Appendix~\ref{sec:integrable}):
\begin{equation}
\begin{gathered}
t\,p_k=a_{k+1}p_{k+1}+a_kp_{k-1},\qquad a_k=\sqrt{\frac{k(k+2\beta)}{(2k+2\beta-1)(2k+2\beta+1)}},\\
p_0=\mu_0^{-1/2},\qquad \mu_0=\int_{-1}^1(1-t^2)^\beta\dd t=\frac{\sqrt\pi\,\Gamma(\beta+1)}{\Gamma(\beta+\frac32)} .
\end{gathered}
\end{equation}

\paragraph{Local terms.} With $\dd x=\half\dd t$ and $x(1-x)=\frac14(1-t^2)$, the Gram, Legendre and mass matrices are
\begin{equation}
G_{jk}=\tfrac12\!\int\!(1-t^2)^{2\beta}p_jp_k,\quad
\cL_{jk}=\tfrac12\!\int\!(1-t^2)^{2\beta-1}Q_jQ_k,\quad
M_{jk}=2\ms^2\!\int\!(1-t^2)^{2\beta-1}p_jp_k ,
\end{equation}
with $Q_k=(1-t^2)p_k'-2\beta tp_k$. These are polynomial integrals against Jacobi weights, evaluated exactly by Gauss--Jacobi quadrature.

\paragraph{Kernel.} Since $-\PV_0^1\dd x'\,\varphi(x')/(x-x')^2=2\,\frac{\dd}{\dd t}q[\varphi](t)$ with $q[\varphi](t)=\PV_{-1}^1\dd s\,\varphi(s)/(t-s)$, the $\gs$-dependent part of the Casimir has matrix elements
\begin{equation}
\mathcal{S}_{jk}=\int_{-1}^1\dd t\,\phi_j(t)\,q_k'(t)-2\int_{-1}^1\dd t\,(1-t^2)^{2\beta-1}p_jp_k,\qquad q_k\equiv q[\phi_k].
\end{equation}
The finite Hilbert transform of the weight is known in closed form \cite{Tricomi:1957,King:2009}:
\begin{equation}
\PV_{-1}^1\dd s\,\frac{(1-s^2)^\beta}{s-t}=\pi\cot(\pi\beta)\,(1-t^2)^\beta-c_\beta\;{}_2F_1\Big(1,-2\beta;1-\beta;\tfrac{1-t}{2}\Big),\qquad c_\beta=\frac{4^\beta\,\Gamma(\beta)\Gamma(\beta+1)}{\Gamma(2\beta+1)} .
\end{equation}
For $\beta=\half$ this gives $-\pi t$, and for $\beta\to0$ it gives $\log\frac{1-t}{1+t}$. Since $s\,p_k(s)=a_{k+1}p_{k+1}+a_kp_{k-1}$ and $\int w\,p_k=\sqrt{\mu_0}\,\delta_{k0}$, the functions $q_k$ obey the inhomogeneous recurrence
\begin{equation}
a_{k+1}q_{k+1}=t\,q_k-a_kq_{k-1}-\sqrt{\mu_0}\,\delta_{k0}.
\end{equation}
We split $q_k=-\pi\cot(\pi\beta)\,w\,p_k+r_k$. Near $t=+1$ the first term carries the entire non-analytic behaviour, and $r_k$ is analytic there. The $r_k$ obey the same recurrence with $r_0=c_\beta\,{}_2F_1(1,-2\beta;1-\beta;\frac{1-t}2)/\sqrt{\mu_0}$, and differentiating the recurrence gives $r_k'$. Using parity, $\phi_j(-t)=(-1)^j\phi_j(t)$ and $q_k'(-t)=(-1)^kq_k'(t)$, we write $\int_{-1}^1\phi_jq_k'=(1+(-1)^{j+k})\int_0^1\phi_jq_k'$. On $[0,1]$ we use Gauss--Jacobi quadrature with weight $(1-t)^{2\beta-1}$ for the singular piece, whose integrand is $(1-t^2)^{2\beta-1}$ times a polynomial, and with weight $(1-t)^\beta$ for $\int_0^1w\,p_j\,r_k'$, whose integrand is $(1-t)^\beta$ times an analytic function. We have checked $q_k$ and $q_k'$ against direct Cauchy-principal-value quadrature, and the kernel matrix is symmetric to $10^{-11}$ before symmetrization.

For small $\beta$ the Gauss--Jacobi exponent $2\beta-1$ approaches $-1$, and library implementations of the rule can lose accuracy. We therefore build the rule with the Golub--Welsch algorithm \cite{Golub:1969}. With this rule the results are independent of the quadrature order to all digits shown in \cref{tab:conv}. The forward recurrences are stable on the interval. Highly excited levels converge more slowly at weak coupling: for the spectra of \cref{fig:largen} ($N_b=420$, $n\le250$) the absolute error, estimated by comparison with $N_b=560$, is below $10^{-3}$ at $\gs=0.3$, below $10^{-4}$ at $\gs=1$ and below $10^{-5}$ for $\gs\ge3$, far below plotting accuracy.

\begin{table}[ht]
\centering\small
\begin{tabular}{cc c cccc}
\toprule
$(\gs,\ms)$ & $N_b$ & quad. & $\Delta_0$ & $\Delta_1$ & $\Delta_2$ & $\Delta_3$\\
\midrule
$(2,\,0.4)$ & 40 & 600  & 2.2429424265 & 4.9559982969 & 6.9943366100 & 8.7487427105\\
            & 60 & 800  & 2.2429424265 & 4.9559982939 & 6.9943365895 & 8.7487426537\\
            & 80 & 1600 & 2.2429424265 & 4.9559982936 & 6.9943365875 & 8.7487426480\\
\midrule
$(10,\,0.05)$ & 40 & 600 & 1.4171776364 & 8.4042681524 & 12.7375557513 & 16.1984833760\\
              & 60 & 800 & 1.4171776364 & 8.4042681508 & 12.7375557383 & 16.1984833404\\
              & 80 & 1600 & 1.4171776364 & 8.4042681507 & 12.7375557372 & 16.1984833371\\
\bottomrule
\end{tabular}
\caption{Convergence in the basis size $N_b$ and the quadrature order. For $(\gs,\ms)=(10,0.05)$ the endpoint exponent is $\beta_0=0.00859$.}
\label{tab:conv}
\end{table}

\paragraph{Rapidity-momentum Nystr\"om method.} As an independent check we discretize the Birman--Schwinger problem \eqref{eq:BS} in the momentum $p$ conjugate to the rapidity. With $\Psi=B^{1/2}\Phi$, the numbers $1/c_n$ are the eigenvalues of the symmetric kernel
\begin{equation}
\widetilde{\BS}(p,k)=\frac{1}{2\pi}\,B(p)^{-1/2}\,\widehat W(p-k)\,B(k)^{-1/2},\qquad \widehat W(q)=\frac{\pi q}{4\sinh(\pi q/2)} ,
\end{equation}
which is the operator $\widetilde{\BS}$ of Appendix~\ref{sec:integrable}. The kernel is analytic in a strip of half-width $2\beta_0$ and decays exponentially away from the diagonal. A trapezoidal rule on $|p|\le\pmax$ therefore converges exponentially in the step size, and at $\pmax=80$, $h=0.025$ the low-lying $c_n$ agree with the Jacobi solver to $10^{-10}$--$10^{-11}$, except for very small $\beta_0$, where the strip is narrow. This method knows nothing about the Hilbert-transform recurrences or the endpoint basis, so the agreement is an independent test of the numerics and of the rapidity representation of \cref{sec:rapidity}. It also handles $\ms>1$ at weak coupling, where the Jacobi basis converges slowly because the solution's dominant endpoint exponent $\ms$ differs from the leading root $\beta_0<1$.

\paragraph{Relaxed solutions, determinants and indicial roots.} The relaxed problems \eqref{eq:relaxed} are solved with the same Nystr\"om matrix, $F_j=(1-c\widetilde{\BS})^{-1}f_j$, on $|p|\le60$ with $h=0.05$ for the checks of the quantum Wronskian and of the trace identities, and on finer grids (up to $\pmax=80$, $h=0.02$) for the converged values quoted in \cref{tab:ratio}. Off the real axis, $\mathcal Q_j(p)$ is obtained from the Fourier representation in \eqref{eq:Qdef}, either directly for $|{\rm Im}\,p|<2$ or, near ${\rm Im}\,p=2$, by first reconstructing $\varphi_j(\theta)$ and integrating in $\theta$. The determinants $D_\pm(c)$ are products over the parity-resolved Nystr\"om eigenvalues. The untwisted trace identity is evaluated with $\mathcal Q_j'$ from the differentiated Fourier representation in \eqref{eq:Qdef}, and the parity-twisted one on a grid offset by $h/2$, which avoids the point $p=0$ where $\tanh(\pi p/2)$ vanishes. Direct evaluations of $\sum_nc_n^{-2}$ on $|p|\le\pmax$ include the large-$p$ tail $\frac1{24\pi}\int_{|p|>\pmax}\dd p\,B^{-2}$, which follows from $\int\dd q\,\widehat W(q)^2=\pi/6$. The indicial roots are computed from $\cot\pi\delta=(s^2-\ms^2+\gs)/(\pi\gs s)$ with $s=j+\delta$, by bracketing for $j\le400$ and by fixed-point iteration beyond; $r_j=2/|I'(\beta_j)|$, with tails estimated from $r_j\simeq2\gs/j^3$. In the series for $F(a,b)$ the combination $\half-x\tilde\beta(x)=-\frac14\int_0^\infty e^{-xt}{\rm sech}^2(t/2)\,\dd t$ is evaluated from its asymptotic expansion for $x>40$ to avoid cancellations, and the alternating sums are accelerated by averaging consecutive partial sums.

\paragraph{Complex scaling for dS$_2$.} For the continued problem \eqref{eq:ds-continuation} we use the same Nystr\"om discretization on the rotated contour $p=s\,e^{\ii\phi}$, $s\in[-\pmax,\pmax]$:
\begin{equation}
B_\alpha(z_i)\,\Phi_i=c\sum_j\frac{w_j\,e^{\ii\phi}}{2\pi}\,\widehat W(z_i-z_j)\,\Phi_j,\qquad z_j=s_j\,e^{\ii\phi},
\end{equation}
symmetrized with $B_\alpha^{-1/2}$ and $w_j^{1/2}$ as above, which gives a complex-symmetric matrix whose eigenvalues are $1/c_n$. The rotation is legitimate because $\widehat W(q)$ and $h(q)$ are analytic in the sectors $|\arg q|<\pi/2$ and $|\arg(-q)|<\pi/2$, and $B_\alpha(s\,e^{\ii\phi})$ has no zeros on the rotated contour, nor in the wedge it sweeps out, for the angles used; we have checked this with the argument principle for $0\le\alpha\le\pi$ (for $B_{\rm dS}$ the zeros we find lie on the real and imaginary axes). We use uniform grids with $\pmax=60$--$80$ and $h=0.04$--$0.05$, and $\phi=0.3$ and $0.45$, and for the pion a $\sinh$-mapped grid $s=L\sinh u$ with $L=0.01$ and step $0.02$ in $u$, which resolves the small endpoint exponent $\beta_0\propto\ms$. Checks: at $\alpha=0$ the AdS$_2$ spectrum is reproduced for any $\phi$; at $\gs=0$, $\alpha=\pi$ the exact values \eqref{eq:ds-free} are reproduced to $10^{-10}$; and at $\alpha=\pi$ the eigenvalues are independent of $\phi$ and of the grid to the digits quoted in \cref{tab:ds}. Spurious eigenvalues, which come from the truncation of the contour, move with $\phi$ and are discarded.

\section{\texorpdfstring{The light-front vacuum on dS$_2$}{The light-front vacuum on dS2}}
\label{app:ds-vacuum}

In \cref{sec:ds-halfspace} we assumed that the light-front Fock vacuum $|0\rangle$, the state with no quarks on the null line, is the de~Sitter-invariant vacuum of the interacting planar theory. In this appendix we collect what can be said about this assumption. In physical terms the question is whether the expansion, which enters the light-front Hamiltonian only through the special-conformal generator $\cK$, pulls quark--antiquark pairs out of the vacuum. We show that it cannot create any pair with non-zero light-front momentum, and we explain why the remaining possibility, an admixture with exactly vanishing light-front momentum, is not settled by our analysis. The analysis is therefore \emph{not complete}: it does not treat the light-front zero-mode sector, which is also where the chiral condensate is encoded, and which we leave for future work.

\paragraph{What has to be shown.} The vacuum must be annihilated by the three charges \eqref{eq:ds-charges}, $H=\Htf-\cK/4R^2$, $D=-\cD$ and $K=-4R^2\cP$. For $K$ and $D$ this holds as in AdS$_2$ (\cref{sec:vacuum}): $K$ is the total light-front momentum, and the pair-creation part of $D$ has no overlap with any normalizable two-particle state. For $\Htf$ it holds as in flat space, up to zero modes, because every term of $\Htf$ conserves the total light-front momentum. Everything therefore hinges on $\cK|0\rangle$. The pair-creation part of $\cK$ is the distribution \eqref{eq:pair-kernel}, which is supported at the single point $k=k'=0$ of the two-particle momentum space. The special-conformal generator can thus only create pairs with exactly vanishing total light-front momentum, and the question is whether this distribution vanishes as an element of the Hilbert space, i.e.\ whether it has zero overlap with all physical two-particle states.

\paragraph{An algebraic constraint.} Suppose that $K|0\rangle=D|0\rangle=0$ and write $\chi\equiv H|0\rangle$. The algebra \eqref{eq:sl2} gives
\begin{equation}\label{eq:vac-alg}
K\chi=-[H,K]|0\rangle=-2\ii D|0\rangle=0,\qquad D\chi=-[H,D]|0\rangle=-\ii\chi .
\end{equation}
The state $\chi$ would therefore carry zero light-front momentum and be an eigenvector of the dilatation $D$ with the imaginary eigenvalue $-\ii$. On the light-front Fock space $D$ is self-adjoint, so for a normalizable $\chi$ the expectation value $\langle\chi|D|\chi\rangle=-\ii\|\chi\|^2$ must be real, which forces $\chi=0$. Hence $H|0\rangle$ is either zero or a non-normalizable object concentrated at zero light-front momentum. In words, \emph{the evolution cannot produce any normalizable state from the vacuum}: no meson, no wave packet of separated quarks, and no pair with non-zero light-front momentum. The explicit kernel \eqref{eq:pair-kernel} is consistent with this: it is supported at zero momentum and scales under dilatations exactly as an eigenvector with eigenvalue $-\ii$. The argument uses only the isometry algebra and the fact that $K$ and $D$ are kinematic, so it holds on AdS$_2$ and on dS$_2$, at any coupling.

\paragraph{The remaining question is coupling independent.} Since $\cK$ is quadratic in the fields and independent of the coupling and of the quark mass, its matrix element $\langle\psi|\cK|0\rangle$ with any two-particle Fock vector $\psi$ is the same in the free and in the interacting theory. What the interaction changes is only which states are physical, and in particular how their wavefunctions behave at small total momentum $P$. For a state whose centre-of-mass wavefunction behaves as $P^a$, the argument of \cref{sec:vacuum} shows that the overlap with $\cK|0\rangle$ vanishes for $a>0$, is a boundary term at $a=0$, and is ill-defined for $a<0$. On AdS$_2$ the physical states have $a=\Delta-1\ge0$, and the only borderline state, the pion, has vanishing overlap. On dS$_2$ two kinds of states have $a\le0$.
\begin{itemize}
\item \emph{The exact $\Delta=1$ states} \eqref{eq:ds-D1}, for which $a=0$. Evaluated naively, with a symmetric cutoff as in \cref{sec:vacuum}, the boundary term is proportional to $\int\dd x\,(1-2x)\varphi$, which vanishes for the charge-conjugation even combination $\cos(p_*\theta)$ but not for the odd one, $\int\dd x\,(1-2x)\sin(p_*\theta)=-\pi p_*^2/4\sinh(\pi p_*/2)$. However, these are literally the same Fock vectors as the free $\Delta=1$ states of quarks of mass $\ms'=p_*/2$, and for free fields the light-front Fock vacuum is the Bunch--Davies state, which is de~Sitter invariant \cite{Kay:1988mu}. The non-vanishing boundary term is therefore an artefact of the naive treatment of the point $P=0$, and since both the operator and the state are the same, the same conclusion holds at $\gs>0$.
\item \emph{The continuum of dissociated pairs}, which carries principal-series representations with ${\rm Re}\,\Delta=\half$. Wave packets in this continuum behave as $P^{-1/2}$ at small momentum, so $a<0$ and their overlap with $\cK|0\rangle$ depends on how the states are completed at $P=0$. This requires information about what enters the light front from $\mathcal I^-$ and leaves it through $\mathcal I^+$, i.e.\ both signs of $K$, which a single null line does not provide (\cref{sec:ds-halfspace}). This is the part of the problem that our analysis leaves open. (The resonances themselves are not normalizable and are not states of the Hilbert space.)
\end{itemize}

\paragraph{Expectation.} Any de~Sitter-invariant state, restricted to the null line, must be annihilated by $K$ and $D$, which fixes it to be the Fock vacuum up to zero-momentum (zero-mode) content. The constraint \eqref{eq:vac-alg} shows that the evolution cannot change this by creating normalizable pairs, and the coupling independence of $\cK$ shows that the free-field result carries over to every normalizable state with ${\rm Re}\,\Delta\ge1$. We therefore expect the light-front Fock vacuum to be the de~Sitter-invariant vacuum of the interacting planar theory, with the zero-mode sector as the only open issue, exactly as in flat space. In this picture the expansion does not pull pairs out of the vacuum. Particle production in the invariant state is observer dependent: a static-patch observer sees a thermal bath at $T_{\rm dS}$, which on the light front is the thermal property of the vacuum restricted to a horizon half-line (\cref{sec:ds-halfspace}). The planar effect of the expansion is instead the dissociation of the mesons computed in \cref{sec:ds-mesons}. The zero-mode sector does respond to the curvature: the chiral condensate is enhanced on AdS$_2$ and suppressed on dS$_2$ (\cref{sec:special,sec:ds-pion}), without any pair creation, as in flat space.

\paragraph{Relation to infrared effects in de~Sitter space.} The open question is a form of the infrared problem of quantum field theory in de~Sitter space. Small light-front momentum means $|\lambda|\to\infty$, i.e.\ $\mathcal I^\pm$, so the light-front zero-mode sector is the image of the late-time, super-horizon region where infrared effects in de~Sitter space arise. Two such effects are relevant here.
\begin{itemize}
\item \emph{Light scalars.} A massless minimally coupled scalar has no de~Sitter-invariant Fock vacuum, because of its zero mode \cite{Allen:1985ux}, and the late-time fluctuations of a light scalar are enhanced by the inverse of its mass squared. The quarks themselves are immune: a massive Dirac fermion on dS$_2$ is always in the principal series \cite{Pethybridge:2021rwf,Letsios:2025pqo}, and there are no light fermions. The only light field is the pion. In the chiral limit it is massless ($\Delta_\pi=1$), but as the phase of the condensate it is a compact scalar, so its zero mode is harmless (\cref{sec:ds-pion}); for small quark mass and $\gs>\gs_c$ it is light, with $1-\Delta_\pi\ll1$. At $N=\infty$ the pion is free and these fluctuations play no role. At finite $N$ we expect its late-time fluctuations, enhanced by $1/(1-\Delta_\pi)$ and suppressed by $1/N$, to smear the phase of the condensate, a de~Sitter counterpart of Coleman's theorem \cite{Coleman:1973ci,Witten:1978qu}, so that a condensate requires $N(1-\Delta_\pi)\gg1$. We have not analysed this.
\item \emph{Interacting massive fields.} Whether interactions destabilize the de~Sitter-invariant vacuum through secular particle production has been debated \cite{Tsamis:1994ca,Polyakov:2007mm,Krotov:2010ma,Akhmedov:2013vka}; for massive fields the stability of the Euclidean (Bunch--Davies) vacuum has been established in perturbation theory \cite{Marolf:2010nz,Hollands:2010pr}. The constraint \eqref{eq:vac-alg} is the light-front counterpart of the statement that de~Sitter invariance forbids such production of normalizable states, and the zero-mode sector is where an infrared effect of this kind would have to reside. Since the planar sector contains no light fields other than the pion, we expect the situation to be as for massive fields.
\end{itemize}

\end{document}